\documentclass[twocolumn,prd,aps,tightenlines,floats,floatfix,eqsecnum,preprintnumbers,nofootinbib]{revtex4}

\def\sp{\kern +3pt}
\def\sm{\kern -3pt}
\def\spQ{\kern +6pt}

\def\bea{\begin{eqnarray}}
\def\eea{\end{eqnarray}}

\def\sfrac#1#2{{\textstyle \frac{#1}{#2}}}

\def\be{\begin{equation}}
\def\ee{\end{equation}}
\def\ba{\begin{eqnarray}}
\def\ea{\end{eqnarray}}

\usepackage{graphics}
\usepackage{graphicx}
\usepackage{epsf}
\usepackage{amsmath}
\usepackage{amssymb}

\usepackage{xcolor}

\begin{document}

\phantom{0}
\vspace{-0.2in}
\hspace{5.5in}

\preprint{{\bf  LFTC-20-2/54}}

\vspace{-1in}

\title
{\bf Covariant model for the Dalitz decay of the $N(1535)$ resonance}
\author{G.~Ramalho$^1$ and 
M.~T.~Pe\~na$^{2,3}$}
\vspace{-0.1in}

\affiliation{$^1$Laborat\'orio de 
F\'{i}sica Te\'orica e Computacional -- LFTC, \\
Universidade Cruzeiro do Sul and Universidade Cidade de S\~ao Paulo, \\
01506-000, S\~ao Paulo, SP, Brazil
\vspace{-0.15in}}
\affiliation{$^2$LIP, Laborat\'orio de Instrumenta\c{c}\~ao e F\'{i}sica 
Experimental de Part\'{i}culas, 
 Avenida Professor Gama Pinto, 1649-003 Lisboa, Portugal
\vspace{-0.15in}}
\affiliation{$^3$Instituto Superior T\'ecnico (IST),
Universidade de Lisboa,
Avenida Rovisco Pais, 1049-001 Lisboa, Portugal}

\vspace{0.2in}
\date{\today}

\phantom{0}

\begin{abstract}
We develop a covariant model for the $\gamma^\ast N \to N(1535)$
transition in the timelike kinematical region, the region where the square momentum transfer $q^2$ is positive.
Our starting point is the covariant spectator quark model
constrained by data in the spacelike kinematical region \mbox{($Q^2 = -q^2 >0$).}
The model is used 
to estimate the contributions of valence quarks
to the transition form factors, and one obtains
a fair description of the Dirac form 
factor at intermediate and large $Q^2$.
For the Pauli form factor there is evidence 
that beyond the quark-core contributions there are also significant contributions 
of meson cloud effects.
Combining the quark-core model 
with an effective description of the meson cloud effects, 
we derive a parametrization of the spacelike data 
that can be extended covariantly to the timelike region.
This extension enabled us to estimate the Dalitz decay widths of the 
$N(1535)$ resonance, among other observables.
Our calculations can help in the interpretation 
of the present experiments at HADES 
($pp$ collisions and others).
\end{abstract}

\vspace*{0.9in}  
\maketitle

\section{Introduction}
\label{secIntro}

The creation and propagation of intermediate nucleonic excitations or $N^\ast$ states, 
followed by virtual photon transitions leading to $N^\ast \to  \gamma^\ast N \to e^+ e^- N$ 
decays~\cite{Timelike,Timelike2,N1520TL,HADES17a,ColeTL,Ramstein18a,MesonBeams,Weil12,Bratkovskaya99,Faessler03,NSTAR} 
can be probed with data on di-electron production from proton-proton ($pp$) 
and proton-nucleus ($pA$) collisions, as well as on inclusive and exclusive pion-nucleus reactions 
provided by secondary pion beams experiments.
Those experiments by the HADES Collaboration 
at GSI~\cite{HADES17a,Ramstein18a,Ramstein18b,HADES-dilepton,HADES-pion,HADES17b}
expand the information from electron-scattering experiments to electromagnetic decay rates,
and provide knowledge on momentum evolution of the electromagnetic couplings of nucleon excitations. 
An example was the recently extraction of the $\Delta(1232)$ Dalitz decay branching ratio by the 
HADES collaboration~\cite{HADES17a}.

Small $Q^2$ photon virtualities 
are sensitive to the more peripherical structure of baryons, 
and may improve the description of several resonances electro-couplings 
based on quark-core constituents alone. 
In theory, general unitarity requirements impose meson-baryon contributions to 
the electromagnetic excitation and decay of the baryons, but in practice 
how to combine both meson-baryon 
and quark-core regimes in electromagnetic reactions is a current challenge of Hadron Physics.
Lattice QCD calculations of transition form factors are not yet available, 
except for a few baryons and when they will be provided, 
the separation of both effects has to rely on models.

The $\gamma^\ast N \to N(1535)$ transition in particular is still a very perplexing
transition from the theoretical point of view,
since at the moment there are no models 
that describe the measured  transverse and longitudinal helicity transition amplitudes 
$A_{1/2}$ and $S_{1/2}$ in the full range of $Q^2$.
In this work it is more convenient to 
discuss directly the transition form factors 
Dirac ($F_1^\ast$) and Pauli  ($F_2^\ast$), 
which can be written as linear combinations of the helicity amplitudes.

Quark models give a partial description of the 
Dirac form factor~\cite{N1535,S11b,S11c,Diaz09},
suggesting that it is dominated 
by valence quark degrees of freedom.
However,  calculations based on chiral models, 
where the baryon states are 
dynamically generated by baryon-meson resonances, 
suggest that the Pauli form factor at low $Q^2$,
is dominated by meson cloud effects~\cite{Jido08,S11b,S11c}.
In addition, results based on a light-front relativistic quark model
indicate that the meson cloud contributions 
to the $\gamma^\ast N \to N(1535)$ transition have 
an isovector character~\cite{Aznauryan17}. 
The transition form factors have also been 
calculated using light cone sum rules, 
based on the distribution amplitudes determined by 
lattice QCD~\cite{Braun09,Anikin15a}.
There are also recent calculations 
based on coupled-channel models~\cite{Burkert04,Kamano16,Liu16},
light-front quark models~\cite{Gutsche19} and AdS/QCD~\cite{Bayona12,Obukhovsky19}.

In this work, we apply the covariant spectator quark model~\cite{Nucleon,Omega,NSTAR-review} 
to this problem, since it has advantageous specific features,  
namely vector dominance of the quark electromagnetic current,
enabling us to consistently expand   
calculations probed in the spacelike regime to the timelike region.
The covariant spectator quark model was tested in the description 
of other resonance sectors~\cite{NSTAR-review,SemiRel,N1535,N1520,SQTM,NDelta,Roper,Delta1600},
and it is here used to describe the $\gamma^\ast N \to N(1535)$ transition in the kinematic region of 
di-electron production, the timelike region,
and we calculate for the first time the Dalitz decay widths in terms 
of the energy of the resonance $W$.
The results for the $N(1535)$ Dalitz decay 
are then compared to the Dalitz decays results
for other resonances~\cite{Timelike,Timelike2,N1520TL}.

In Ref.~\cite{N1535}, we presented the first results for this resonance in the spacelike regime.
However, in that work
the validity of our valence quark model 
was limited to the $Q^2> 2$ GeV$^2$ region.
As meson cloud effects are naturally more important in the vicinity of the $Q^2=0$ point, 
the search for those effects requires that this restriction is lifted --- 
which is an important objective 
accomplished in this work.

The restriction to the large $Q^2$ region 
was a consequence
of the difficulty of the covariant spectator quark model 
in defining a covariant wave function of the 
$N(1535)$ compatible with the orthogonality of the states, 
and with a gauge invariant transition current.
This happens because the baryon wave functions that 
we use are constructed by using symmetries alone, 
and not obtained from a dynamical calculation.
In the $Q^2= -q^2  \to 0 $ limit,  
because of the difference of masses between the initial and final baryons, covariance makes 
the three-vector ${\bold q}$  non-zero, 
and the initial and final state become non-orthogonal (in the non-relativistic sense). 
The consequence of the non-orthogonality is that 
the transition current violates gauge invariance
and consequently the transition form factors ($F_1^\ast$ and $F_2^\ast$) 
are not well defined.
In these conditions the helicity amplitudes are also not well defined at low $Q^2$, 
and our estimates cannot be compared with experimental data.
For large $Q^2$, the impact of the gauge invariance breaking 
is small and the form factors and helicity amplitudes 
can be computed without restrictions~\cite{N1535}.

To fix the problem above, we treat here the baryon 
transitions within what  
we call the {\it semirelativistic approximation}, 
introduced in Ref.~\cite{SemiRel} and seen 
to be compatible with the construction 
of the wave function from symmetry principles alone.
This approach allows us to obtain 
the correct behavior of the form factors and of the
helicity amplitudes and to satisfy gauge invariance exactly.
Similarly to what is done in 
heavy-baryon chiral perturbation theory~\cite{Jenkins91},
the mass difference between the baryons is neglected in a first approximation, 
such that the orthogonality of the wave functions in the non-relativistic 
sense is preserved, while 
the covariance of the model is kept at the same time. 
Notice that the mass difference is not neglected in the kinematic factors 
in the formulas of the helicity amplitudes 
as combinations of the transition form factors.

Not only the analytic expressions for the transition form factors
are simpler when we consider the semirelativistic approximation~\cite{SemiRel}, 
but also in that approximation the radial wave function of the resonance ($\psi_R$)
can be taken with the same form of the wave function of the nucleon ($\psi_N$) 
without destroying gauge invariance.
Then the  only input into our model is the parametrization of the quark form factors 
and of the nucleon radial wave function, both determined in the study of the nucleon
electromagnetic structure~\cite{Nucleon}.

In this work, we conclude that the 
contributions of the valence quarks degrees of freedom 
are insufficient to describe the two transition form factors 
in the range $Q^2=0$--4 GeV$^2$.
Therefore we extracted also some phenomenological parametrizations
of the meson cloud contributions for the transition form factors.
Those contributions are seen to be negligible when
compared with the valence quark contributions at large $Q^2$. Also,
although the meson cloud contributions seem to be dominated 
by the isovector component, 
we tested the role of a non negligible contribution 
from the isoscalar component.

The first part of this article  
includes the calibration 
of the meson cloud contribution by the physical data 
in the spacelike region.
With the valence quark and the meson cloud 
contributions fixed in the spacelike domain, 
we proceed in the second part 
to perform their extension to the timelike region.
We present results for the two isospin cases, i.e., reactions 
with proton or neutron targets,
for which HADES experimental data can be provided.

This article is organized as follows:
In the next section we review the formalism 
associated with the $\gamma^\ast N \to N(1535)$ transition.
The covariant spectator quark model and 
the theoretical expressions for the transition 
form factors are presented in Sec.~\ref{secCSQM}.
In Sec.~\ref{secTimelike}, we present the results 
of the extension of our model to the timelike region.
The formalism associated with  $N(1535)$ Dalitz decay
is given in Sec.~\ref{secDalitz}.
The numerical results related to the $N(1535)$ Dalitz decay
are presented in  Sec.~\ref{secResults}.
Outlook and conclusions are given in  Sec.~\ref{secConclusions}.
Additional information is included in the appendices.

\section{$\gamma^\ast N \to N(1535)$ transition}
\label{secN1535}

We present here the different parametrizations 
of the electromagnetic structure 
between  a state $J^P= \frac{1}{2}^+$
(spin 1/2, positive parity),
and a resonance $J^P= \frac{1}{2}^-$
(spin 1/2, negative parity).

The $\gamma^\ast N \to N(1535)$ transition current 
can be written, in units of elementary charge ($e$), as~\cite{N1535,Aznauryan12a}
\ba
J^\mu = \bar u_R
\left[F_1^\ast 
\left( \gamma^\mu -  \frac{{\not \! q} q^\mu}{q^2}
\right)
 + F_2^\ast \frac{i \sigma^{\mu \nu} q_\mu}{M_R + M_N} \right] \gamma_5 u_N,
\nonumber \\
\label{eqJS11}
\ea
where $u_R$ and $u_N$ are the resonance and nucleon spinors, respectively,
and $M_R$ and $M_N$ are the masses of the resonance 
and the nucleon, respectively.
Equation (\ref{eqJS11}) defines the elementary form factors,
Dirac ($F_1^\ast$) and Pauli ($F_2^\ast$)~\cite{NSTAR,SemiRel,N1535}.
Due to gauge invariance, we can conclude 
that $F_1^\ast \propto Q^2$ near $Q^2=0$~\cite{Aznauryan12a,Compton} 
(a simple way to see this is to notice that
the $\frac{{\not q} q^\mu}{q^2}$ term in (\ref{eqJS11}) 
would not be finite unless $F_1^\ast \propto q^2$).
In the calculations, we distinguish between the form factors 
of the proton and neutron targets.

The empirical data associated 
with the electromagnetic structure of 
the $\gamma^\ast N \to N(1535)$ transition
are usually represented in terms of the 
helicity amplitudes in the resonance rest frame.
In this frame
the momentum transfer is
\ba
q= \left(\frac{M_R^2 -M_N^2 -Q^2}{2 M_R}, {\bf q} \right).
\ea 
Here ${\bf q}$ is the photon three-momentum, with magnitude
\ba
|{\bf q}| = \frac{\sqrt{Q_+^2 Q_-^2}}{2M_R},
\label{eqQR}
\ea
with
\ba
Q_\pm^2 &=& (M_R \pm M_N)^2 + Q^2 \nonumber \\  
       &=& (M_R \pm M_N)^2 - q^2.
\label{eqQpm}
\ea
 
Since the magnitude of the photon three-momentum $|{\bf q}|$ 
is non-negative by construction,
the analysis of the helicity amplitudes and 
transition form factors is restricted to the region 
$Q_-^2 \ge 0$, or equivalently $q^2 \le (M_R -M_N)^2$. 
The point $q^2=   (M_R -M_N)^2$, when $Q_-^2=0$, 
is usually referred to as the pseudothreshold~\cite{Devenish76,Siegert1,Siegert4}.
Experiments based on electron-nucleon scattering probe 
only the spacelike region ($Q^2 \ge 0$)~\cite{MesonBeams,NSTAR,ColeTL}.

The explicit forms for the transverse ($A_{1/2}$) 
and longitudinal ($S_{1/2}$) amplitudes 
in the resonance rest frame are~\cite{N1535,Siegert1,Aznauryan12a,Note1}:
\ba
A_{1/2} &= & {\cal B} \left[ F_1^\ast + \eta F_2^\ast \right], 
\label{eqA12s}\\
S_{1/2} &= &- 
\frac{ {\cal B}}{\sqrt{2}}  (M_R + M_N) \frac{|{\bf q}|}{Q^2} 
\nonumber \\ 
 & & \times \left[\eta F_1^\ast  - \tau F_2^\ast\right],
\label{eqS12s}
\ea
where 
${\cal B} = \frac{e}{2} \sqrt{\frac{Q_+^2}{M_N M_R K}}$,
$K=\frac{M_R^2-M_N^2}{2M_R}$, 
$\eta = \frac{M_R -M_N}{M_R + M_N}$,
and $e$ is the elementary electric charge 
($\alpha \equiv \frac{e^2}{4\pi}\simeq 1/137$).
The amplitudes for the proton targets are
represented by $A_{1/2}^p$, $S_{1/2}^p$;
the amplitudes associated with neutron 
targets are represented by $A_{1/2}^n$, $S_{1/2}^n$.

For the calculations in the timelike region (Sec.~\ref{secDalitz}),
it is convenient to introduce 
the electric ($G_E$) and Coulomb ($G_C$) 
transition form factors:
\ba
& &
G_E =  F_1^\ast + \eta F_2^\ast, \label{eqGE0} \\
& &
G_C = 
-\frac{M_R}{2}  \frac{(M_R+N_N)}{Q^2} 
\left[\eta F_1^\ast  - \tau F_2^\ast\right].
\label{eqGC0}
\ea
The previous definitions 
of $G_E$ and $G_C$ are non-standard, and differ 
from other forms in the literature, 
by multiplicative factors~\cite{Devenish76,Krivoruchenko02}.
The conversion to alternative representations 
is presented in Appendix~\ref{appS11}.

The form factors $G_E$ and $G_C$ are related to the 
helicity amplitudes from Eqs.~(\ref{eqA12s})--(\ref{eqS12s}) by 
\ba
G_E = \frac{1}{\cal B} A_{1/2},
\hspace{.7cm}
G_C = \frac{1}{\sqrt{2}{\cal B}} \frac{M_R}{|{\bf q}|}S_{1/2},
\label{eqGC00}
\ea
and have the advantage of being dimensionless,
contrarily to other definitions~\cite{Devenish76,Krivoruchenko02}.
Details related to the $\gamma^\ast N  \to N(1535)$
form factors and the helicity amplitudes 
are presented in Appendix~\ref{appS11}.

The available data for the $\gamma^\ast N  \to N(1535)$ transition 
for the amplitudes $A_{1/2}$ and $S_{1/2}$
are mainly from CLAS at JLab~\cite{CLAS1}.
For large $Q^2$ ($Q^2 > 5$ GeV$^2$) 
there are measurements of the $A_{1/2}$ amplitude 
(neglecting the effect of $S_{1/2}$)
from  JLab/Hall C~\cite{Dalton09}.
There are also some estimates of the helicity amplitudes 
from MAID~\cite{MAID1,MAID2} based on data from different experiments 
(including CLAS).
Our calculations are preferentially compared  
with the CLAS data,
well distributed in the range $Q^2=0$--4 GeV$^2$,  
and Particle Data Group (PDG) at $Q^2=0$~\cite{PDG}.

\subsection*{Brief review of the literature}

There are estimates of the valence quark 
contributions to the  $\gamma^\ast N  \to N(1535)$ form factors
based on the EBAC/Argonne-Osaka
coupled-channel dynamical model~\cite{Diaz09,Burkert04,Kamano16}.
The hybrid structure (baryon core combined to meson cloud) 
of the  $N(1535)$ is also supported 
by Hamiltonian Field Theory applications to 
lattice QCD simulations~\cite{Liu16}.

The results from EBAC~\cite{Diaz09} are very close to the 
valence quark estimates based on the 
covariant quark model~\cite{N1535,SemiRel}. 
However, there is some evidence that the $\gamma^\ast N \to N(1535)$
transition form factors at low $Q^2$
cannot be described only on the basis of the valence quark structure,
as discussed in Ref.~\cite{SemiRel}.
Calculations based on the chiral unitary model~\cite{Jido08}, 
which use meson-baryon resonance states as effective degrees of freedom, 
also indicate that the meson cloud effects can be significant, in 
particular to $F_2^\ast$.  
Those calculations show that the meson cloud  
contributions are comparable in magnitude to
the estimates from the covariant spectator quark model 
but differ in sign~\cite{S11c}.
This result provides a possible explanation 
to the small magnitude of the experimental data for $F_2^\ast$, for $Q^2 > 2$ GeV$^2$,
as discussed in the following sections (see also Ref.~\cite{S11b}).

An alternative explanation for the results for 
$F_2^\ast$ come from  light-front sum rules
in next-to-leading order~\cite{Anikin15a}.
The calculations suggest that the $p$-state three-quark wave functions 
give important contributions to $F_2^\ast$.
A recent light-front quark model calculation 
predicts that the quark-core contributions for $F_2^\ast$ 
are significant at low $Q^2$~\cite{Obukhovsky19}.
There is, however, some disagreement with the $S_{1/2}$ 
and $G_C$ data at low $Q^2$.

Calculations based on a light-front relativistic quark model~\cite{Aznauryan17}
indicate that the transition form factors 
can be explained as a combination of 
the valence quark and meson cloud contributions.
The authors use the model and the data 
to estimate meson cloud contributions and conclude that 
the relative contribution is 16\%.
They conclude also that the meson cloud contributions 
are dominated by isovector components~\cite{Aznauryan17}.

There are also calculations of transition form factors based on 
AdS/QCD~\cite{Gutsche19,Bayona12}.
Reference~\cite{Gutsche19} shows that the data 
can be described assuming significant contribution 
of higher order Fock states, namely from $q\bar q$ 
and $(q \bar q)(q \bar q)$ contributions.

Given the success of the covariant spectator quark model 
in the description of other resonances  both in the spacelike and 
timelike regime, we investigate here the valence quark and meson cloud 
contributions to the $N(1535)$ excitation within that model.


\section{Covariant spectator quark model}
\label{secCSQM}

The covariant spectator quark model is based 
on the covariant spectator theory~\cite{Gross}.
In this framework, the baryons can be described 
as quark-diquark systems, where the diquark 
is on-mass-shell with an effective mass $m_D$.
The electromagnetic interaction with the baryon 
is described by the photon coupling with 
a single quark at a time (impulse approximation). This coupling is characterized by
constituent quark forms factors which take 
into account the gluon and quark-antiquark dressing effects 
of the quarks~\cite{Nucleon,Nucleon2,Omega,NSTAR-review}.

The covariant spectator quark model has been applied to the 
study of the structure of the 
nucleon~\cite{NucleonDIS,Axial,Medium},
to the electromagnetic structure of 
several nucleon excitations~\cite{Roper,NDelta,N1535,N1520,SQTM,DeltaFF,Delta1600,Lattice,LatticeD},
as well as to the electromagnetic structure of 
octet and decuplet baryons~\cite{OctetFF,OctetDecuplet,DecupletDecayTL,Medium,Hyperons}.
An overview of the results of the covariant spectator quark model 
for several nucleon resonances can be found in Ref.~\cite{NSTAR-review}.

The nucleon wave function was obtained
in Ref.~\cite{Nucleon} and the wave function of the 
resonance $N(1535)$  in Ref.~\cite{N1535}.
Those wave functions describe only the 
valence quark content of those baryons allowing
estimates of those contributions to electromagnetic transitions.
In this work we combine the 
covariant spectator quark model with 
the semirelativistic approximation~\cite{SemiRel}, 
which guarantees the orthogonality between the initial and final baryon states, and provide a significant simplification 
in the transition form factor.
Our quark model estimates are then used 
to obtain  a consistent parametrization 
of the meson cloud contributions,
including the isoscalar and isovector components,
from the constraints imposed by the data.
The combined parametrization of the two effects 
is presented at the end.

We start by discussing the general formalism 
developed for the study of the spacelike region $Q^2=-q^2 \ge 0$.

\subsection{Formalism}

The constituent quark electromagnetic
current in the $SU(2)$ sector is written 
as the sum of a Dirac and a Pauli component, as 
\ba
j_q^\mu(q) & =& \left( \frac{1}{6} f_{1+} + \frac{1}{2} f_{1-} \tau_3 
\right)  \gamma^\mu  + \nonumber  \\
& &     \left( \frac{1}{6} f_{2+} + \frac{1}{2} f_{2-} \tau_3 
\right) \frac{i \sigma^{\mu \nu} q_\nu}{2M_N},
\label{eqJq}
\ea
where $\tau_3$ is the Pauli matrix that acts on the   
(initial and final) baryon isospin states,
$f_{i\pm} (q^2)$ are the quark isoscalar/isovector form factors.
Those form factors are parametrized with analytical formulas 
consistent with the vector meson dominance 
(VMD) mechanism~\cite{Nucleon,Lattice,LatticeD}.
This dominance in the quark-photon vertex is very useful for
generalizations of the dynamics from spacelike 
to the timelike region~\cite{Timelike,Timelike2,N1520TL,DecupletDecayTL}.
The covariant spectator quark model explicit formulas
for the quark form factors $f_{i\pm}$ ($i=1,2$) 
in the timelike region can be found in Ref.~\cite{N1520TL}.

Since in our calculation within the covariant spectator 
quark model we use the relativistic impulse approximation,
the transition current can be written
in terms of nucleon wave function ($\Psi_N$) 
and the resonance wave function ($\Psi_R$) both expressed in terms of
the single quark and quark-pair states,
specified by the adequate 
flavor, spin, orbital  angular momentum 
and radial excitations of the quark-diquark states defined by the baryon 
quantum numbers~\cite{NSTAR,Nucleon,Nucleon2,Omega,OctetFF}.

In the impulse approximation the electromagnetic baryon transition 
current reads~\cite{Nucleon,Omega,Nucleon2}
\ba
J^\mu=
3 \sum_{\Gamma} 
\int_k \bar \Psi_R (P_R,k) j_q^\mu \Psi_N(P_N,k),
\label{eqJmu}
\ea  
where $P_R$, $P_N$, and $k$ are  
the resonance, the nucleon, and the diquark momenta, respectively.
The previous equation is the result of integrating over 
the internal relative motion of the quarks in the diquark.
The index $\Gamma$ labels
the intermediate diquark polarization states,
the factor 3 takes into account the contributions from
all different quark pairs,  and the integration
symbol represents the covariant integration over the
diquark on-mass-shell momentum.
In the study of the inelastic transitions 
we use the Landau prescription to ensure
current conservation~\cite{SemiRel,N1520,Kelly98,Gilman02,Batiz98}.

The radial wave function of the nucleon $\psi_N(P_N,k)$
in the covariant spectator quark model
is taken as a function of the dimensionless variable~\cite{Nucleon}:
\ba
\chi = \frac{(M_N -m_D)^2 -(P_N-k)^2}{2 M_N m_D}.
\ea  
This representation is possible because the baryons and the diquark 
are both on-mass-shell~\cite{Nucleon}.
The explicit form for $\psi_N$ is 
\ba
\psi_B (P_N,k) = \frac{N_0}{m_D(\beta_1 + \chi) (\beta_2 + \chi)},
\label{eqPsi-scalar}
\ea
where $N_0$ is a normalization constant 
and the parameters $\beta_1= 0.049$ and $\beta_2 =0.717$ 
are parameters determined by the fit to the 
nucleon electromagnetic form factor data~\cite{Nucleon}. 
They effectively represent two different momentum ranges
that have to be described by the radial wave function.
In the next sub-section we discuss the radial wave function 
of the resonance $\psi_R(P_R,k)$.

To represent the transition form factors 
it is convenient to  use the symmetric ($S$) and anti-symmetric ($A$) 
combination of quark currents,  
which read as combinations of 
quark form factors~\cite{Nucleon,OctetFF,Medium} ($i=1,2$):
\ba
& &
j_i^S= \frac{1}{6} f_{i+} +  \frac{1}{2} f_{i-} \tau_3, 
\label{eqjiS}\\
& &
j_i^A= \frac{1}{6} f_{i+} -  \frac{1}{6} f_{i-} \tau_3.
\label{eqjiA}
\ea

The $\gamma^\ast N \to N^\ast$ transition current (\ref{eqJmu}), 
where $N^\ast$ is a $J^P=\frac{1}{2}^-$ or a $J^P=\frac{3}{2}^-$ state,
becomes proportional to
the following overlap integral~\cite{SemiRel,N1520},
\ba
{\cal I}_R (Q^2)= \int_k \frac{k_z}{|{\bf k}|} \psi_R (P_R,k) \psi_N(P_N,k).
\label{eqIR}
\ea 
The integral (\ref{eqIR}) is frame invariant 
and can be evaluated in any frame.
For simplicity, we write the integral (\ref{eqIR}) 
in the resonance rest frame. 
The general expression for ${\cal I}_R$ 
can be found in Refs.~\cite{N1535,N1520}.


\subsection{Semirelativistic approximation}
\label{secSRapp}

We consider now the results for the $\gamma^\ast N \to N(1535)$ 
transition~\cite{SemiRel} within the covariant spectator quark model
in the semirelativistic approximation.

The semirelativistic approximation is based on 
two assumptions~\cite{SemiRel}
\begin{itemize}
\item
The difference of mass between the nucleon and the resonance
can be neglected in the calculation of the 
Dirac and Pauli form factors from Eq.~(\ref{eqJS11}).
\item
One takes
$\psi_R \equiv \psi_N$, i.e., 
the radial structure 
of the resonance to be the same as the radial structure of the nucleon,
with no need to introduce 
additional parameters for the structure of the resonance
(in $\psi_R$ we replace the mass and momentum by 
$M_R$ and $P_R$). 
\end{itemize}

We implement the semirelativistic approximation   
replacing the dependence on $M_R$ and $M_N$ by  $M$, where
\ba
M \equiv \sfrac{1}{2}(M_N + M_R),
\ea
in the calculation of the overlap integral (\ref{eqIR}),
and use the result to estimate the Dirac and Pauli form factors.
The final expressions for the transition form factors 
and helicity are, however, sill covariant~\cite{SemiRel,NSTAR-review}.
The label {\it semirelativistic approximation} 
is motivated the condition of no mass difference, 
as in the non-relativistic limit.

From the previous assumptions,
one can conclude that~\cite{SemiRel,N1535}
\ba
{\cal I}_R \propto |{\bf q}| \propto Q,
\label{eqIR1}
\ea 
where the last relation is a consequence 
of the form for $|{\bf q}|$, 
in the semirelativistic approximation  
\ba
|{\bf q}| = Q \sqrt{1 + \tau},
\ea
with $\tau= \frac{Q^2}{4 M^2}$.

The consequence of (\ref{eqIR1}) is that the overlap integral (\ref{eqIR})
vanishes, ensuring the orthogonality between the states~\cite{SemiRel}.
The final expressions for the transition form factors depend 
on the quark form factors and on the radial wave functions.
Those formulas have no adjustable parameters,
since the quark current was previously determined 
from the study of the nucleon electromagnetic form factors~\cite{Nucleon}.
Therefore those formulas provide predictions from assuming 
that nucleon and resonance have basically the same radial wave functions.
We present next the expressions for
the valence quark contributions to the  $\gamma^\ast N \to N(1535)$
transition, and 
after that, we will discuss the parametrizations 
of the meson cloud contributions that we indirectly extract from the data.

It is convenient at this moment to discuss the range of application 
of the semirelativistic approximation.
A consequence of the approximation is that 
$M_R - M_N \simeq 0$,
and the variables (\ref{eqQpm}) become $Q_-^2=Q^2$
and $Q_+^2 = 4M^2 + Q^2$.
%
This prevents the direct calculation of 
transition form factors for $Q^2 < 0$, since $Q_-^2 \ge 0$.
The minimal value for $|{\bf q}|$ is then obtained when $Q^2=0$ ($|{\bf q}|=0$).
This is an important difference 
between this work and the previous applications of 
the covariant spectator quark model~\cite{Timelike,Timelike2,N1520TL,DecupletDecayTL} 
that did not use the semirelativistic approximation and could access 
the $Q^2 < 0$ region directly.
Instead, we perform here a numerical extrapolation
of the spacelike results into the region $-(M_R -M_N)^2 \le Q^2 < 0$.
This process is discussed in Sec.~\ref{secTimelike}.

\subsection{Valence quark contributions}
\label{secN1535bare}

In the semirelativistic approximation, we obtain the 
following final results from the valence quark contributions 
to the transition form factors~\cite{SemiRel}:
\ba
& &
F_1^{\rm B}(Q^2) = \frac{1}{2}(3 j_1^S + j_1^A) Z {\cal I}_R,
\label{eqF1a}
\\
& &
F_2^{\rm B}(Q^2) = -\frac{1}{2}(3 j_2^S - j_2^A) 
Z {\cal I}_R,
\label{eqF2a}
\ea
where the factor $Z \propto \sqrt{Q^2}$, 
introduced in the present work for the first time, 
is discussed next.
The upper index B labels the bare contribution.
For a detailed discussion of Eqs.~(\ref{eqF1a})--(\ref{eqF2a}), 
check Refs.~\cite{SemiRel,NSTAR-review}.

As discussed in Refs.~\cite{SemiRel,NSTAR-review},
the equations with $Z=1$ are derived 
for the case $M_R - M_N \simeq 0 $ and thus
do not include any dependence on the mass difference $M_R - M_N$.
The consequence, then would be that  
the form factors $F_1^{\rm B}$ and $F_2^{\rm B}$ go with $Q$ near $Q^2=0$.
In those conditions, we fail to obtain 
the expected result $F_1^\ast  \propto Q^2$ needed for gauge invariance.
The form $F_1^{\rm B}  \propto Q$ would also change
the expected behavior of the helicity amplitudes.
In particular the amplitude $S_{1/2}$ defined by Eq.~(\ref{eqS12s}) would
diverge at $Q^2= 0$, unless we replace 
$|{\bf q}|$ by its equal mass limit value
$|{\bf q}| = Q \sqrt{1 + \tau}$
(see discussion in Ref.~\cite{SemiRel}).

To have the correct $Q^2$ behavior of $F_1^\ast$ near the origin, 
we then define $Z$ as
\ba
Z = \sqrt{\frac{Q^2}{\Lambda_R^2 + Q^2}},
\label{eqZ}
\ea
where $\Lambda_R$ is a momentum scale.
This scale should be small compared to 
the nucleon and resonance masses, 
in order to preserve the good results at intermediate 
and large $Q^2$ of the valence quark model~\cite{SemiRel},  
but should not be too small in order to avoid singularities 
within the range of its extension to the timelike region.


With the inclusion of $Z$,
we recover the expected behavior near $Q^2=0$,
$F_1^{\rm B} \propto Q^2$.
When we consider a  moderate scale for $\Lambda_R$, 
and keep the results for intermediate $Q^2$ almost unchanged.

A particular good choice for this scale is $\Lambda_R = m_\rho$ (rho mass).
It allows the extension of 
our spacelike results into the region $0 < q^2 \le (W-M_N)^2$
for a given resonance energy $W$, provided that $W  \le  M_N  + m_\rho$.
We will turn to this point later in more detail.

\begin{figure*}[t] 
\vspace{.45cm}
\centerline{\mbox{
\includegraphics[width=7.5cm]{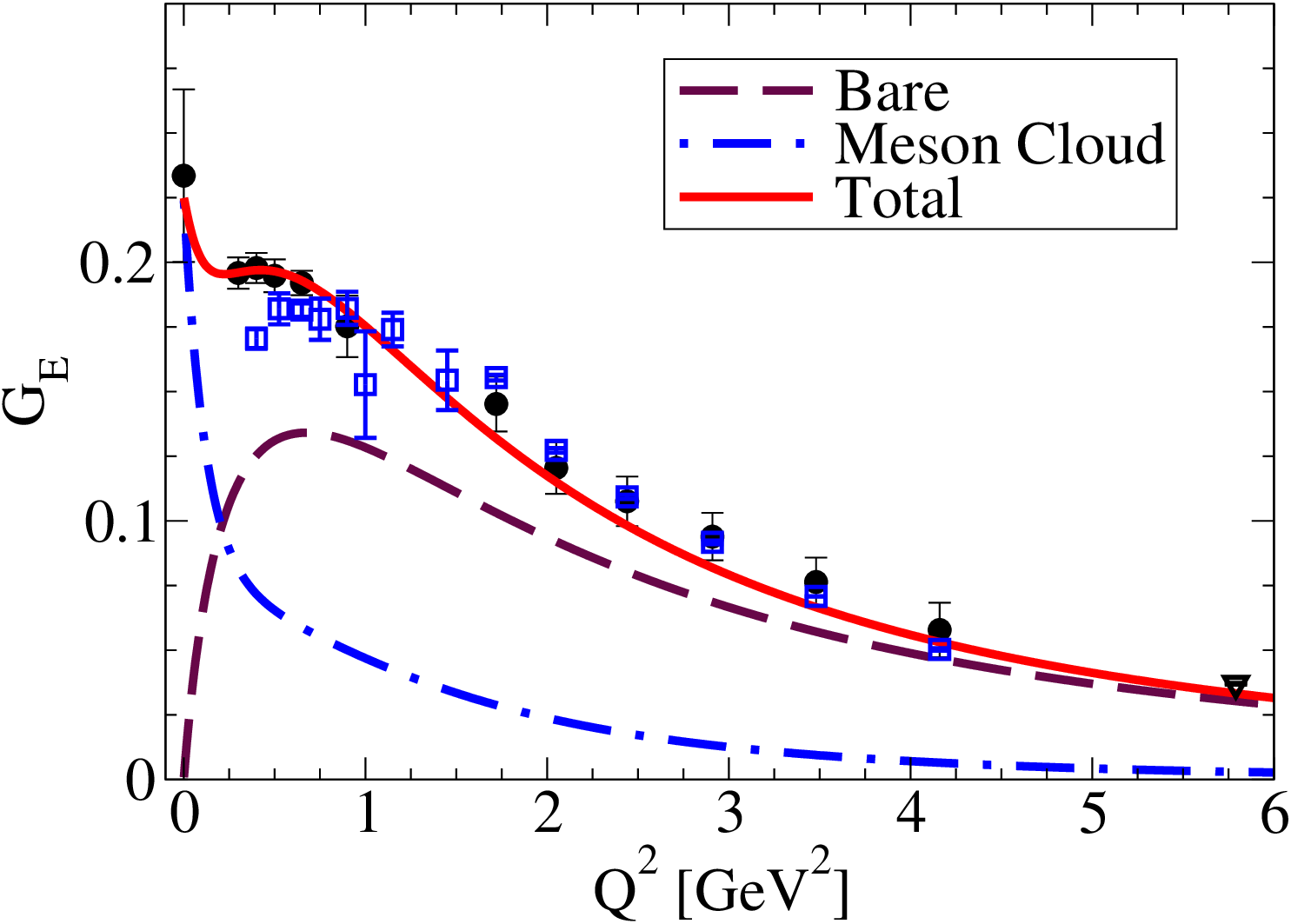}}  \hspace{1.cm}
\mbox{
\includegraphics[width=7.5cm]{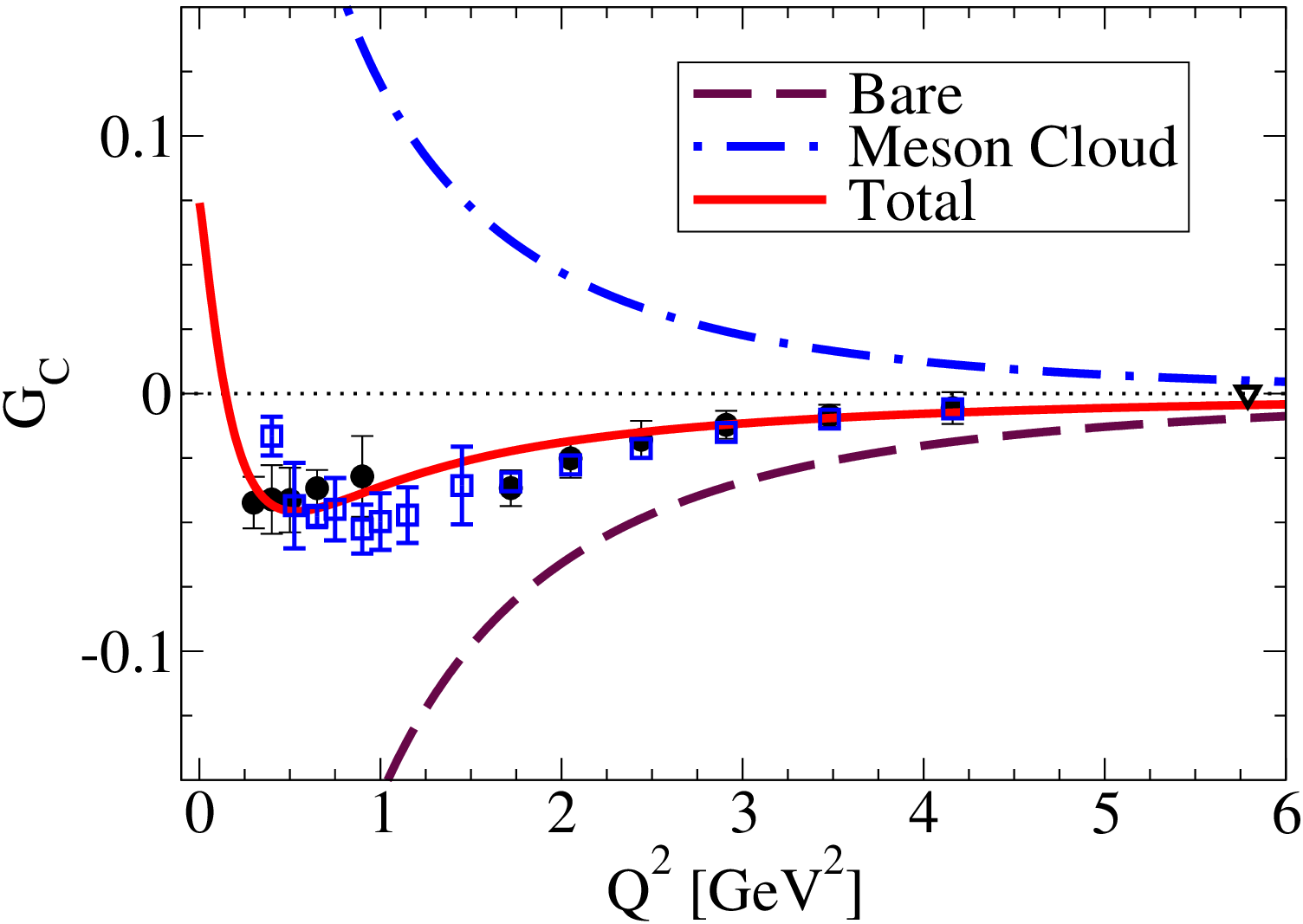}}}
\caption{\footnotesize
$\gamma^\ast N \to N(1535)$ transition form factors $G_E$ and $G_C$ 
for proton target.
Data from  CLAS~\cite{CLAS1},
(circles) MAID~\cite{MAID1} (squares), 
 and  JLab/Hall C~\cite{Dalton09} (triangles).
The data at $Q^2=0$ is from PDG~\cite{PDG}.} 
\label{figN1535R-2}       
\end{figure*}

\begin{figure*}[t] 
\vspace{.45cm}
\centerline{\mbox{
\includegraphics[width=7.5cm]{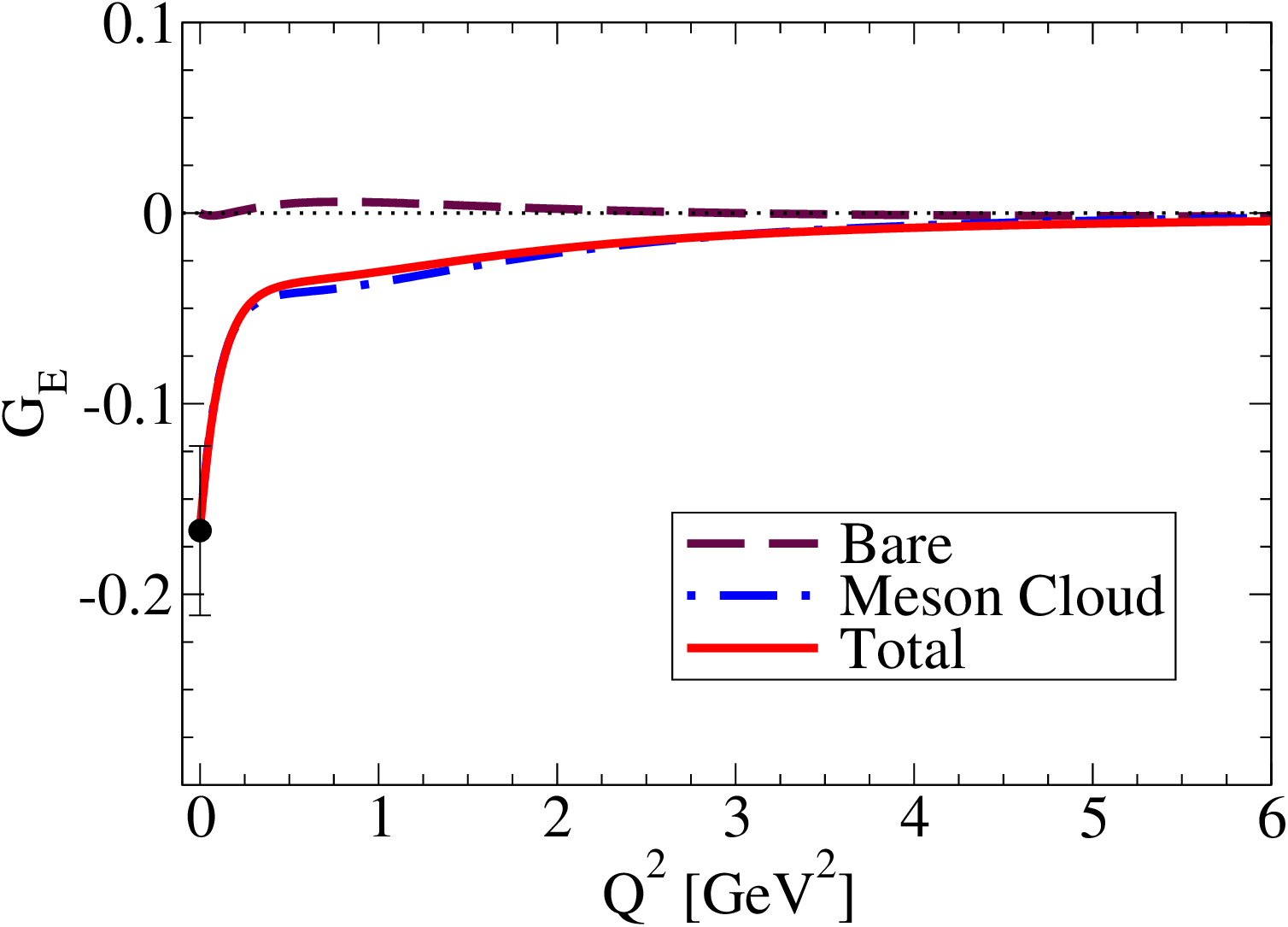}}  \hspace{.5cm}
\mbox{
\includegraphics[width=7.5cm]{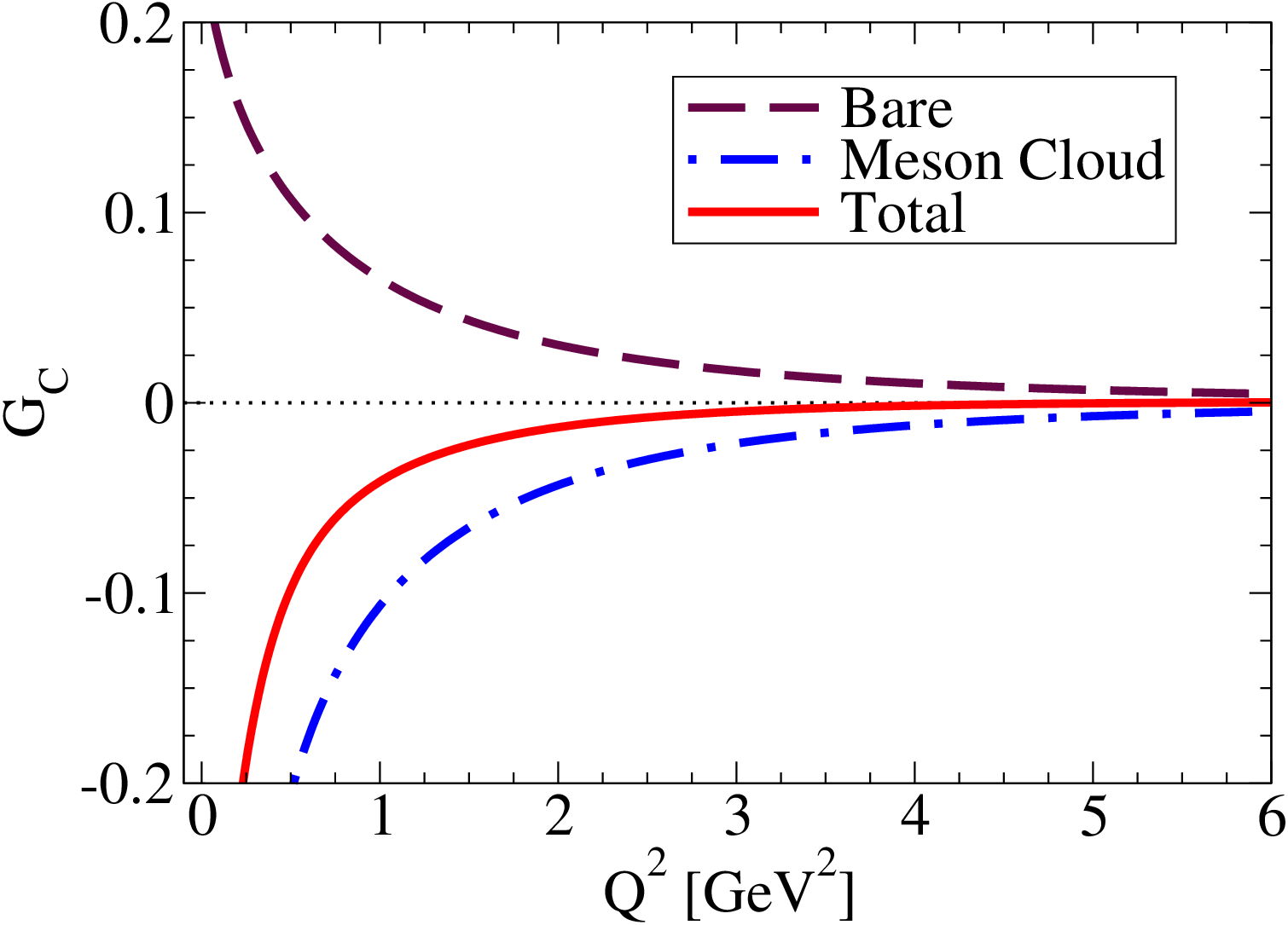}}}
\caption{\footnotesize
$\gamma^\ast N \to N(1535)$ transition form factors 
$G_E$ and $G_C$ for neutron target.
The data are from PDG~\cite{PDG}.} 
\label{figN1535R-4}       
\end{figure*}

The valence quark contributions 
to the form factors $F_1^\ast$, $F_2^\ast$
and $G_E$, $G_C$, defined by Eqs.~(\ref{eqGE0})--(\ref{eqGC00}), 
are presented in Figs.~\ref{figN1535R}
and \ref{figN1535R-2} for the proton target, by the dashed-lines.
The results for the neutron target 
are presented in Figs.~\ref{figN1535R-3} and \ref{figN1535R-4},
also by the dashed-lines.
Note in this case the small magnitude 
of the valence quark contributions 
for $F_1^\ast$ and in particular for $G_E$.

From  Fig.~\ref{figN1535R}, we can conclude 
that in the semirelativistic approximation 
the valence quark contributions alone provide 
a fair description of the data at large $Q^2$~\cite{SemiRel},
particularly for $F_1^\ast$, 
but it does not properly describe the low-$Q^2$ region.

\subsection{Meson cloud contributions}
\label{secMesonCloud}

The failure of the quark model at low $Q^2$
indicates 
the importance of the meson cloud excitations 
on baryons bare cores, probed in the low-$Q^2$ regime.
In order to improve the description of the data, 
we consider here effective parametrizations which mimic 
the effects not included in our valence quark model.
It is necessary to identify three 
different contributions associated with 
the Dirac and Pauli form factors.
Two (isoscalar and isovector) components for the Pauli form factor 
and one isovector component for the Dirac form factor.


\begin{figure*}[t] 
\vspace{.5cm}
\centerline{\mbox{
\includegraphics[width=7.5cm]{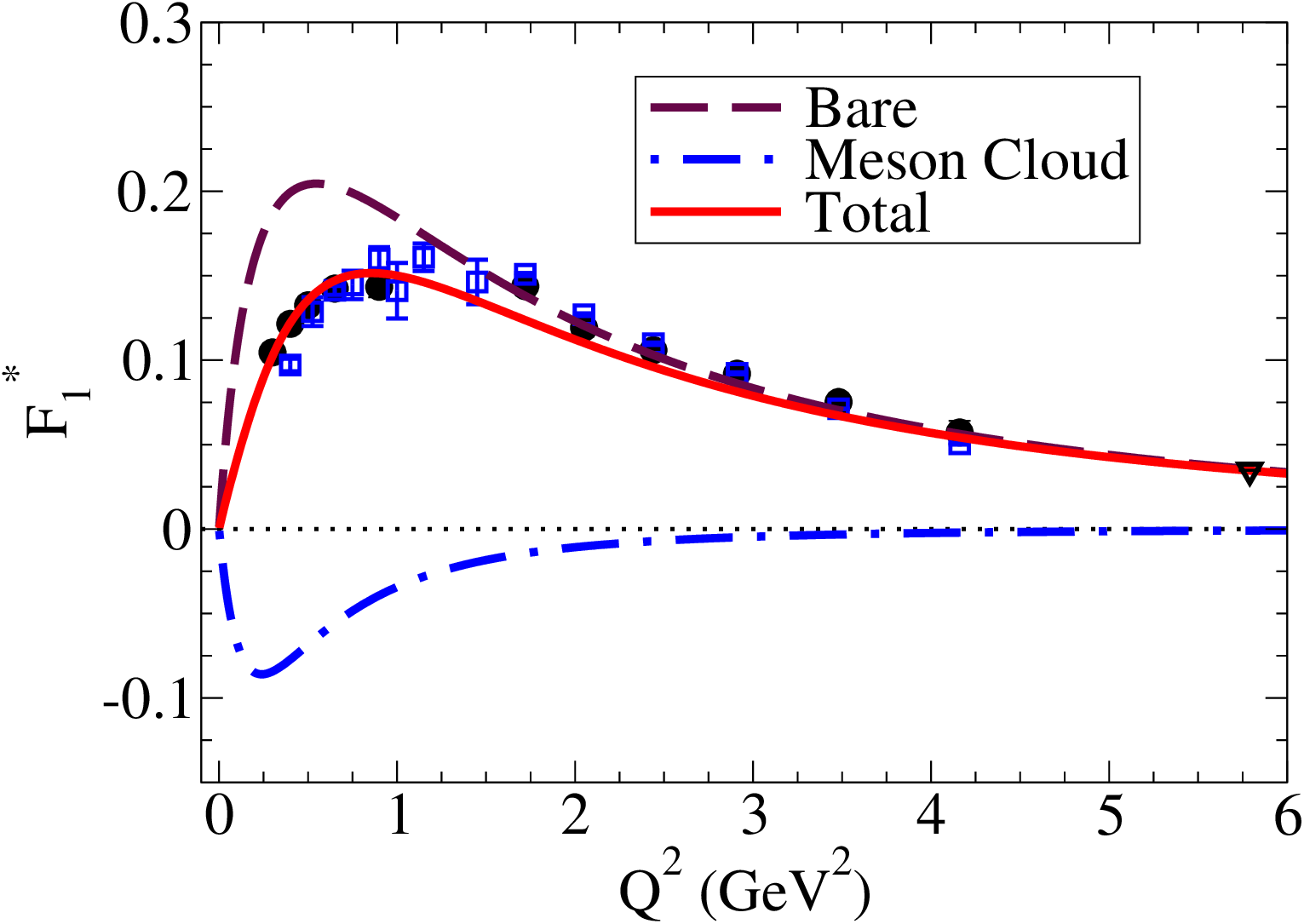}}  \hspace{1.cm}
\mbox{
\includegraphics[width=7.5cm]{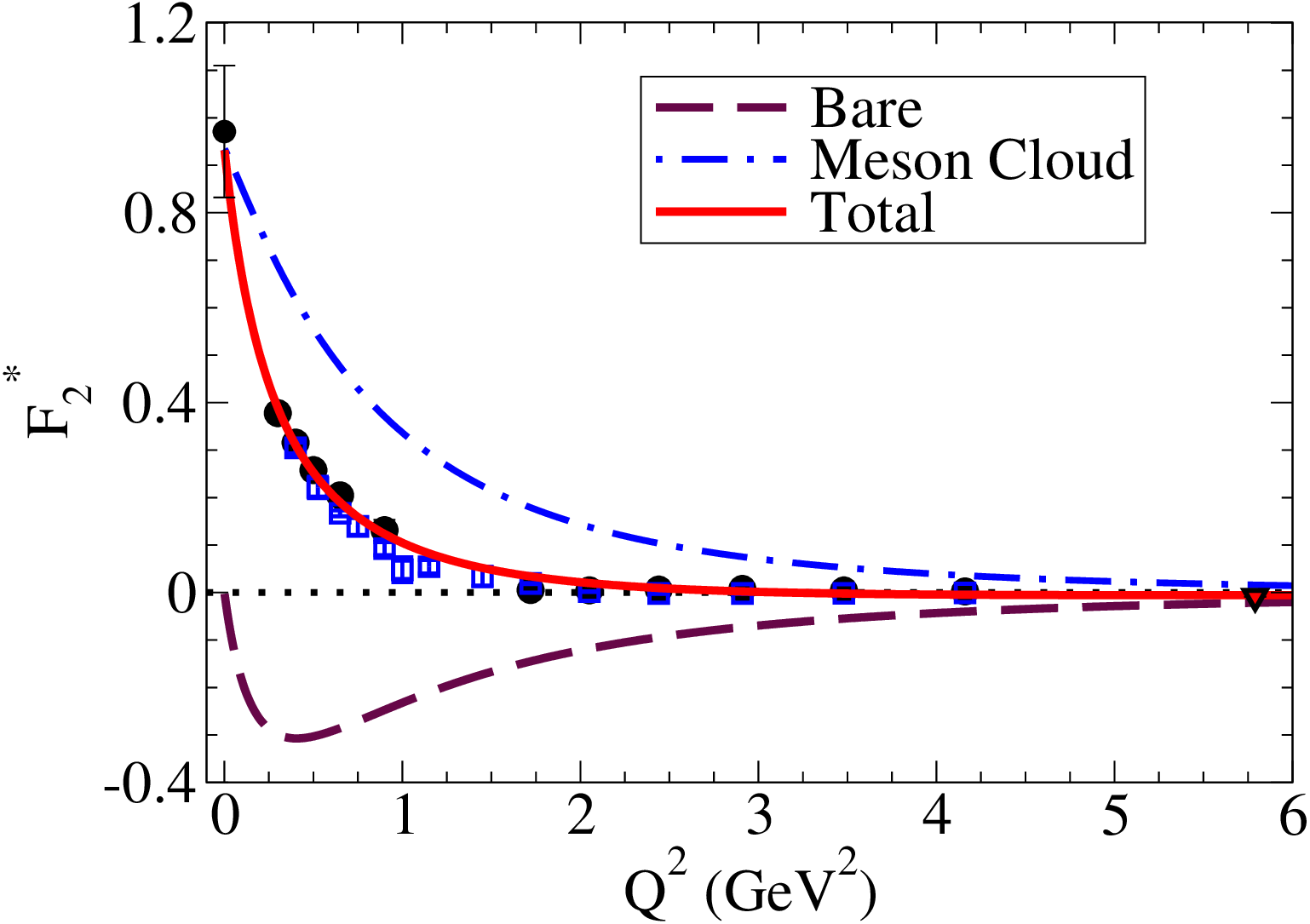}}}
\caption{\footnotesize
$\gamma^\ast N \to N(1535)$ transition form factors for proton target.
Data from  CLAS~\cite{CLAS1},
(circles) MAID~\cite{MAID1} (squares), 
 and  JLab/Hall C~\cite{Dalton09} (triangles).
 The data at $Q^2=0$ is from PDG~\cite{PDG}.} 
\label{figN1535R}       
\end{figure*}

\begin{table}[t]
\begin{center}
\begin{tabular}{c c c c}
\hline
\hline
   &   $A_{1/2} (0)$  & $F_2^\ast(0)$ & $A(0),B(0)$\\
\hline
\hline
$p$  &  0.105$\pm$0.015  &  0.97$\pm$0.14  &  0.14$\pm$0.12  \\  
$n$  & $-0.075\pm$0.020  & $-0.69\pm$0.19  &  0.83$\pm$0.12  \\
\hline
\hline
\end{tabular}
\end{center}
\caption{\footnotesize
Amplitude $A_{1/2} (0)$ and results for $F_2^\ast(0)$
for the $\gamma^\ast N \to N(1535)$ transition.
$A_{1/2}(0)$ is in units GeV$^{-1/2}$.
Data from PDG~\cite{PDG}.
In the last column, the first line 
refers to $A(0)$ and the second line to B(0),
defined by Eqs.~(\ref{eqA0}) and (\ref{eqB0}).}
\label{tabN1535-1}
\end{table}
%


To prepare the following discussion, it is important to notice 
that $F_1^\ast(0)=0$ by construction, 
and that the amplitude $S_{1/2}$ cannot be measured 
at the photon point since there are no real photons 
with longitudinal polarization.
Therefore, the direct information about 
the form factors at $Q^2=0$ come only from $A_{1/2}$ and $F_2^\ast$.
Those functions are seen to be related at $Q^2=0$ by
\ba
A_{1/2}(0)= {\cal C} F_2^\ast(0),
\ea
where ${\cal C}= {\cal B}_0 \eta$ 
where ${\cal B}_0$ is the value of ${\cal B}$,
defined in Eq.~(\ref{eqS12s}) at $Q^2=0$.
One obtains then ${\cal C} = \frac{e}{2} \frac{M_R -M_N}{\sqrt{M_N M_R K}}$.
A summary of the $A_{1/2}$ and $F_2^\ast$ data at $Q^2=0$ 
is presented in Table~\ref{tabN1535-1}.

Having in mind that at low $Q^2$ the meson cloud excitations are important,
we decompose the transition form factors into a bare term
(labeled with superscript $\rm B$) and a meson cloud term
(labeled with superscript $\rm mc$):
\ba
& &
F_1^\ast = F_1^{\rm B} + F_1^{\rm mc}, 
\label{eqF1decomp}\\
& &
F_2^\ast = F_2^{\rm B} + F_2^{\rm mc},
\label{eqF2decomp}
\ea
where the bare contributions are determined by  
Eqs.~(\ref{eqF1a})--(\ref{eqF2a}) of the 
covariant spectator quark model in the semirelativistic approximation. 
The meson cloud terms $F_1^{\rm mc}$ and $F_2^{\rm mc}$
are to be extracted indirectly from the data.

\begin{figure*}[t] 
\vspace{.5cm}
\centerline{\mbox{
\includegraphics[width=7.5cm]{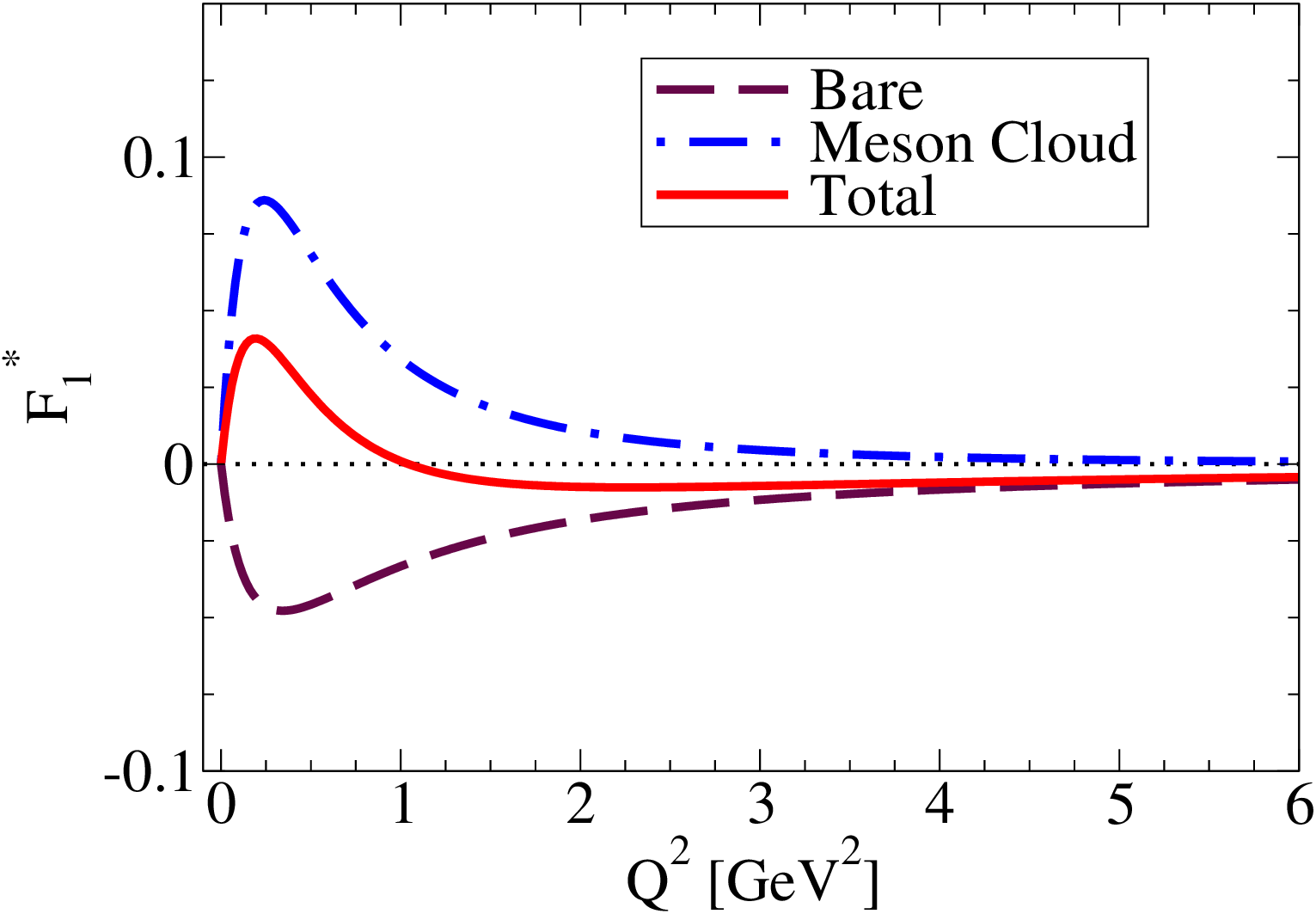}}  \hspace{.5cm}
\mbox{
\includegraphics[width=7.5cm]{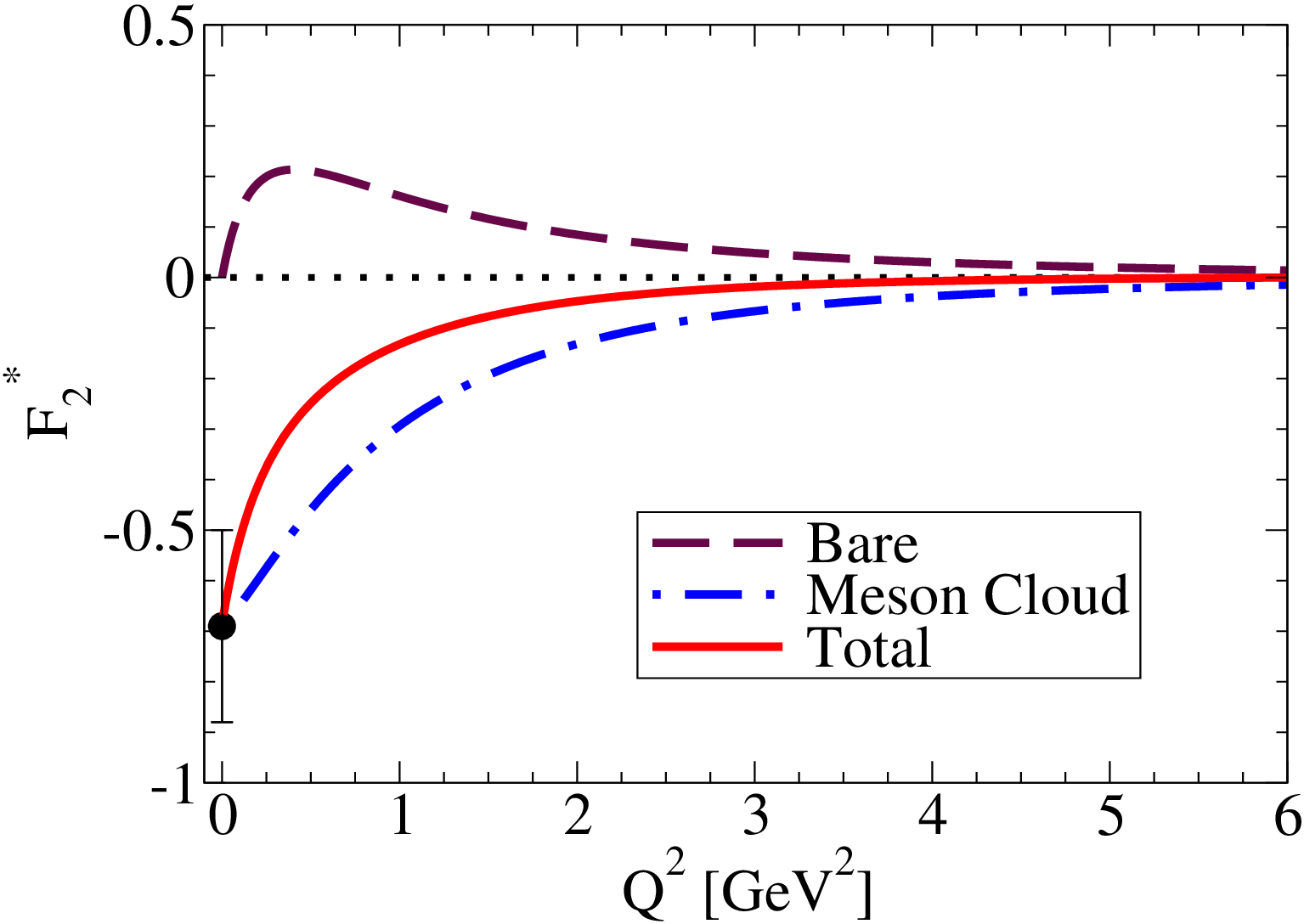}}}
\caption{\footnotesize
$\gamma^\ast N \to N(1535)$ transition form factors
for neutron target. 
The data are from PDG~\cite{PDG}.} 
\label{figN1535R-3}       
\end{figure*}

We start our discussion with $F_2^\ast$.
Since $ F_2^{\rm mc}$ is a function of $Q^2$,
we can use the general decomposition  
\ba
F_2^{\rm mc}(Q^2) = A(Q^2) + B (Q^2) \, \tau_3,
\label{eqF2mc1}
\ea
where $A$  represent the 
isoscalar contribution and  $B$ 
represent the isovector contribution.
Note that, we cannot parametrize 
the functions $A$ and $B$ simultaneously 
because the empirical data for finite $Q^2$
are restricted to proton targets.
Only for $Q^2=0$, there are data for neutron targets.
In the last case then we can extract 
the contributions for $A$ and $B$ using 
\ba
& &A(0) = \frac{1}{2{\cal C}}[A_{1/2}^p(0) + A_{1/2}^n(0) ], 
\label{eqA0} \\
& &B(0) = \frac{1}{2{\cal C}}[A_{1/2}^p(0) - A_{1/2}^n(0) ].
\label{eqB0}
\ea
The results of $A(0)$ and $B(0)$ extracted from the 
experimental data for $A_{1/2}^{p,n}(0)$ for proton and neutron targets 
are presented in Table~\ref{tabN1535-1}. 

We conclude from Table~\ref{tabN1535-1}
that the isovector component dominates, 
near $Q^2=0$, since $B(0) \gg A(0)$.
Although the results from Table~\ref{tabN1535-1}
suggest that $A(0)$ is almost compatible with zero,
consistent with the isovector dominance 
of the meson cloud contribution,
the upper limit of $A(0)$, could also be as large as about 1/4 of $B$.
This is why we include the isoscalar term $A$ in our parametrization
of the meson cloud.

We now discuss the function $F_1^\ast$.
Since $F_1^\ast(0)=0$ and the bare contribution 
also vanishes at $Q^2=0$, we conclude that 
the meson cloud contribution should also vanish at $Q^2=0$.
From the difference between the data 
and the valence quark contributions (dashed-line)
in Fig.~\ref{figN1535R}, we infer that 
the meson cloud contributions 
for $F_1^\ast$ can be significant, below $Q^2=1$ GeV$^2$.
Those contributions are important to
the amplitude $A_{1/2}$ (proportional to $G_E)$
which dominates the structure of the resonances at small $q^2$
in the timelike region, as discussed in Sec.~\ref{secDalitz}.
To parametrize the meson cloud contribution to $F_1^\ast$,
we consider the form
\ba
F_1^{\rm mc}(Q^2) = C(Q^2) \; \tau_3,
\label{eqF1mc}
\ea
where $C$ is a function proportional to $Q^2$, near $Q^2=0$.
With this parametrization, 
we assume the isovector character of the meson cloud term, 
motivated by the evidence of the 
isovector dominance in the amplitude $A_{1/2}$ at $Q^2=0$
seen in Table~\ref{tabN1535-1}.
However, the isovector character of $F_1^{\rm mc}$
cannot be tested with the present data, since  $F_1^{\rm mc} =0$ at $Q^2=0$, and 
there are at the moment no available data for the amplitude $A_{1/2}$ 
with neutron targets, for nonzero $Q^2$.
Our description of $F_1^{\rm mc}$ is then based 
on an ansatz that can only be tested  in the future, 
once helicity amplitude data for the neutron for finite $Q^2$
became also available.

In summary, we can parametrize the meson cloud 
contributions to the form factors $F_1^\ast$
and $F_2^\ast$ using three functions ($A$, $B$ and $C$).
The parametrization of 
$F_2^{\rm mc}$ near $Q^2=0$ is fixed by the experimental results for 
the amplitudes $A_{1/2}^{p,n}(0)$,
while the general $Q^2$-dependence is determined only by the 
combination $A+ B$ (proton target), 
since there are no data yet for $A-B$ 
(neutron targets) for finite $Q^2$.

The available data support the dominance 
of the isovector component of the meson cloud 
on the form factors  $F_1^\ast$ and $F_2^\ast$.
This effect can be observed in 
the results for the proton targets (Fig.~\ref{figN1535R}) 
and neutron targets (Fig.~\ref{figN1535R-3}), 
where one can notice that the combination 
of the bare contributions with the meson cloud contributions 
(solid-lines) based on the isovector dominance 
provide a good description of the data.
Recall that also calculations based on 
light front relativistic quark models~\cite{Aznauryan17}
conclude that the meson cloud contributions to both Dirac and Pauli form factors
are dominated by the isovector component.

To define a parametrization of the meson cloud effects, 
we have looked at the possible decays of the $N(1535)$ state.
There are two main channels for these decays,
the $\pi N$ channel and the $\eta N$ channel 
with about 50\% contribution from each component.
Minor contributions came from $\pi \Delta$,
$\sigma N$ and $\pi N(1440)$.
We ignore these last contributions, since the combined effect 
of those channels is at most 14\%~\cite{PDG}.
We can then assume that the electromagnetic interaction
with the meson cloud is dominated by the 
$\pi N$ and the $\eta N$ states.
Since the $\eta$ meson has no charge, we conclude 
that the $\eta N$ states contribute to the isoscalar 
component of the meson cloud, and therefore to the function $A$,
while the $\pi N$ states contribute to the isovector 
component of the meson cloud, and therefore to the function $B$.

For the isoscalar component, 
we take a parametrization of the form $A \propto F_\eta$,
where $F_\eta$ is the $\eta$ electromagnetic form factor
and the additional factors is a multipole type function
with a phenomenological cutoff.
Since $F_\eta$ is not known, we consider 
the simplest case where all the structure 
is simulated by a single multipole function,
\ba
A (Q^2)= A(0) \left(\frac{ \Lambda_A^2}{\Lambda_A^2 + Q^2}
\right)^5,
\label{eqA1}
\ea
and where $\Lambda_A$ is a cutoff parameter.
Importantly, this choice was made to be consistent with 
perturbative QCD (pQCD) estimates where for very large $Q^2$, one has 
$A \propto 1/Q^{10}$ \cite{Carlson}\footnote{According 
to the pQCD analysis, the leading order 
contribution to $F_2^\ast$ comes from the $N=3$ contribution
(3 constituents) 
and has the form $F_2^\ast \propto 1/Q^2 \cdot 1/Q^{2(N-1)} = 1/Q^6$.
The next leading order contribution associated
with a $q \bar q$  excitation implies that $N =5$
(5 constituents), which corresponds to a correction of the previous estimate 
by a factor $1/Q^4$, leading to the estimate 
$F_2^{\rm mc} \propto 1/Q^{10}$.}.


As for the isovector component,
the coupling with the  $\pi N$ states 
is in first approximation determined 
by the photon coupling with the pion,
which is given by the pion electromagnetic form factor $F_\pi$.
Then one expects that the function $B$
in Eq.~(\ref{eqF2mc1})
to have the form
\ba
B (Q^2) \propto F_\pi (Q^2).
\nonumber
\ea
The omitted multiplicative functions in this relation 
are structure functions that determine the extension
of the nucleon and resonance cores.
One then writes this structure in an effective way as 
\ba
B (Q^2)= B(0)
\left(\frac{\Lambda_B^2}{\Lambda_B^2 + Q^2}  \right)^4
(1 + c \, Q^2) F_\pi (Q^2), \nonumber \\
\label{eqB2}
\ea
where the $\Lambda_B$ is a short-range (large $Q^2$) regulator
and $c$ is an adjustable coefficient.
The factor $(1 + c\, Q^2)$  was included to 
improve the quality of the fit by smoothening the variation with $Q^2$
in the low-$Q^2$ region: a multipole function alone is incompatible 
with a smooth behavior near $Q^2=0$
for $F_2^\ast$ and $G_E$.
The power of the multipole function is chosen 
in order to mimic the falloff of pQCD at very large $Q^2$.
In a model with $F_\pi \propto 1/(Q^2 \log Q^2)$,
as the one that we consider here, 
we obtain then $B \propto 1/(Q^{8} \log Q^2)$,
close $1/Q^{10}$, expected from pQCD.
Note, however, that for the purpose of the present study,
the exact power of the multipole in Eq.~(\ref{eqB2}), (power 3 or 4), 
is not very relevant, 
since $\Lambda_B^2$ only cuts the large the momentum $Q^2$ region, 
and the behavior of $F_2^{\rm mc}$ is more sensitive to 
the low-$Q^2$ scale included in $F_\pi(Q^2)$.


To describe $F_1^{\rm mc}$, we consider 
the parametrization
\ba
C(Q^2) = - C_0 \frac{Q^2}{\Lambda_C^2}
\left( \frac{\Lambda_C^2}{\Lambda_C^2 + Q^2}
\right)^3 
F_\pi (Q^2),
\label{eqC1}
\ea 
where $C_0$ is a positive constant, 
and $\Lambda_C$ is an adjustable cutoff.
At large $Q^2$, $F_1^{\rm mc}$ goes with $1/(Q^6 \log Q^2)$,
closer to the falloff $1/Q^{8}$ estimated by 
pQCD\footnote{The leading order  
form factors ($F_1^\ast$) are ruled by the $1/Q^{2(N-1)}= 1/Q^4$ falloff.
The meson cloud contribution changes $N=3$ to $N=5$ 
(extra $q \bar q$ pair) leading to a 
$F_1^{\rm mc} \propto 1/Q^8$ falloff.}.
The factor $F_\pi$ is included due to 
the isovector form associated with  $F_1^{\rm mc}$
discussed before.

For $F_\pi$ we use the parametrization 
already tested in the case of the $\Delta(1232)$~\cite{Timelike2}
\ba
F_\pi (Q^2) &=& \frac{\alpha}{\alpha + Q^2 + \frac{1}{\pi} 
\beta Q^2 \log \frac{Q^2}{m_\pi^2}},
\nonumber \\
&=& \frac{\alpha}{\alpha -q^2 - \frac{1}{\pi} 
\beta q^2 \log \frac{q^2}{m_\pi^2} + i \beta q^2},
\label{eqFpi}
\ea
where $\alpha =0.696$ GeV$^2$, $\beta = 0.178$, 
and $m_\pi$ is the pion mass.

The previous parametrization was derived 
in Ref.~\cite{Timelike2} based on analytic 
expressions that take into account 
the effects of the pion loop contributions 
to the $\rho$-meson propagator.
The original form~\cite{Iachello,Iachello73,Frohlich10} 
included the effect of the two-pion threshold 
expressed by a dependence on  $(q^2 - 4 m_\pi^2)$.
We consider the approximation $q^2  \gg 4 m_\pi^2$ and 
obtain a smoother description of the imaginary 
components without loss of accuracy~\cite{Timelike,Timelike2}.

\vspace{.5cm}

\subsection{Combination of valence quark and 
meson cloud contributions}
\label{secTotal}


The parameters of meson cloud contributions to the transition form factors
$F_1^\ast$ and $F_2^\ast$ can be determined by the 
fit of the parameters of the 
expressions (\ref{eqA1}), (\ref{eqB2}) and (\ref{eqC1}),
to the $F_{1}^\ast$ and  $F_{2}^\ast$ form factor data 
for proton targets, and the $F_{2}^\ast(0)$ data for neutron targets.
An alternative, is to fit those 
parametrizations directly to the form factors $G_E$ and $G_C$.

We choose the second option for two main reasons:
our final goal is to derive 
parametrizations for the multipole form 
factors in the timelike region; 
the $F_1^\ast$ and $F_2^\ast$ data are represented 
by very sharp functions near $Q^2=0$.
By contrast the form factors $G_E$ and $G_C$
have a softer shape at low $Q^2$.

In the fit, we considered an additional constraint:
we imposed that  the ratio $A(Q^2)/B(Q^2)$ should not increase 
in the region of study ($Q^2 < 5$ GeV$^2$)
in order to be consistent with the isovector dominance 
observed at the photon point ($Q^2=0$),
and supported by independent calculations~\cite{Aznauryan17}.

\begin{table}[t]
\begin{center}
\begin{tabular}{l |c}
\hline
\hline
$A(0)$  &  0.125    \\  
$\Lambda_A^2$ (GeV$^2$) & 2.384  \\
\hline
$B(0)$   &   0.810 \\
$c$ (GeV$^{-2}$) & 2.040  \\
$\Lambda_B^2$ (GeV$^2$) &  3.365  \\
\hline
$C_0$ &  0.873\\
$\Lambda_C^2$ (GeV$^2$) & 0.785  \\
\hline
\hline
\end{tabular}
\end{center}
\caption{\footnotesize
Parameters of the meson cloud parametrizations.
The numerical results for the functions $A$, $B$ and $C$ are presented
in Fig.~\ref{figN1535-AB}.}
\label{tabParameters}
\end{table}

The parameters associated for the best fit to $G_E$ and $G_C$
with the described constrains
are displayed in Table~\ref{tabParameters}.
In the following, we represent the meson cloud contributions
associated with the fit by the dashed-dotted-lines.
The valence quark (bare) contributions are 
represented by dashed-lines.
The result of the combination of the valence quark and the 
meson cloud contributions 
is represented in the same graph by the solid-lines.
We start our discussion with the results for proton targets.
The final results for the $F_1^\ast$ and $F_2^\ast$ form 
factors for proton target are presented in Fig.~\ref{figN1535R}.
The corresponding results for $G_E$ and $G_C$
are presented in Fig.~\ref{figN1535R-2}.


In the figures, one notices a sharp 
variation of the  functions $G_E$ and $G_C$
at low $Q^2$, more particularly in the range $Q^2=0$--0.3 GeV$^2$.
Those results are a consequence 
of the fit to the low-$Q^2$ data 
and of the lack of data in the region $Q^2=0$--0.3 GeV$^2$.
New data in that region are necessary 
for more definitive conclusions relative to 
the shape of $G_E$ and $G_C$ at low $Q^2$~\cite{Compton,Siegert4}.
It is interesting to notice, however,
that the shape of  $G_C$ near $Q^2=0$ is similar 
to the shape estimated with the constraints 
from Siegert's theorem~\cite{Siegert1,Compton}.

The results of the form factors for neutron targets
are presented in Figs.~\ref{figN1535R-3} and \ref{figN1535R-4}.
From the figures, we can conclude that the magnitudes 
of the $F_1^\ast$ and  $F_2^\ast$ are smaller than those
for proton target.
As for the $G_E$ form factor 
it is interesting to notice that $G_E$ is very small 
(except for $Q^2 < 0.25$ GeV$^2$) as a consequence of the 
cancellation between valence quark and meson cloud contributions for the neutron case.
As for $G_C$, one notices that it is larger 
in magnitude than in the case of the proton,
as the consequence a less significant cancellation 
between valence quark and meson cloud contributions.

Our result of $G_C$ for the proton target 
at low $Q^2$ (Fig.~\ref{figN1535R-2}) requires some extra discussion.
In the graph the function 
changes sign, when $Q^2$ approaches the photon point.
But due to the lack of data below $Q^2=0.3$ GeV$^2$, 
we cannot say that this change of sign is imposed by the data.
Other parametrizations suggest that $G_C$ is small
near $Q^2=0$~\cite{MAID1,Compton,Siegert4}, but
the present data are unable to determine the exact sign. In our model, the magnitude of $G_C$ near $Q^2=0$
is related to the regularization of the 
form factors $F_i^\ast (Q^2)$, specifically 
the factor $\sqrt{\frac{Q^2}{\Lambda_R^2 + Q^2}}$ from Eq.~(\ref{eqZ}).
In appendix~\ref{appGC0}, we explicitly demonstrate  this by decomposing $G_C(0)$ 
into three terms, two positive in sign,
associated with $F_2^\ast(0)$ and with the meson cloud 
contribution to $F_1^\ast$,  and one negative, 
proportional to $1/\Lambda_R^2$.
Then,  a small value for $\Lambda_R$,
leads to a large cancellation of terms and a small value for $G_C(0)$.
A larger value for $\Lambda_R$, like  $\Lambda_R = m_\rho$,
reduces the magnitude of that cancellation
and increases the value of $G_C(0)$.
In summary, the value for $G_C(0)$ 
is a consequence of the value $\Lambda_R$ chosen to estimate 
the bare contribution of the transition form factors,
and it is not well constrained by the data.

On the other hand, the combined fit to the 
proton and neutron data for $G_E(0)$
is weakly dependent on the parameter $B(0)$.
This is a consequence of the small error bars 
of the $F_2^\ast$ data for the proton, 
for finite $Q^2$, and the large error bars at $Q^2=0$ of the data 
for the proton and neutron (PDG data).
For that reason the finite $Q^2$ points have a stronger 
impact on the fit, leaving less room for the much less constrained data at $Q^2=0$.

\begin{figure}[t] 
\vspace{.5cm}
\centerline{\mbox{
\includegraphics[width=7.5cm]{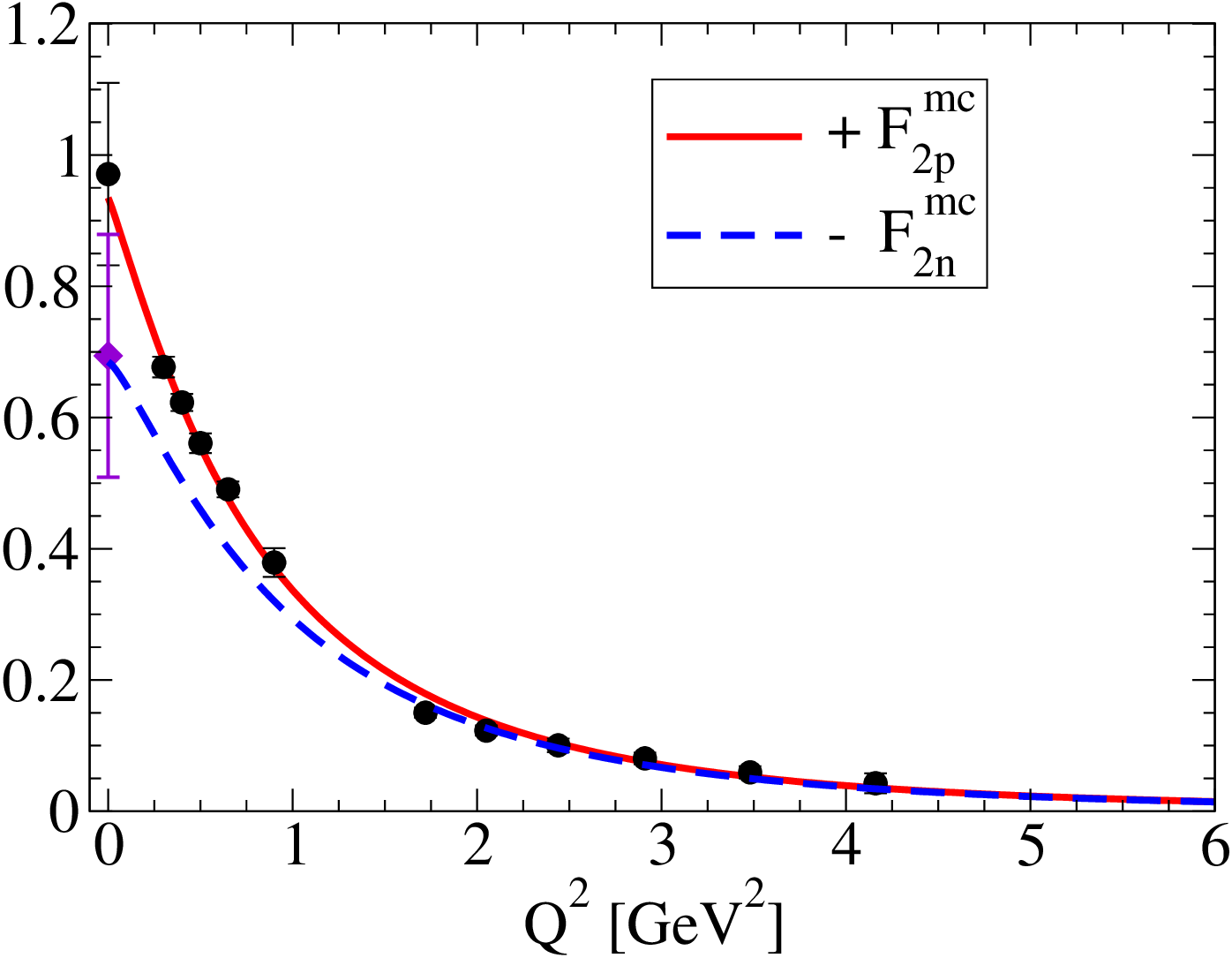}}}  
\caption{\footnotesize
Data estimate of the functions $F_{2p}^{\rm mc}$ 
(circles) and $-F_{2n}^{\rm mc}$ (diamond) for the  
$\gamma^\ast N \to N(1535)$ transition.}
\label{figF2mc}       
\end{figure}

To exhibit the control on the uniqueness of 
the parameters obtained in the fitting procedure we show in Fig.~\ref{figF2mc} the 
results of $F_2^{\rm mc}$ for the proton and neutron cases 
in comparison with their parametrizations by $A + B$ and $A-B$ respectively.
The $F_2^{\rm mc}$ {\it experimental points}  are determined by 
$F_2^{\rm mc} = F_2^\ast - F_2^{\rm B}$, 
where $F_2^\ast$ is the experimental value and $F_2^{\rm B}$ the model result, 
for both proton and neutron targets
(there is a single  {\it experimental point} for the neutron case).

Those {\it experimental points} are included just to guide the eye,
since the meson cloud parametrization are determined 
by the direct fit to the $G_E$ and $G_C$ data.
If we neglect the isoscalar component, we obtain  $F_{2n}^{\rm mc} = - F_{2p}^{\rm mc}$, 
or in other words, the difference between the red and blue curves
of Fig.~\ref{figF2mc} is due to that component alone.

Figure~\ref{figF2mc} shows also that increasing or decreasing the estimate for $F_{2n}^{\rm mc}$ 
by about 0.05 (one third of the error bar)
the result for $F_{2n}^{\rm mc}(0)$,
is still consistent with its experimental limits.
This result shows, that  $B(0)$ can vary within a certain range without changing 
the results for $F_{2p}^{\rm mc}$ provided 
that the value for $A(0)$ is redefined in order 
to keep the result for  $F_{2p}^{\rm mc}(0)  \simeq 0.97$
(estimated based on experimental data, Table~\ref{tabN1535-1}).
In summary, we obtain a solid estimate 
for the proton data, but 
a poorer estimate of the neutron data.
This happens because the neutron data is constrained only by
one data point ($Q^2=0$) with a large error bar.

\begin{figure}[t] 
\vspace{.5cm}
\centerline{\mbox{
\includegraphics[width=7.5cm]{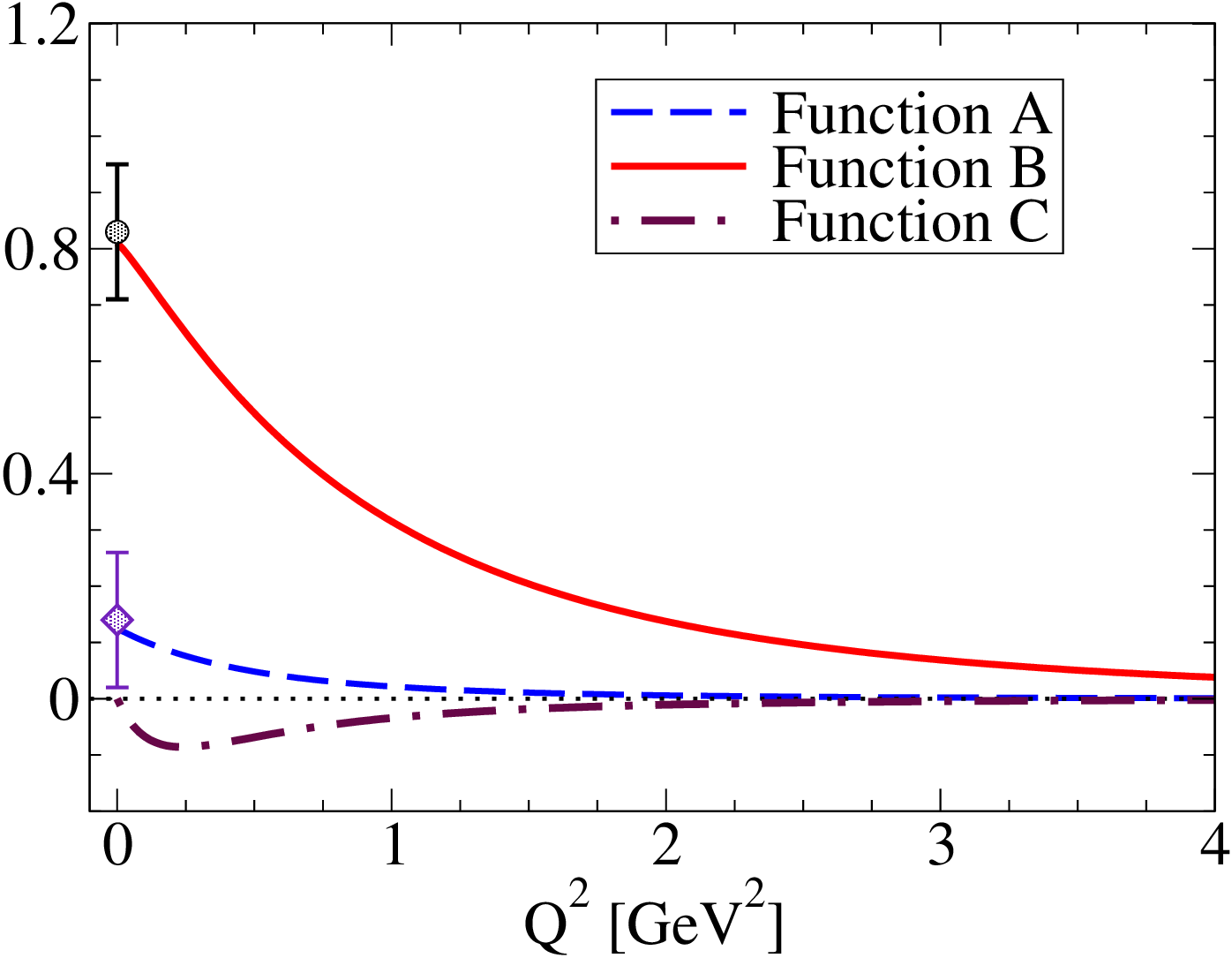}}}  
\caption{\footnotesize
$\gamma^\ast N \to N(1535)$ transition:
functions $A$, $B$ and $C$ used to parametrize 
the meson cloud contributions to the transition form factors 
$F_1^{\rm mc}$ and $F_2^{\rm mc}$.}
\label{figN1535-AB}       
\end{figure}

We represent separately in Fig.~\ref{figN1535-AB},
the functions $A$, $B$ and $C$, parametrizing the meson cloud effects.
In the case of the functions $A$ and $B$,
we include also their experimental limits at $Q^2=0$ presented in Table~\ref{tabN1535-1}.
We recall that i) by construction, and to enforce the isovector dominance for larger values of $Q^2$
the ratio $A/B$ is smaller than
the ratio at $Q^2=0$ (about $0.15/0.82 \approx  0.2$); ii) there 
are no experimental constraints except for \mbox{$C(0)=0$.}

We emphasize that the isovector 
character of $F_1^{\rm mc}$, from Eq.~(\ref{eqF1mc}) is an ansatz.
No empirical information is available at the moment 
that allows us to test this assumption.

From the study of the proton and neutron form factors, 
we conclude that is possible to obtain 
an accurate description of the proton target data.
The results for the neutron target, however, are poorly constrained.
More precise calculations of the transition 
form factors with neutron targets 
are possible only with more accurate constraints from the neutron sector.

It is worth mentioning, 
that we can obtain almost equivalent descriptions 
of the proton and neutron target data 
with a small modification of the parameter $A(0)$,
provided that $B(0)$, is readjusted 
such that $B(0) \simeq [F_{2p}^\ast(0)]_{\rm exp} - A(0)$ holds,
and keeping all the remaining parameters unchanged.
The results for the proton target remain almost unchanged 
and the results for $F_2^\ast$ for the neutron target are modified 
according with the new values for $A(0)$ and $B(0)$.
This property may be very useful in future works,
allowing us to investigate the sensitivity of the  
neutron transition form factors for different classes of 
parametrizations characterized by different $A(0)$.

\subsection{Summary of the results in the spacelike region}

Combining the parametrizations of the meson cloud contribution 
with results from the valence quark contribution
from the covariant spectator quark model
we obtain a good description of the proton target data 
for $G_E$ and $G_C$ and 
at the same time a  good description of 
the form factors $F_1^\ast$ and  $F_2^\ast$ in the spacelike region.
Also, we have identified the isoscalar and isovector the contributions, 
based on the Pauli-Dirac representation, 
in a model that accounts for transitions both
with proton and neutron targets. 

Next, in the extension of their results to the timelike region,
we will focus on the form factors $G_E$ and $G_C$, 
since timelike formulas for decay widths are more readily
expressed in terms of those form factors.

\begin{figure*}[t] 
\vspace{.5cm}
\centerline{\mbox{
\includegraphics[width=7.5cm]{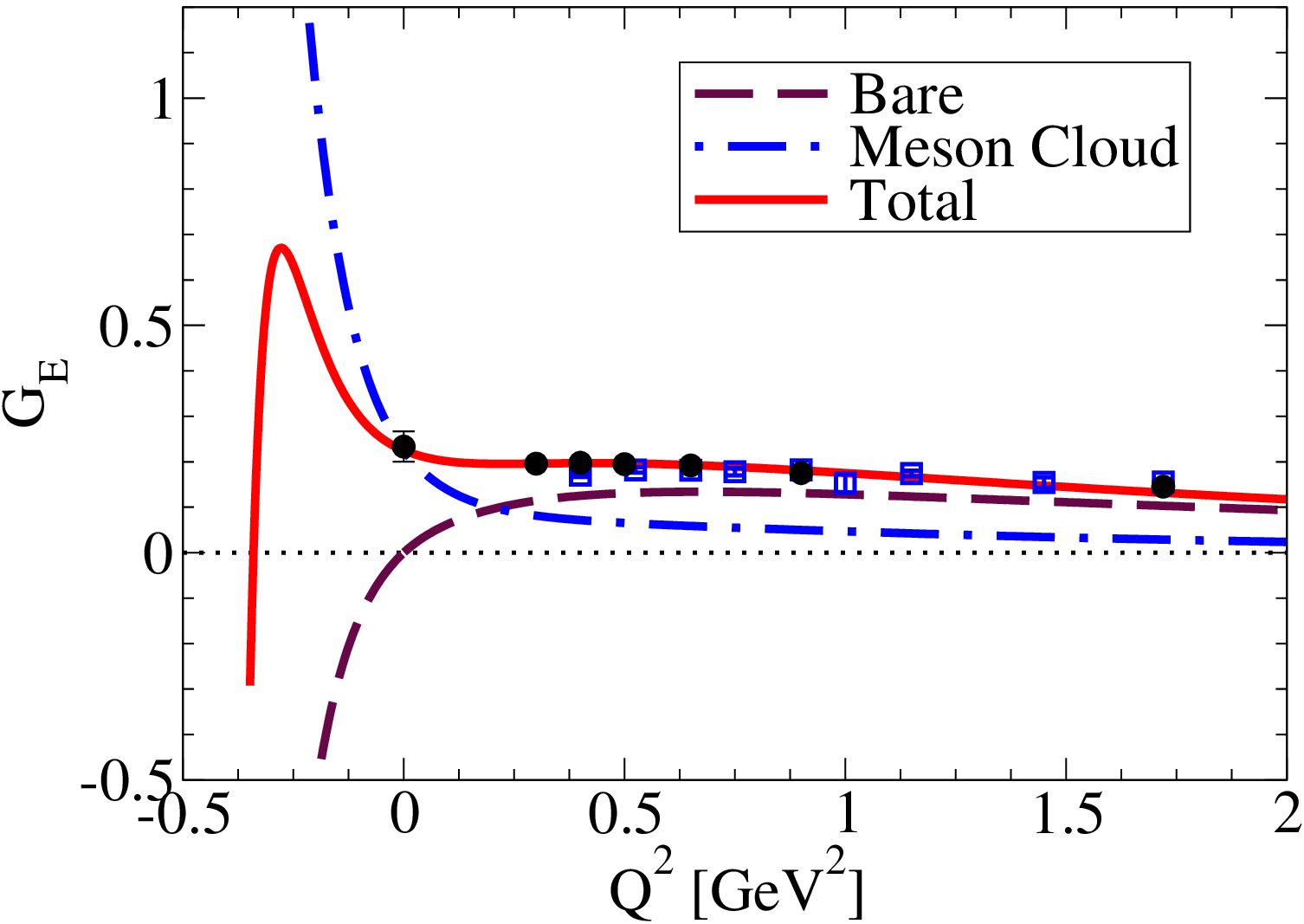}}  \hspace{1.cm}
\mbox{
\includegraphics[width=7.5cm]{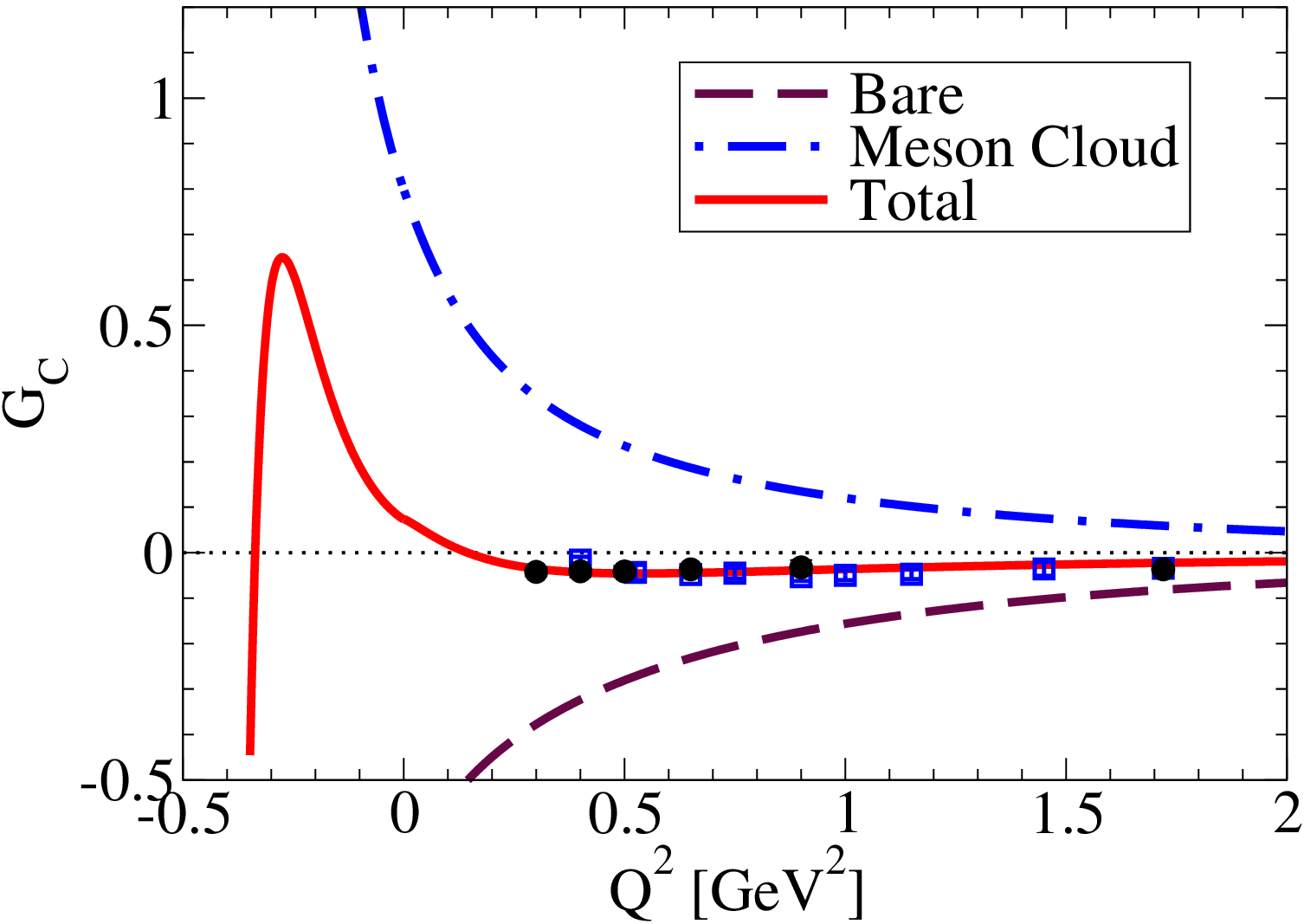}}}
\caption{\footnotesize
Real part of $\gamma^\ast N \to N(1535)$ transition form factors 
in the spacelike and timelike region,
for proton target for $W=1.535$ GeV.
Same data as in Fig.~\ref{figN1535R}.} 
\label{figN1535R-TL3}       
\end{figure*}
\begin{figure*}[t] 
\vspace{.5cm}
\centerline{\mbox{
\includegraphics[width=7.5cm]{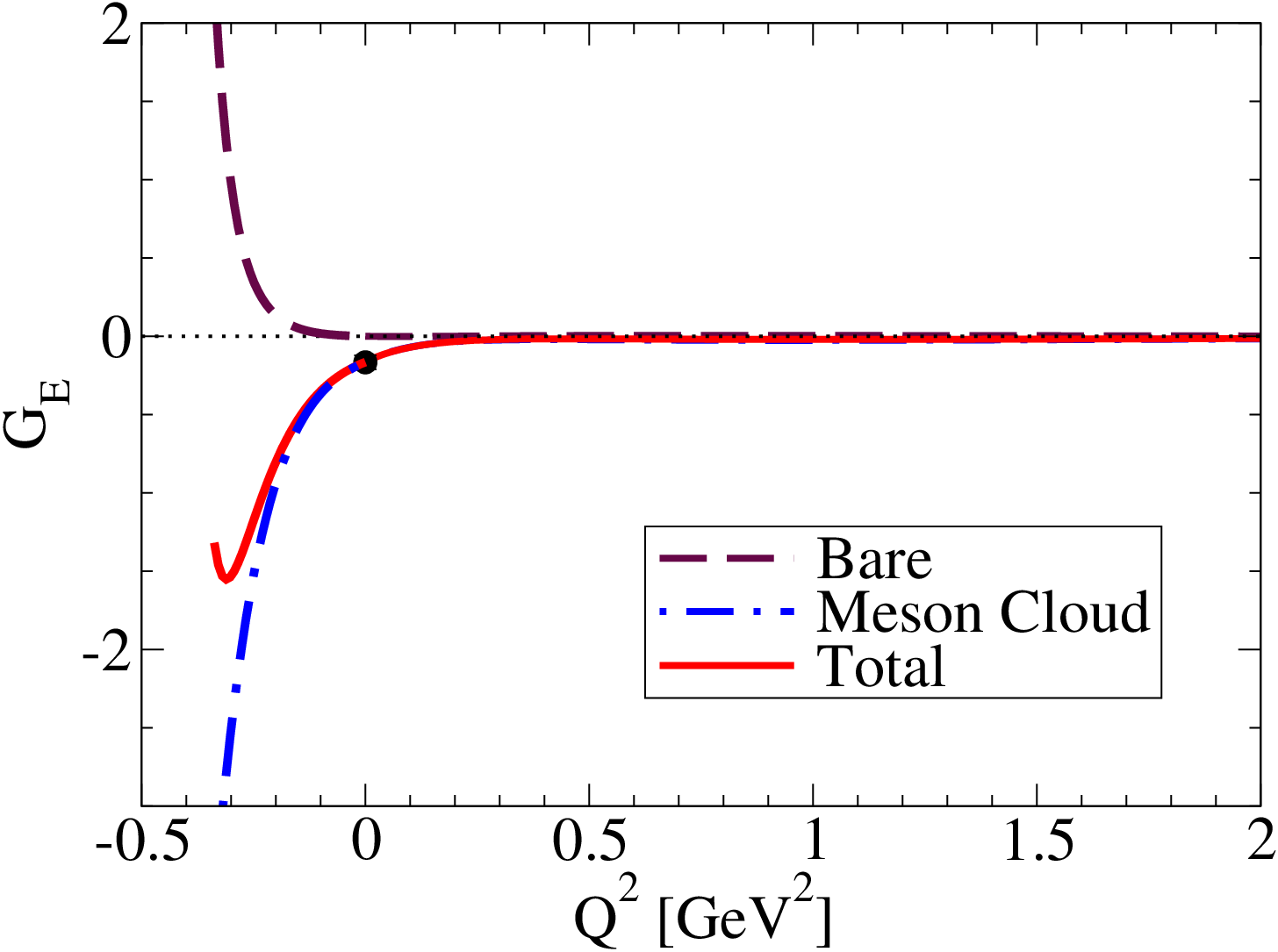}}  \hspace{.5cm}
\mbox{
\includegraphics[width=7.5cm]{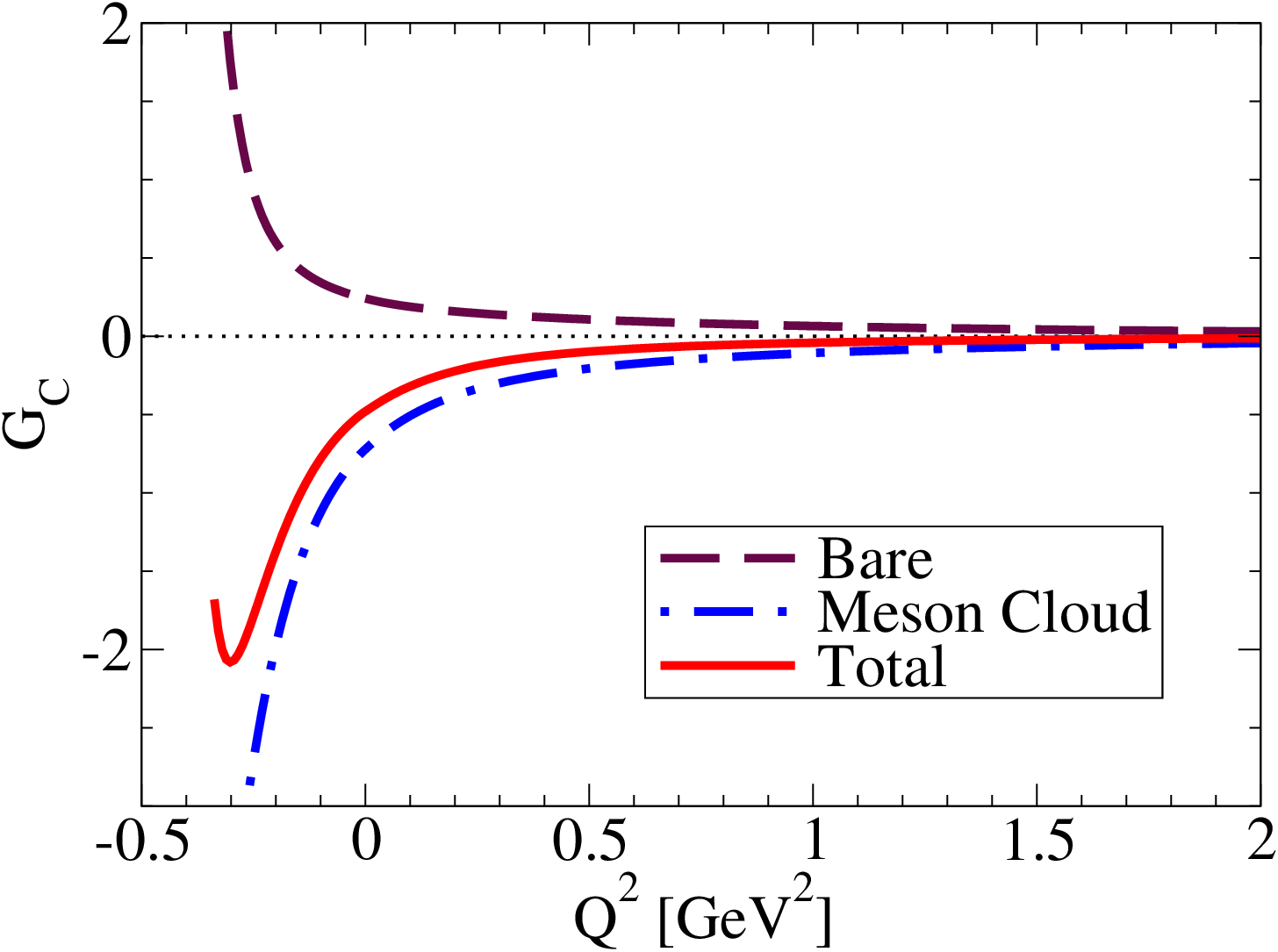}}}
\caption{\footnotesize
Real part of 
$\gamma^\ast N \to N(1535)$ transition form factors
in the spacelike and timelike region,
for neutron target for $W=1.535$ GeV.
The data at $Q^2=0$ is from PDG~\cite{PDG}.}
\label{figN1535R-TL-n}       
\end{figure*}

\begin{figure*}[t] 
\vspace{.5cm}
\centerline{\mbox{
\includegraphics[width=7.5cm]{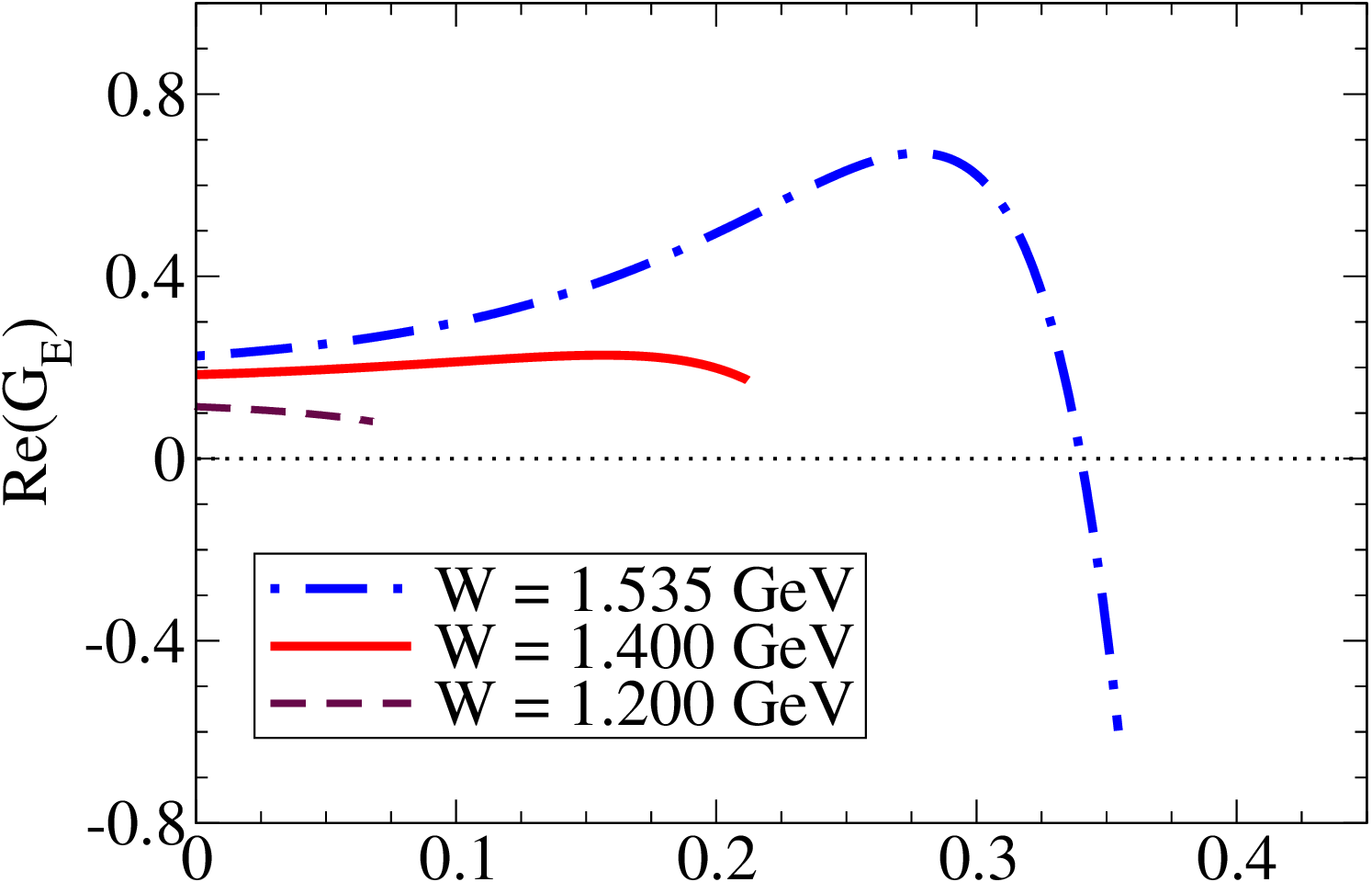}}  \hspace{1.cm}
\mbox{
\includegraphics[width=7.5cm]{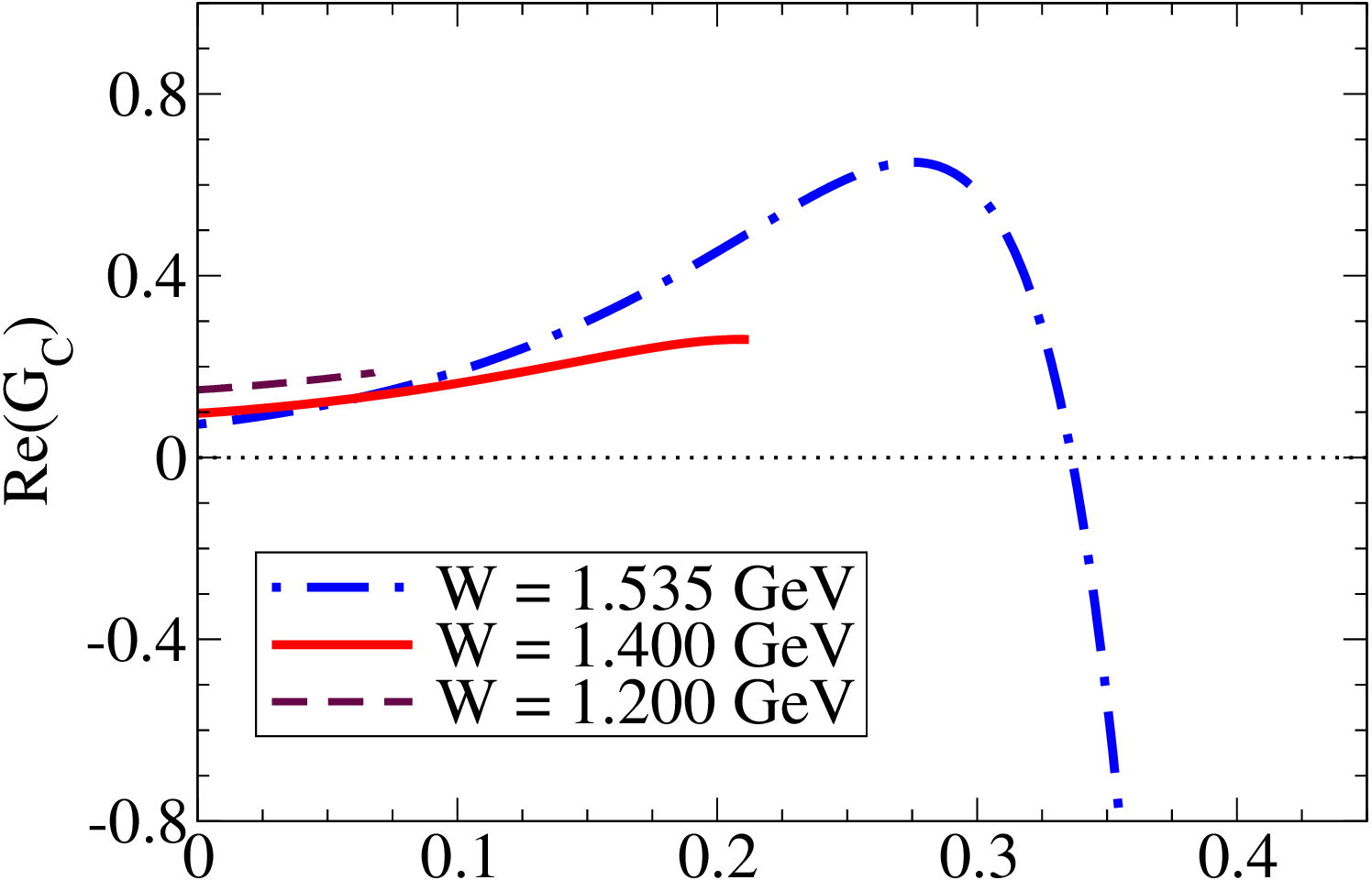}}}
\vspace{.94cm}
\centerline{\mbox{
\includegraphics[width=7.5cm]{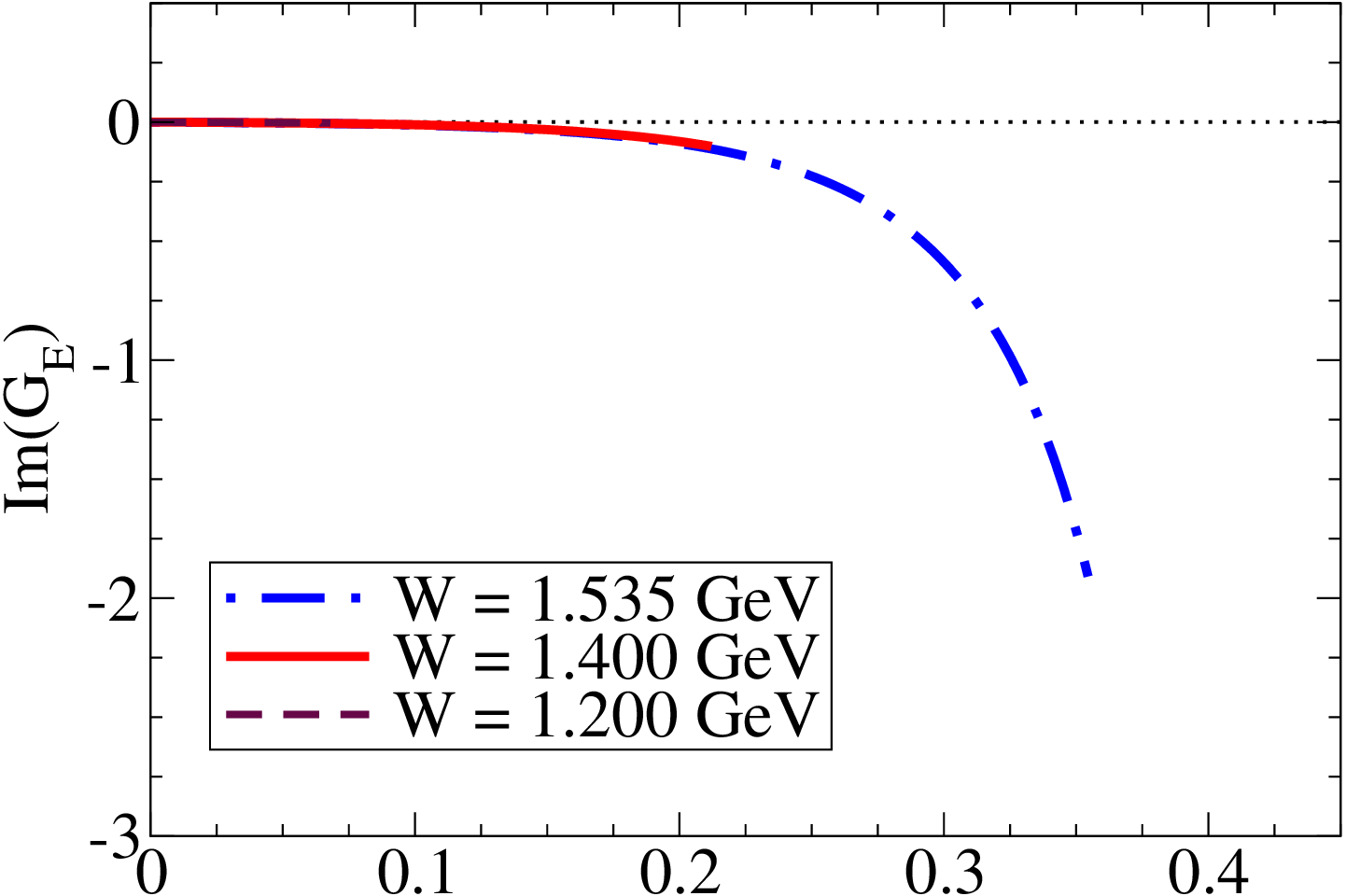}}  \hspace{1.cm}
\mbox{
\includegraphics[width=7.5cm]{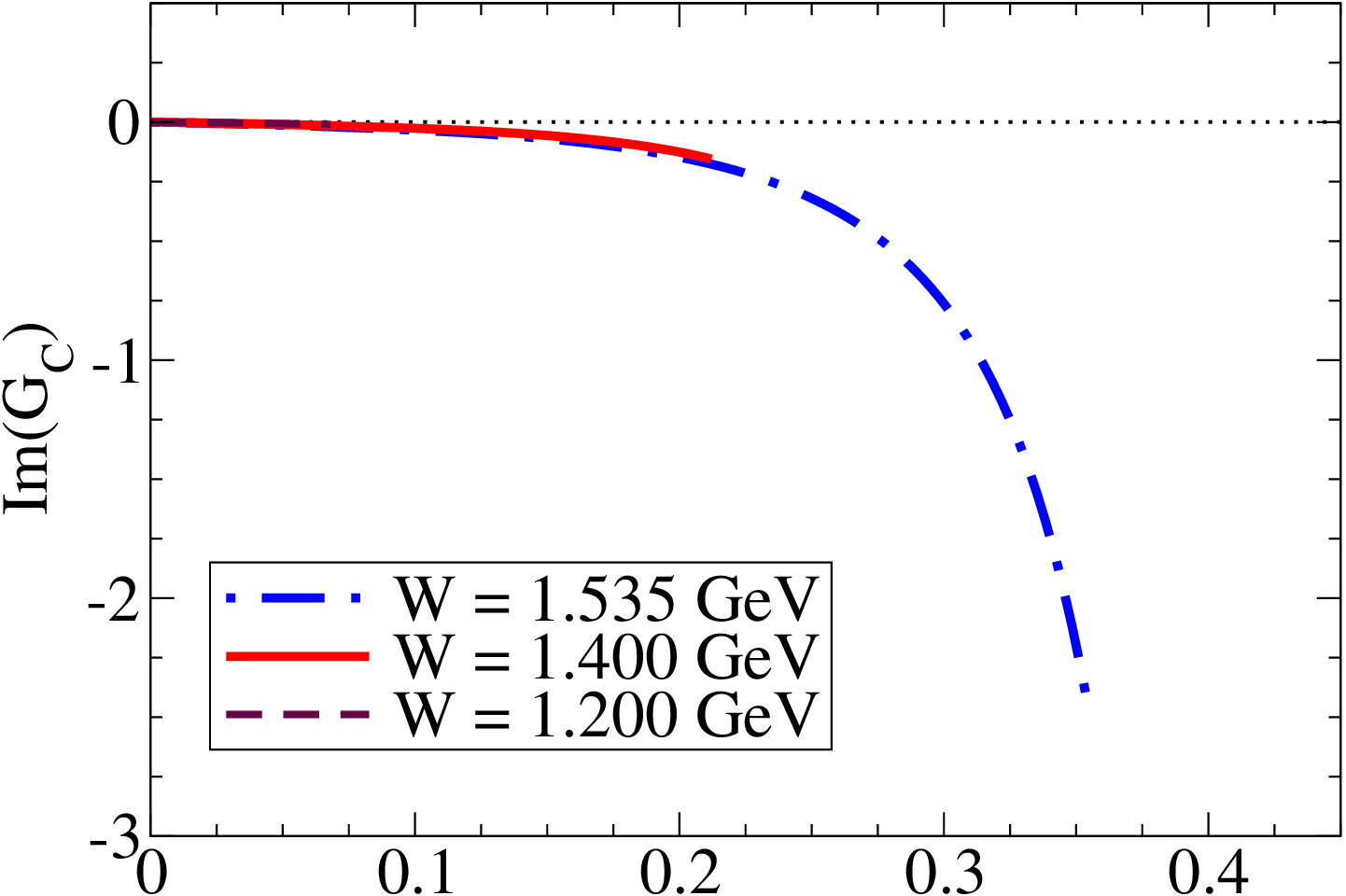}}}
\vspace{.9cm}
\centerline{\mbox{
\includegraphics[width=7.5cm]{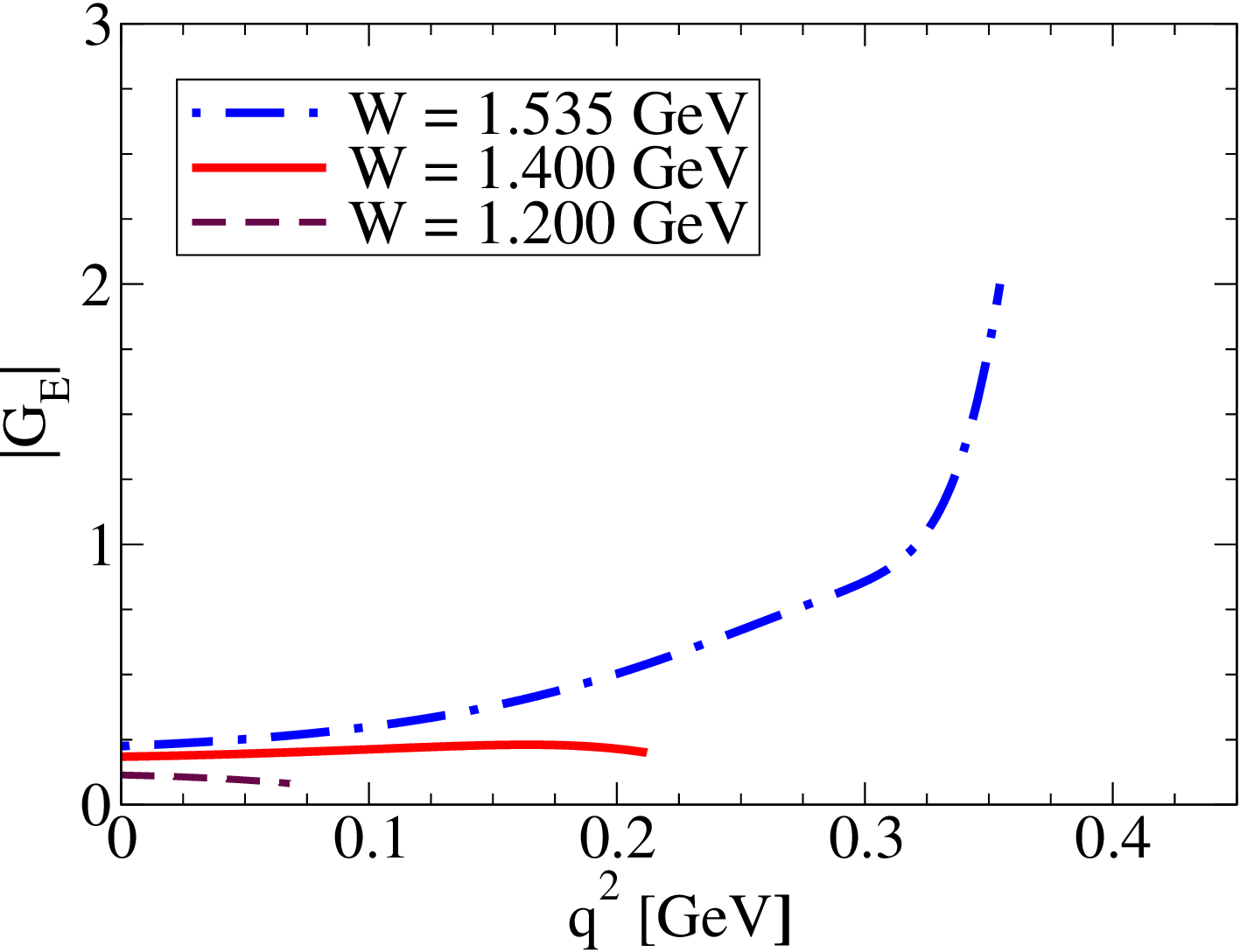}}  \hspace{1.cm}
\mbox{
\includegraphics[width=7.5cm]{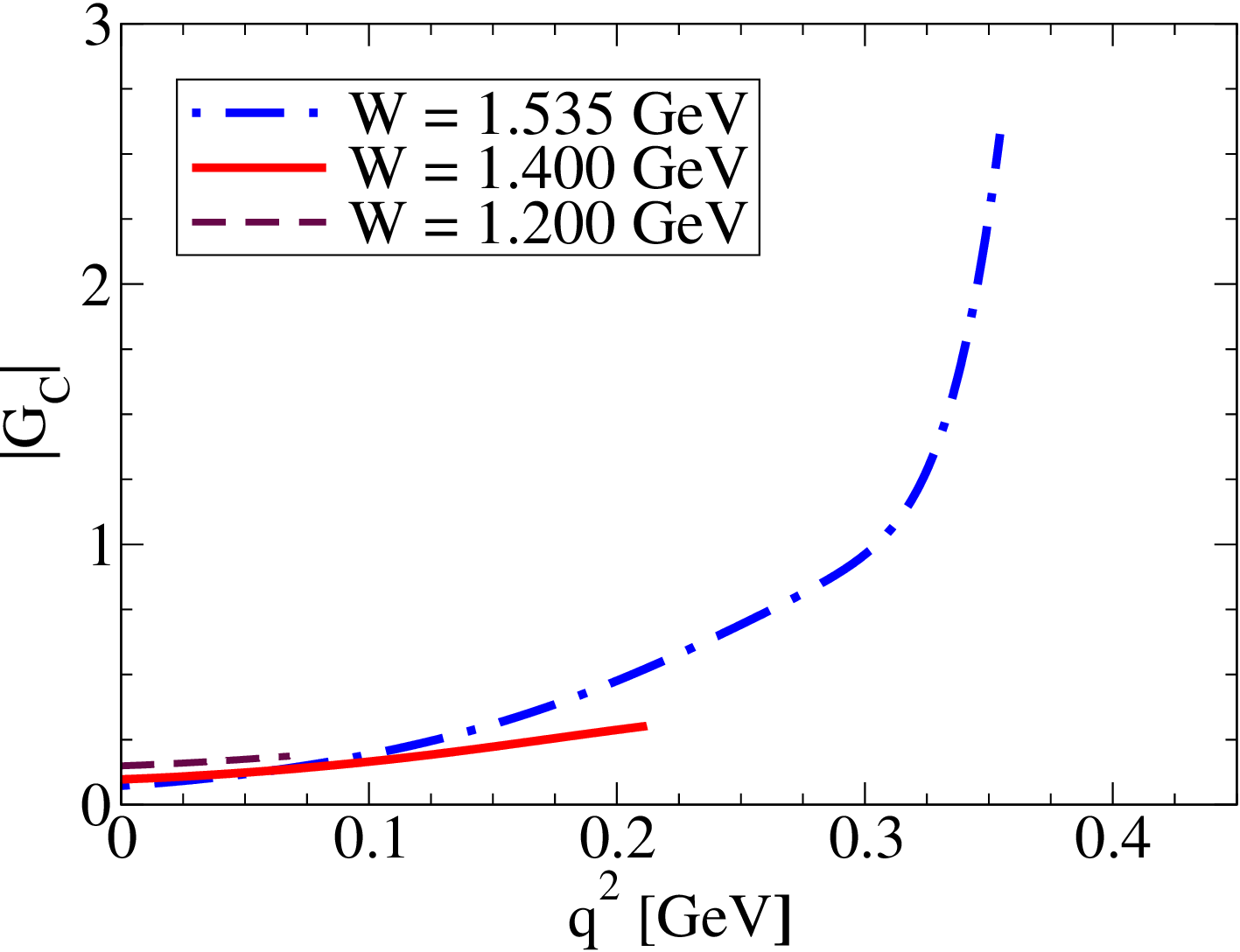}}}
\caption{\footnotesize
$\gamma^\ast N \to N(1535)$ transition form factors
for proton target for different values of $W$.} 
\label{figN1535R-TL4}       
\end{figure*}

\begin{figure*}[t] 
\vspace{.5cm}
\centerline{\mbox{
\includegraphics[width=7.5cm]{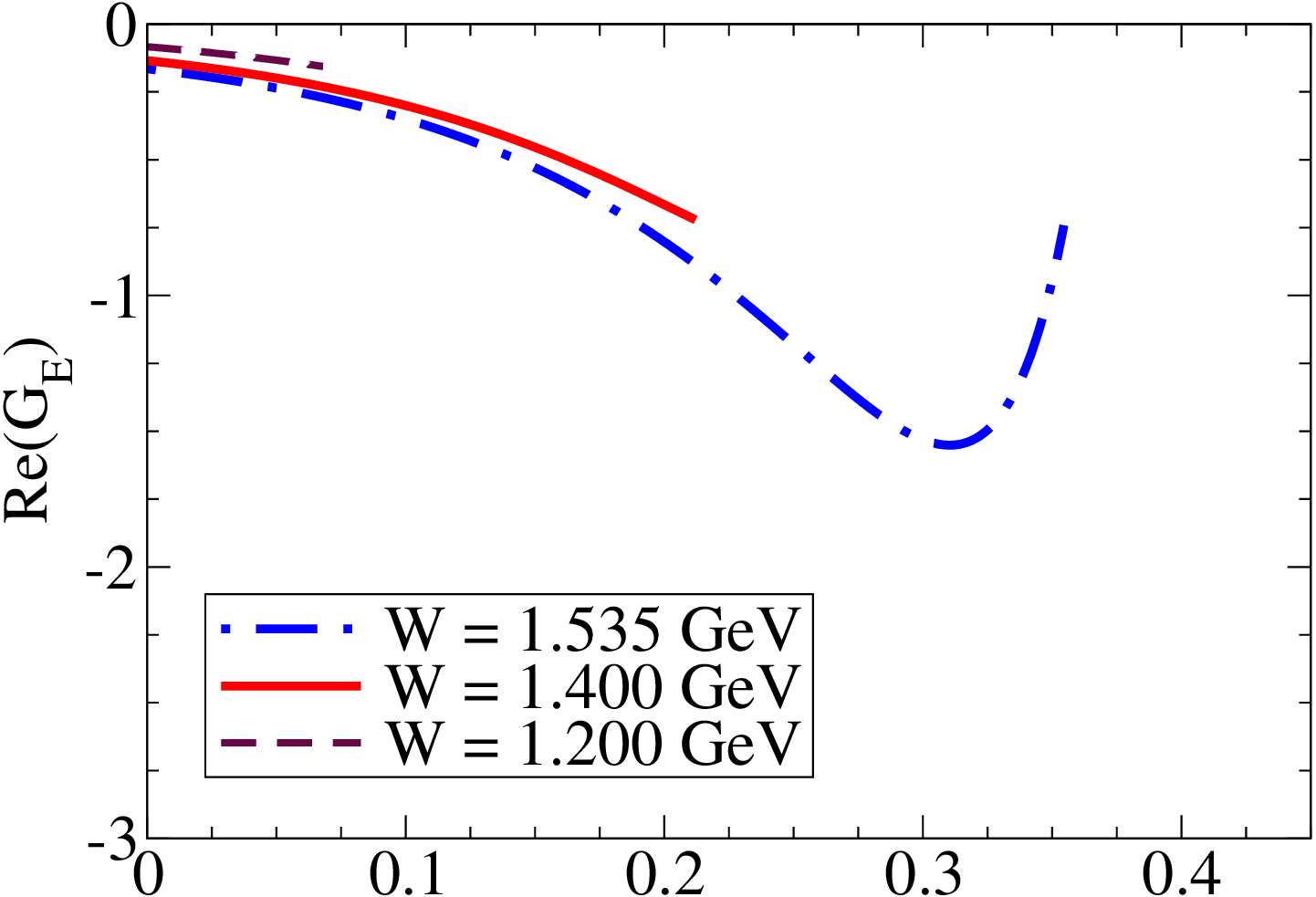}}  \hspace{1.cm}
\mbox{
\includegraphics[width=7.5cm]{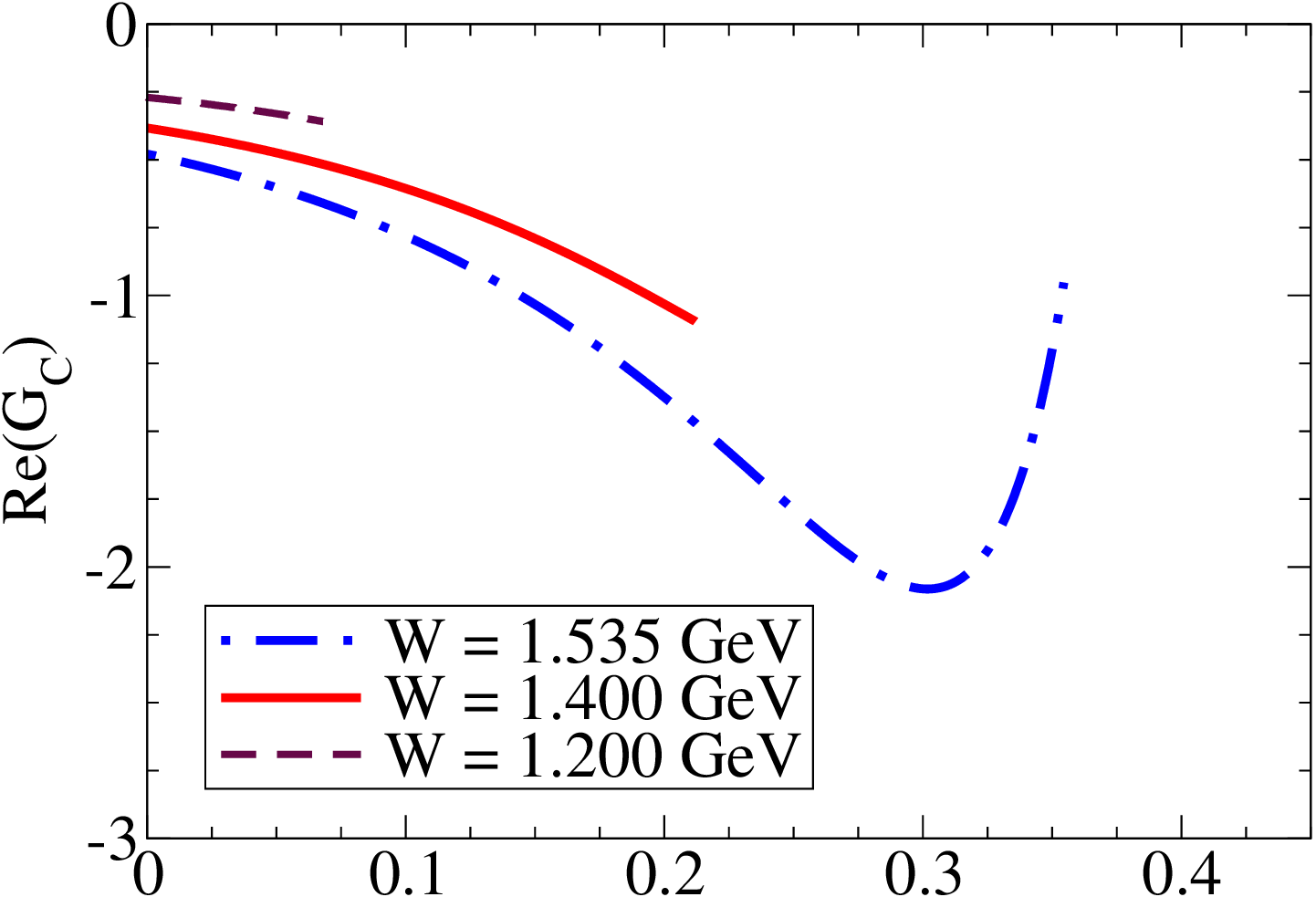}}}
\vspace{.9cm}
\centerline{\mbox{
\includegraphics[width=7.5cm]{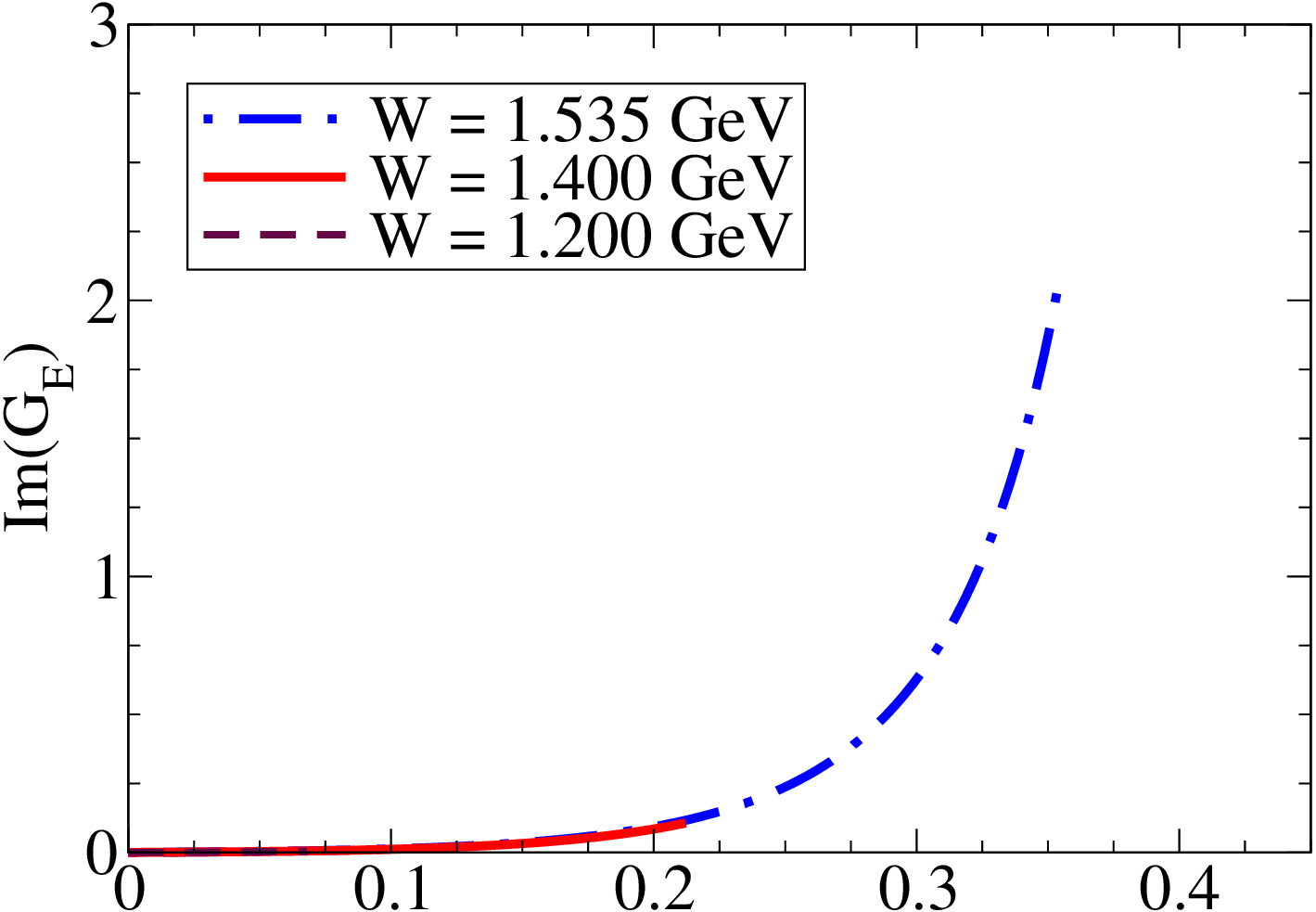}}  \hspace{1.cm}
\mbox{
\includegraphics[width=7.5cm]{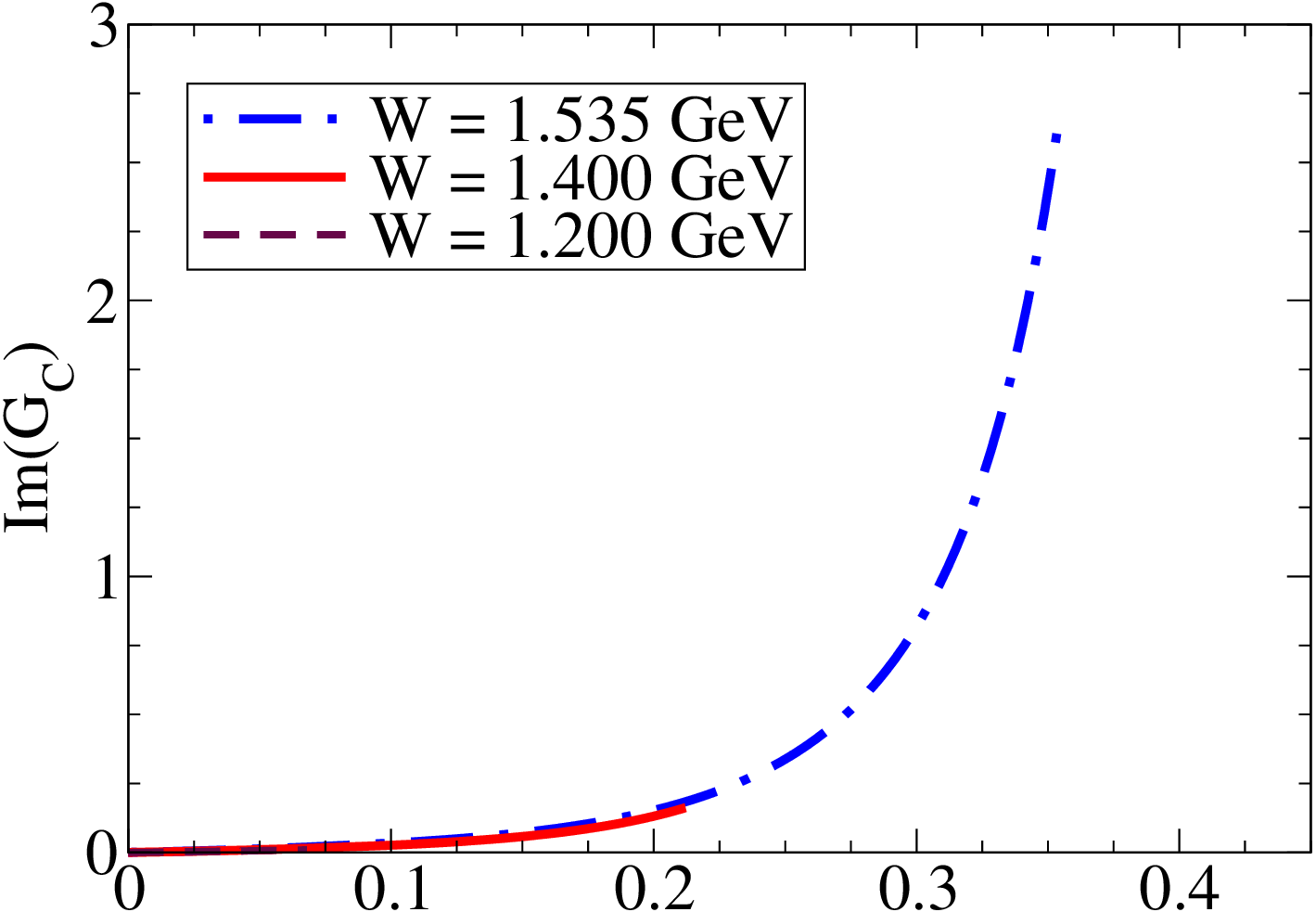}}}
\vspace{.9cm}
\centerline{\mbox{
\includegraphics[width=7.5cm]{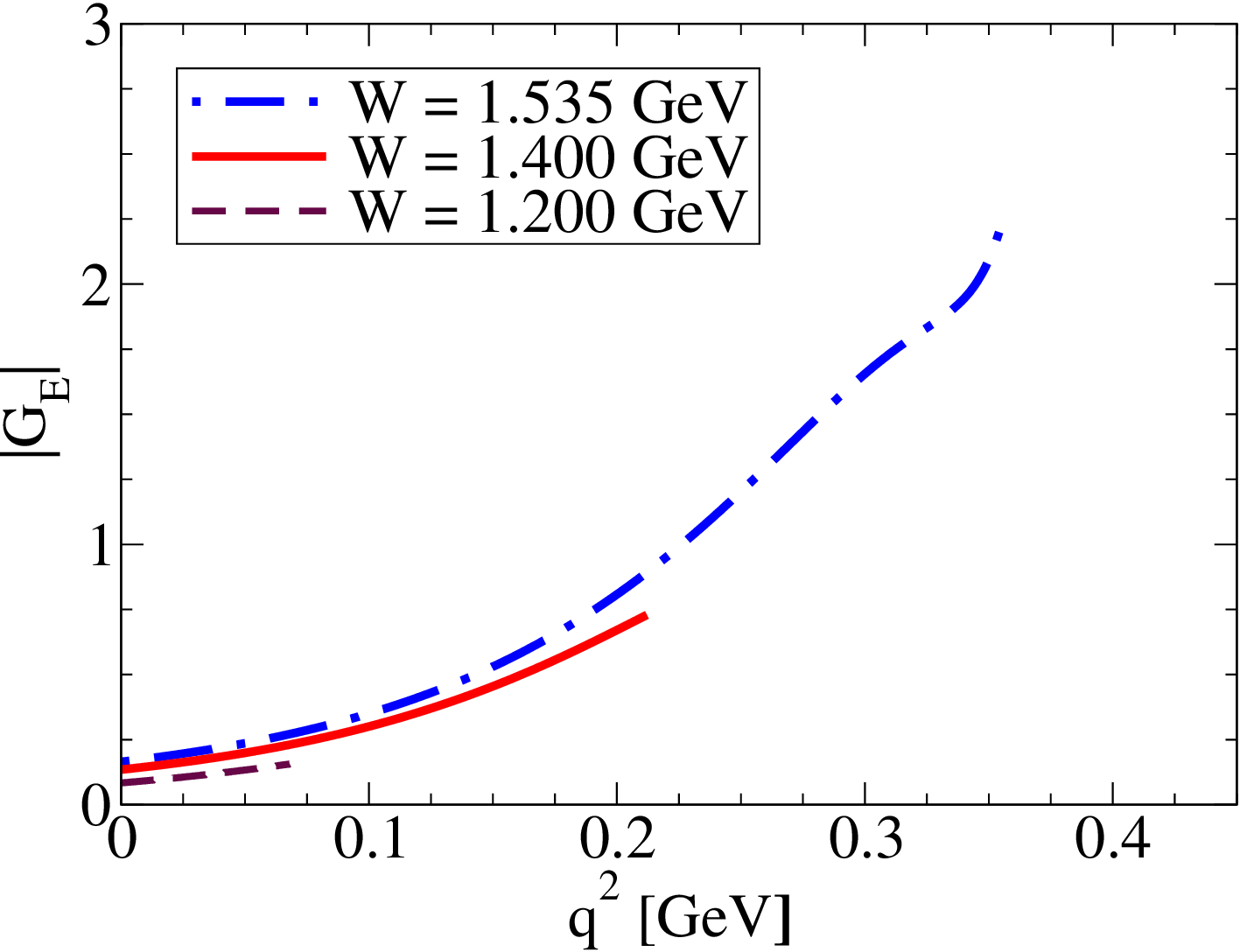}}  \hspace{1.cm}
\mbox{
\includegraphics[width=7.5cm]{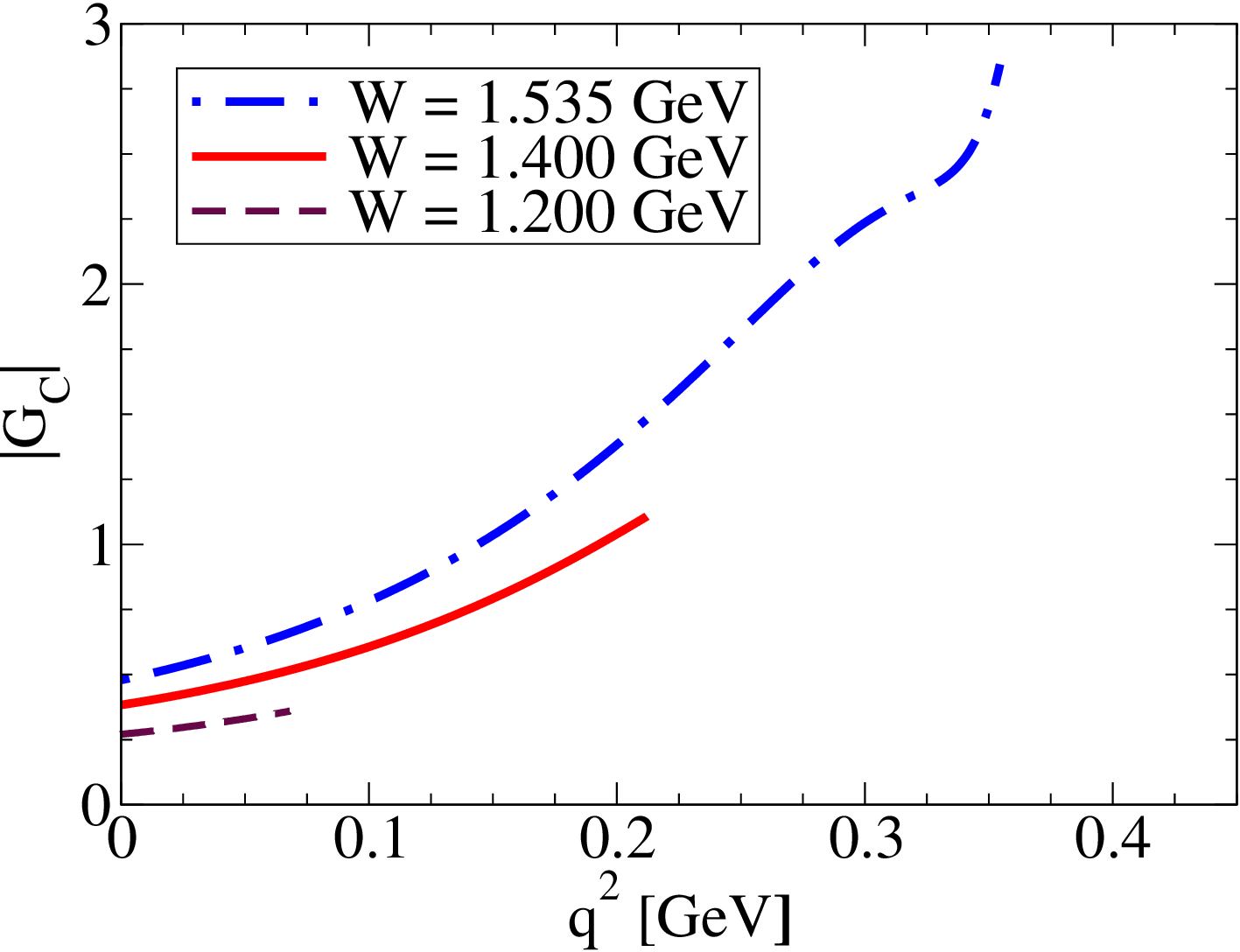}}}
\caption{\footnotesize
$\gamma^\ast N \to N(1535)$ transition form factors
for neutron target for different values of $W$.} 
\label{figN1535R-TL4-n}       
\end{figure*}

\begin{figure*}[t] 
\vspace{.5cm}
\centerline{\mbox{
\includegraphics[width=7.5cm]{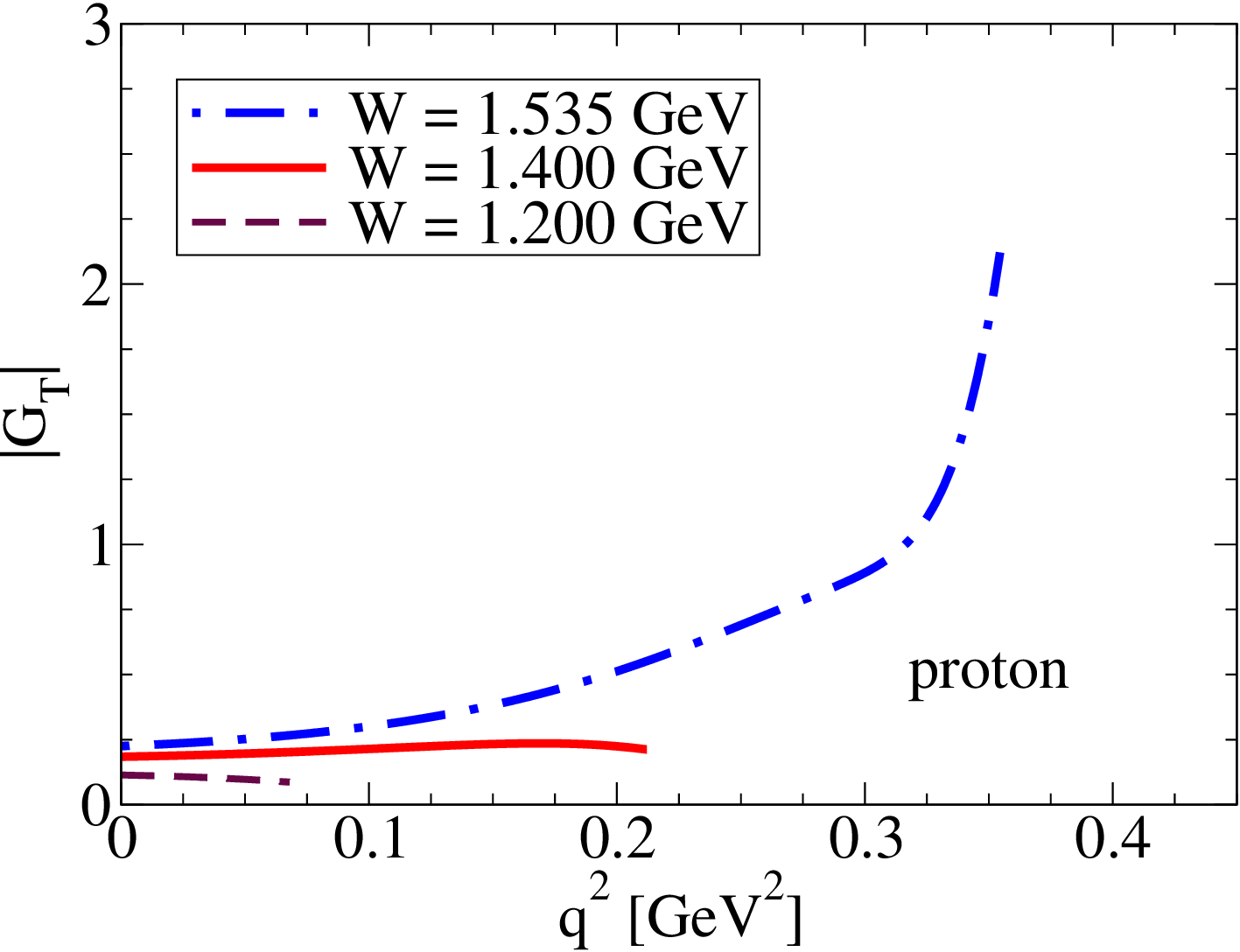}}  \hspace{1.cm}
\mbox{
\includegraphics[width=7.5cm]{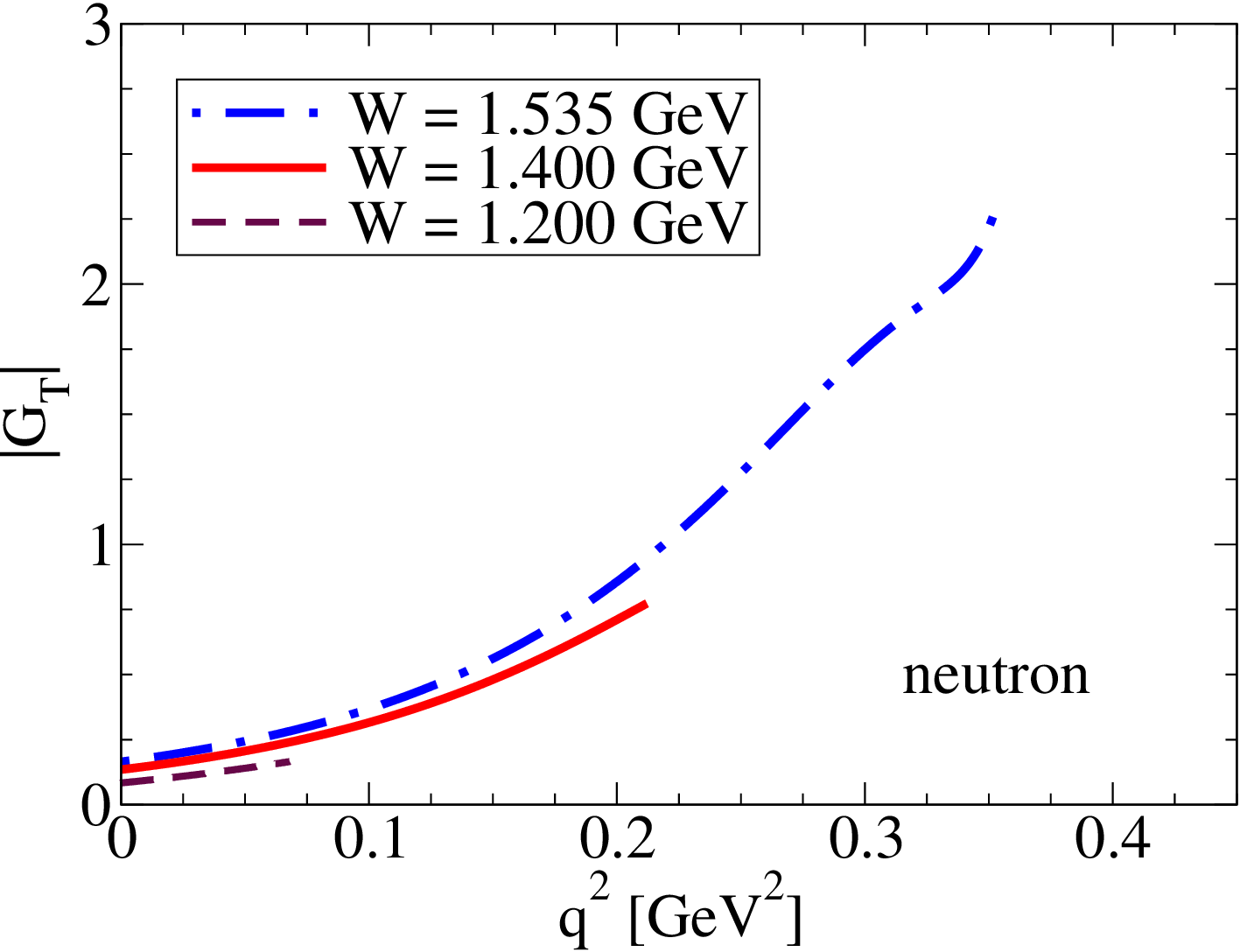}}}
\caption{\footnotesize
Effective form factor function $|G_T|$ 
for different values of $W$.
Comparison between proton and neutron results.}
\label{figN1535R-TL6}       
\end{figure*}


\section{Extension to the timelike region}
\label{secTimelike}

We describe now how the extension 
of the model from the spacelike region 
for the timelike region.
In the timelike region, we vary the energy $W$
of the $\gamma^\ast N$ system, which may differ from the $N(1535)$ 
resonance mass ($M_R$).
In transforming the spacelike formulas to the 
timelike region we then replace $M_R$ by $W$.
As done in the spacelike region, we decompose 
the form factors into the bare and meson cloud 
contributions, according to Eqs.~(\ref{eqF1decomp})--(\ref{eqF2decomp}).

A note about the range of application of our framework is in order.
We are aiming at the region of $W$ 
accessible in the present day experiments, 
in particular the range of $W$ probed 
at HADES, which is restricted to typical values 
of $W \simeq 1.490$ GeV~\cite{HADES17a,HADES17b}.
Since, according to the kinematic relations 
(\ref{eqQR}) and (\ref{eqQpm}), the values 
of $q^2$ are restricted to $q^2 \le (W -M_N)^2$,
one concludes that the square invariant moment
of the dilepton are limited to $q^2 \le 0.3$ GeV$^2$.

The formalism described in the present work 
for the valence quark component of the model is
constrained by the scale $\Lambda_R = m_\rho$ 
and therefore restricted to the upper limit $W= M_N + m_\rho \simeq 1.7$ GeV.
The numerical approximations discussed below, however, 
are also limited to not very large values for $q^2$,
reducing the range of application of $W$ to the order 1.6 GeV, 
still within the window covered by the HADES experiments.
In comparison with other resonances 
described by the covariant spectator quark model,
$\Delta(1232)$ and $N(1520)$~\cite{Timelike,Timelike2,N1520TL},
the $N(1535)$ resonance is the one lying closer to the $\rho$-pole.
Thus, for large $W$, there is the possibility 
of enhancement of the transition form factors.

\subsection{Valence quark contributions }

In the present work, we use the timelike 
extension of the quark electromagnetic form factors
of the covariant spectator quark model defined 
by the study of the $\gamma^\ast N \to N(1520)$ transition 
in the timelike region~\cite{N1520TL}.
Note that differently to the $\gamma^\ast N \to \Delta(1232)$ transitions
the $\gamma^\ast N \to N(1535)$ and $\gamma^\ast N \to N(1520)$ 
transitions require isovector and isoscalar components. 

The extension to timelike of the valence quark contributions is based 
on the analytic expressions for $F_1^{\rm B}$ and $F_2^{\rm B}$
in the spacelike region,
given by Eqs.~(\ref{eqF1a})--(\ref{eqF2a}).
In those expressions we convert $Q^2 \to -q^2$, 
and replace the physical mass $M_R$ by $W$.

One can factorize the bare form factors 
(\ref{eqF1a})--(\ref{eqF2a})
into two leading factors:
One first factor includes all the functions $j_i^{A,S}$ ($i=1,2$) comprising the
quark form factors, according to Eqs.~(\ref{eqjiS})--(\ref{eqjiA}).
There, the quark isoscalar and isovector form factors 
contain the vector meson poles, including the mesons $\rho$ and $\omega$,
and are then naturally defined in the timelike region 
with the introduction of $q^2$-dependent widths.

The other factor is the product of $Z$ from Eq.~(\ref{eqZ})
and the integral ${\cal I}_R$ over the radial wave functions in 
Eq.~(\ref{eqIR}), and reads
\ba
{\cal I}_R^\prime = {\cal I}_R \sqrt{ \frac{Q^2}{m_\rho^2 + Q^2}}.
\label{eqIRp1}
\ea

Remember that, as discussed in 
the context of the semirelativistic approximation,
the overlap integral ${\cal I}_R$ of Eq.~(\ref{eqIR}) cannot
be evaluated below $Q^2=0$ because 
the region $- (W -M_N)^2 \le Q^2 < 0$ cannot be accessed. 
One can, however, use a numerical extrapolation of 
the results in the spacelike region to the timelike region,
using an analytic continuation for $Q^2 < 0$.

The analytic continuation of ${\cal I}_R$ is based on the observation 
that the in the spacelike region the function ${\cal I}_R/|{\bf q}|$
is well described by a dipole form for small values of $Q^2$.
One uses then the replacement:
\ba
\frac{{\cal I}_R}{|{\bf q}|}  \to \frac{G_D(q^2)}{M_N},
\label{eqGD}
\ea
where $G_D$ is a dipole function 
with $G_D(0)$ and a cutoff $\Lambda_D$ 
determined by the values of $Q^2$ close to $Q^2=0$.
The details of this procedure are presented in Appendix~\ref{appIntR}.

Combining the analytic extension of the two factors 
from Eq.~(\ref{eqIRp1}), we obtain  (see Appendix~\ref{appIntR}):
\ba
{\cal I}_R^\prime  \simeq - \frac{q^2  G_D(q^2)}{M_N(W + M_N)}
\sqrt{ \frac{(W + M_N)^2 - q^2}{m_\rho^2 - q^2}}. 
\label{eqIntRp}
\ea

It is worth noticing that this analytic 
extrapolation is not free of uncertainties 
and that specially an extension for very large values 
of $q^2$ may not be very accurate.
Nonetheless, since we are restricted 
to $q^2 \le  (W -M_N)^2$, the approximation is 
justified as far as we restrict our study 
to not very large values for $W$.  
For $W=1.535$ GeV, we obtain at most $q^2\simeq  0.35$ GeV$^2$
or $\sqrt{q^2} \simeq 0.6$ GeV.

The singularities presented on (\ref{eqIntRp}), 
one associated to the dipole factor, and another 
with the $\rho$-pole, can be regularized:
to each regulating scale $\Lambda$ ($\Lambda_D$ or $m_\rho$)
we associate finite width $\Gamma_\Lambda$.
For a given power of $n$ (integer) we then use the replacement
\ba
\left( \frac{\Lambda^2}{\Lambda^2 -q^2}\right)^n
\to
 \left( \frac{\Lambda^4}{
(\Lambda^2 -q^2)^2 + \Lambda^2 [\Gamma_\Lambda (q^2)]^2}\right)^{\frac{n}{2}}.
\label{eqRegulator}
\ea
This way, we consider the absolute value of the multipole.
The same method was used in previous works~\cite{N1520TL,Timelike2}.
The explicit expression for the effective width $\Gamma_\Lambda (q^2)$
is presented in Appendix~\ref{appRegulariza}.
In the case of ${\cal I}_R^\prime$ in Eq.~(\ref{eqIntRp}), 
we extend the previous expression  to  half-integers, $n= \frac{1}{2}$. 
With the procedure (\ref{eqRegulator}), we simplify the expressions
in the timelike region.

\subsection{Meson cloud contributions}

The extension of the meson cloud component to the timelike region
is straightforwardly
based on Eqs.~(\ref{eqF2mc1})--(\ref{eqC1}).
The parametrizations of the meson cloud
contributions are determined by the calibration 
of the form factors $F_1^\ast$ and $F_2^\ast$ at the physical mass ($W=M_R$).
Although the  meson cloud parametrizations for $F_1^\ast$ and $F_2^\ast$ 
are independent of $W$, since the calculations $G_E$ and $G_C$ 
are done through Eqs.~(\ref{eqGE0}) and (\ref{eqGC0})
where the coefficients now depend on $W$,
the meson cloud contributions for those form factors depend on $W$.

The relations~(\ref{eqF2mc1})--(\ref{eqC1}) 
used in the parametrizations of the meson cloud contributions 
are automatically converted to the timelike region 
with the replacement $Q^2 \to -q^2$.
To regularize the multipole functions we use 
the procedure from Eq.~(\ref{eqRegulator}).
In the pion form factor $F_\pi (q^2)$ the imaginary component 
is generated naturally for $q^2 >0$ (see Eq.~(\ref{eqFpi})).
Due to the magnitude of the regulators presented 
on Eqs.~(\ref{eqA1}), (\ref{eqB2}) and (\ref{eqC1}) 
only the the function $C$, given by Eq.~(\ref{eqC1}) 
requires in fact the use of the regularization (\ref{eqRegulator}), 
because $\Lambda_C^2 \simeq 0.785$ GeV$^2$ 
is closer to the region of study $q^2= (M_R -M_N)^2 \simeq 0.36$ GeV$^2$.

\subsection{$\gamma^\ast N \to N(1535)$ form factors 
in the timelike region}

We present now the results in the timelike region
for the form factors $G_E$ and $G_C$.
We start with the
results for the case $W=M_R$ that
expand
Figs.~\ref{figN1535R-2} and \ref{figN1535R-4}
into the region $-(M_R-M_N)^2 \le Q^2 \le 0$.
Later on, we study the dependence 
of the form factors (real and imaginary parts)
for several values of $W$.

The results in the timelike region 
are presented in Fig.~\ref{figN1535R-TL3} for proton targets 
and  in Fig.~\ref{figN1535R-TL-n} for neutron targets.
Although the transition form factors became 
complex in the timelike region,
in the figures, for now, we show only the real part 
of the form factors in order to be able to compare 
the results directly with the physical spacelike data.

The results from Fig.~\ref{figN1535R-TL3} 
(proton target) indicate that the real part of $G_E$ 
changes sign below $Q^2=0$.
This is a consequence of the combination 
$F_1^\ast$ and $F_2^\ast$ 
and may have been anticipated from the results from Fig.~\ref{figN1535R}:
since those functions have opposite sign below $Q^2=0$,
then $G_E = F_1^\ast + \eta F_2^\ast$ may vanish at some point below $Q^2=0$.
Notice that $F_1^\ast$ and $F_2^\ast$
are both finite at the pseudothreshold~\cite{Siegert1}. 
The zero in the real part of $G_C$ then occurs as a consequence 
of the constraint from Siegert's theorem,
which states that $G_C \propto G_E$ 
near the pseudothreshold.
More details can be found in Appendix~\ref{appS11}.

The results for the real part of 
the transition form factors for neutron targets
in Fig.~\ref{figN1535R-TL-n}
show a significant enhancement of the bare 
and meson cloud contributions below $Q^2=0$ 
for both form factors.
Recall that the value of $G_E$ at $Q^2=0$
is the only physical constraint used in the 
model parametrization.
The results for neutron target are then 
the result of the extrapolation of our 
model parametrization to the valence quark contribution 
and our phenomenologically motivated parametrization 
of the meson cloud effects.
Concerning the final result (solid-line) 
for $G_E$ and $G_C$, the negative bump 
observed in both functions is the consequence 
of two main effects: the enhancement of 
the function ${\cal I}_R$ due to the dipole shape, 
and the vicinity the $\rho$-meson pole 
which suppresses the real part of the form factors, and is
present on both the bare and meson cloud components
(since $F_\pi$ peaks near $q^2 \approx m_\rho^2$).

The results for the form factors for the values $W=1.2$, 1.4 and 1.535 GeV 
are presented in Figs.~\ref{figN1535R-TL4}
and \ref{figN1535R-TL4-n} for proton and neutron targets, respectively.  
The upper value of $q^2$ for each value of the energy resonance energy $W$  corresponds 
to the pseudothreshold point \mbox{$q^2=(W-M_N)^2$.}
The figures show both the real part 
(upper panel) and the imaginary part (middle panel),
and present also the results for the 
absolute values $|G_E|$ and $|G_C|$ (lower panel).
For convenience, we present the results only for
the timelike region, i.e.~$q^2 >0$.

In Figs.~\ref{figN1535R-TL4} and \ref{figN1535R-TL4-n}
the results near the pseudothreshold, $q^2=(W-M_N)^2$,
are modified compared to a model 
which ignores the regularization 
of the singularities
(poles $\Lambda_D^2$, $m_\rho^2$, $\Lambda_C^2$, etc.).
In general, the regularization reduces 
the magnitude of the form factors in the vicinity of the pseudothreshold.
We can anticipate here that although the effective results 
for the real and imaginary parts of the form factors depend 
on the regularization, in particular
on the width associated with the dipole function 
in Eq.~(\ref{eqGD}), the final results 
for the $N(1535)$ Dalitz decay widths
(integrated in $q$), presented in Sec.~\ref{secResults},
 have a very weak dependence 
on the regularization parameters.

From the previous figures, one can conclude 
that $G_E$ and $G_C$ have similar magnitudes for their 
real and imaginary components. 
The relevant function for the timelike calculations, 
discussed in the next section, is, however, the effective form factor 
defined by the combination of $|G_E|^2$ and $|G_C|^2$
\ba
|G_T(q^2,W)|^2= |G_E(q^2,W)|^2 + \frac{q^2}{2 W^2} |G_C(q^2,W)|^2,
\nonumber \\
\label{eqGTS}
\ea
where the form factors $G_E$ and $G_C$
are defined by Eqs.~(\ref{eqGE0})--(\ref{eqGC0}),
with the replacement $M_R \to W$.

Notice in Eq.~(\ref{eqGTS}) that the contribution of $|G_C|$
is suppressed at low $q^2$ by the factor $q^2/(2W^2)$.
One expects then that  $|G_C|$ becomes relevant 
only for large $q^2$ values, 
corresponding also to large values of $W$, since $q^2 \le (W-M_N)^2$.

The results for the form factor 
 $|G_T|$ for the values $W=1.2$, 1.4 and 1.535 GeV 
are presented in  Fig.~\ref{figN1535R-TL6}
for proton and neutron targets.
Note that the magnitude of $|G_T|$ becomes larger 
for neutron targets when $q^2 > 0.1$ GeV$^2$.
In both channels the dominant effect comes 
from the electric form factor.
The contribution of the Coulomb form 
factors increases the function  $|G_T|$ in 10\% at most.

In the next section we present the formalism associated 
with the $N(1535)$ Dalitz decay.
In Sec.~\ref{secResults}
we use our results for  $|G_T|$
to calculate the Dalitz decay functions.


\section{$N(1535)$ Dalitz decay}
\label{secDalitz}

We discuss now the formalism associated 
with the  $N(1535)$ Dalitz decay ($N^\ast \to e^+ e^- N$).
As in the previous section, $W$ represent 
the  mass of the resonance.

Our starting point is the calculation 
of the function $\Gamma_{\gamma^\ast N}(q,W)$, which 
determine the decay width of state with mass $W$
into a photon with virtuality $q^2 >0$.
The variable $q$ is then defined by $q= \sqrt{q^2}$.

The function $\Gamma_{\gamma^\ast N}(q,W)$
is defined according to Ref.~\cite{Krivoruchenko02} 
\ba
\Gamma_{\gamma^\ast N} (q,W) 
= \frac{\alpha}{2 W^3} 
\sqrt{y_+ y_-} y_+ |G_T(q^2,W)|^2, \nonumber \\
\label{eqGammaS1}
\ea
where  $|G_T(q^2,W)|$ is defined by Eq.~(\ref{eqGTS}) 
and 
\ba
y_\pm = (W \pm M_N)^2 -q^2.
\ea

From Eq.~(\ref{eqGammaS1}) one concludes
that the impact of the form factors 
in the Dalitz decay functions
is determined by the function $|G_T|$, given by Eq.~(\ref{eqGTS}).

Once the function $\Gamma_{\gamma^\ast N} (q,W)$ is defined, 
one can calculate the dilepton decay rate 
using the derivative~\cite{Krivoruchenko02} 
\ba
\Gamma^\prime_{e^+ e^- N} (q,W) & \equiv & 
\frac{d \Gamma_{e^+ e^- N}}{d q} (q,W) \nonumber \\
&  = &  \frac{2 \alpha}{3 \pi q^3}  (2 \mu^2 + q^2) 
\sqrt{1 - \frac{4 \mu^2}{q^2}}  \Gamma_{\gamma^\ast N}(q,W), \nonumber \\
& & 
\label{eqDGamma}
\ea
where $\mu$ is the electron mass.

The Dalitz decay width can then be determined 
by the integral of $ \Gamma^\prime_{e^+ e^- N} (q,W)$ 
in the region $4 \mu^2 \le q^2 \le (W -M_N)^2$:
\ba
 \Gamma_{e^+ e^- N} (W) = 
\int_{2 \mu}^{W -M_N} \Gamma^\prime_{e^+ e^- N} (q,W) \, dq.
\label{eqGeeNint}
\ea

The radiative decay, $N^\ast \to \gamma N$, 
is calculated from the function $\Gamma_{\gamma^\ast N}(q,W)$,
in the limits $q^2=0$ and $W = M_R$.
Using Eq.~(\ref{eqGammaS1}), one obtains
\ba
\Gamma_{\gamma N} = \frac{\alpha}{M_R^2} (M_R + M_N)^2 K |G_E(0,M_R)|^2.
\label{eqGN1}
\ea
The previous result is consistent with 
the general expression in terms of helicity amplitudes 
for a resonance with spin $J=\frac{1}{2},\frac{3}{2}$~\cite{PDG97,Aznauryan12a}:
\ba
\Gamma_{\gamma N} = 
\frac{2}{(2J +1) \pi} K^2 \frac{M_N}{M_R}
\left[ |A_{1/2}|^2 + |A_{3/2}|^2
 \right],
\label{eqGamma1}
\ea
where $A_{1/2}$,  $A_{3/2}$ represent 
the transverse helicity amplitudes (at resonance rest frame) for $Q^2=0$.
As before $K= \frac{M_R^2- M_N^2}{2 M_R}$.
In the present case ($J=\frac{1}{2}$),  
one has $A_{3/2} \equiv 0$.


\section{Results for the radiative and Dalitz decay widths}
\label{secResults}


\begin{figure}[t] 
\vspace{.5cm}
\centerline{\mbox{
\includegraphics[width=7.5cm]{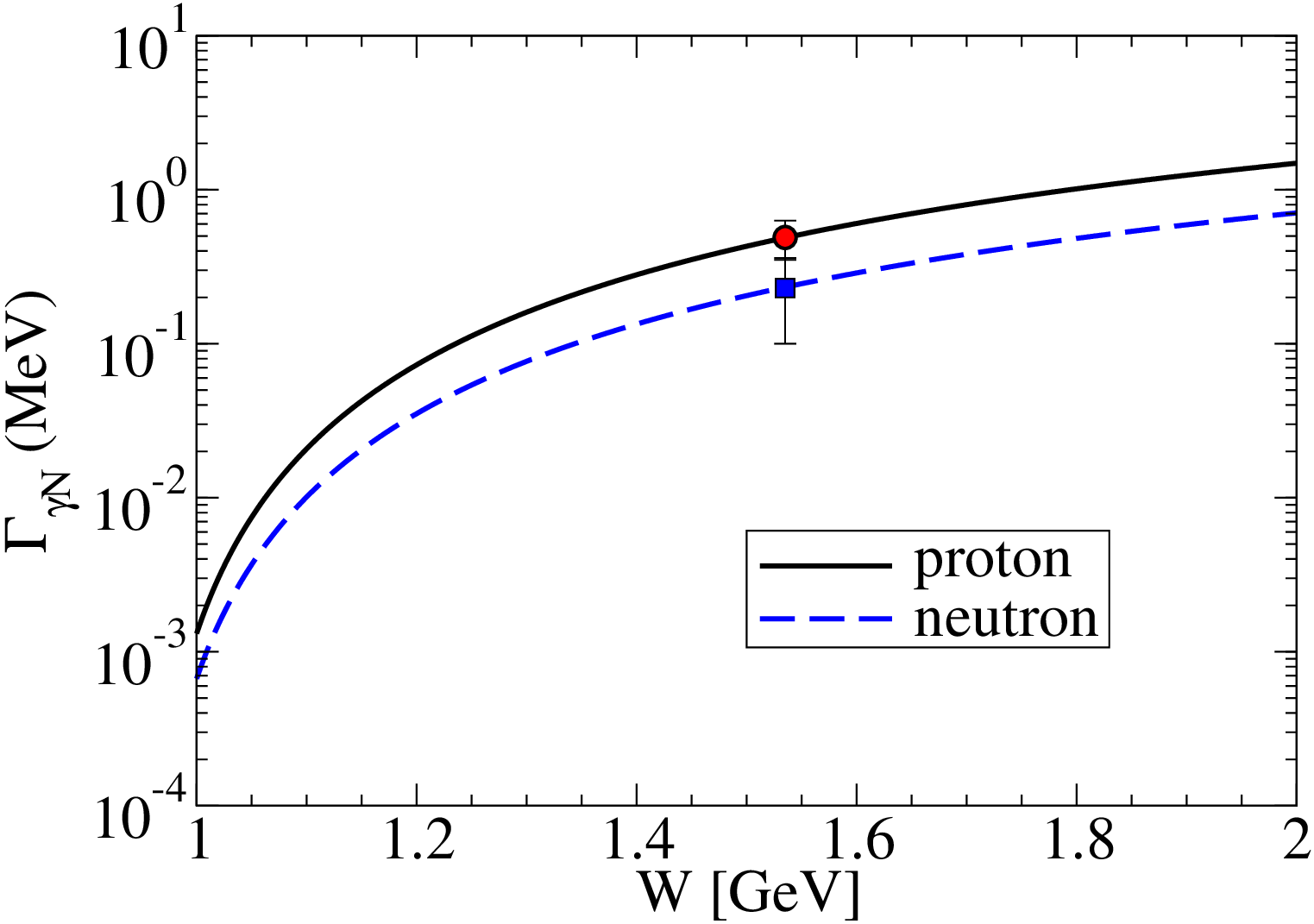}}} 
\caption{\footnotesize
Radiative decay width as function of $W$
for the proton and neutron cases. 
The data ($W=M_R$) are determined 
from the PDG data for the amplitude $A_{1/2} (0)$.}
\label{figGamma-gN}       
\end{figure}
\begin{figure}[t] 
\vspace{.5cm}
\centerline{\mbox{
\includegraphics[width=7.5cm]{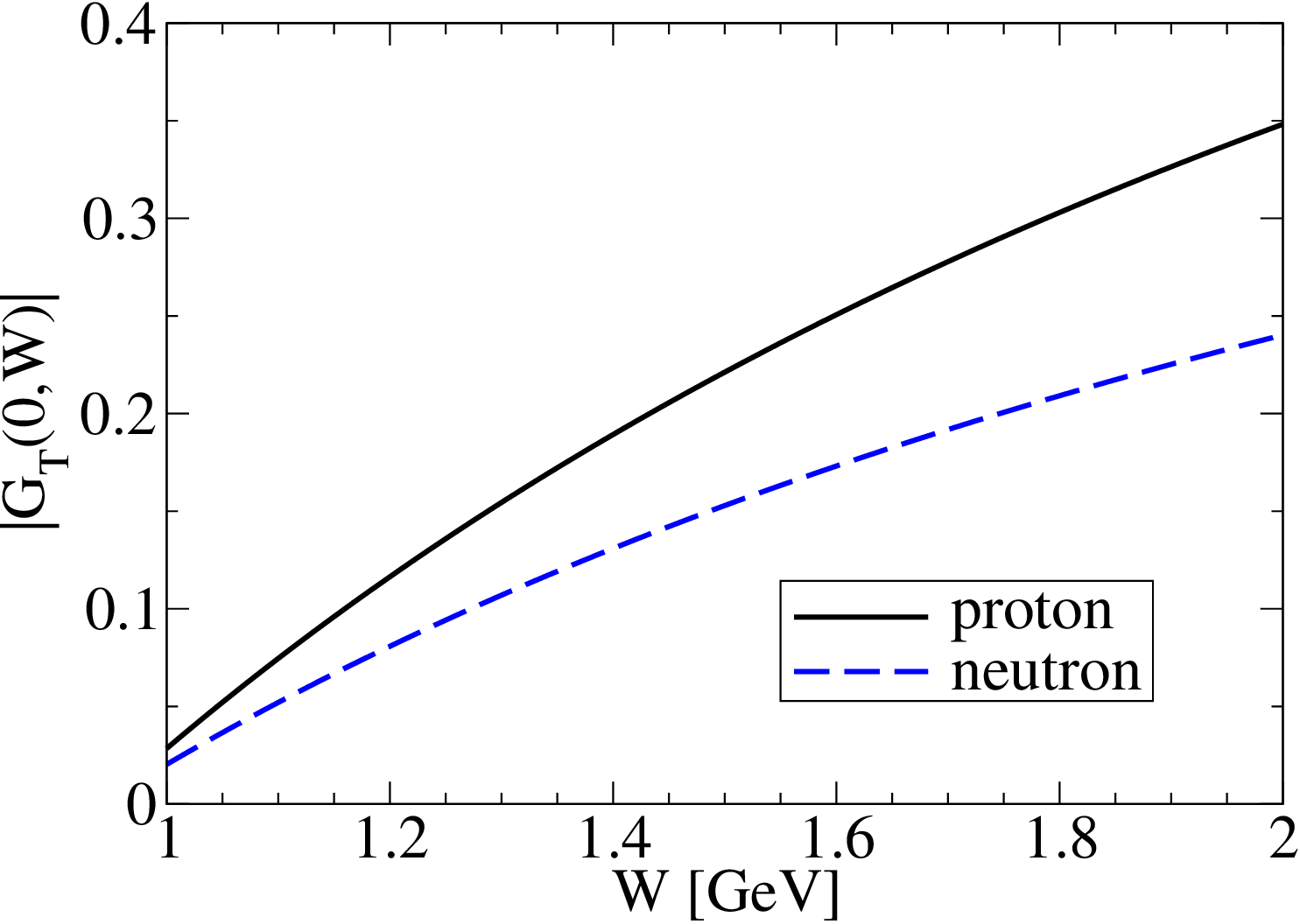}}} 
\caption{\footnotesize
Effective form factor function $|G_T (0,W)|$ 
for the proton and neutron cases.}
\label{figGT}       
\end{figure}

\begin{figure*}[t] 
\vspace{.5cm}
\centerline{\mbox{
\includegraphics[width=7.5cm]{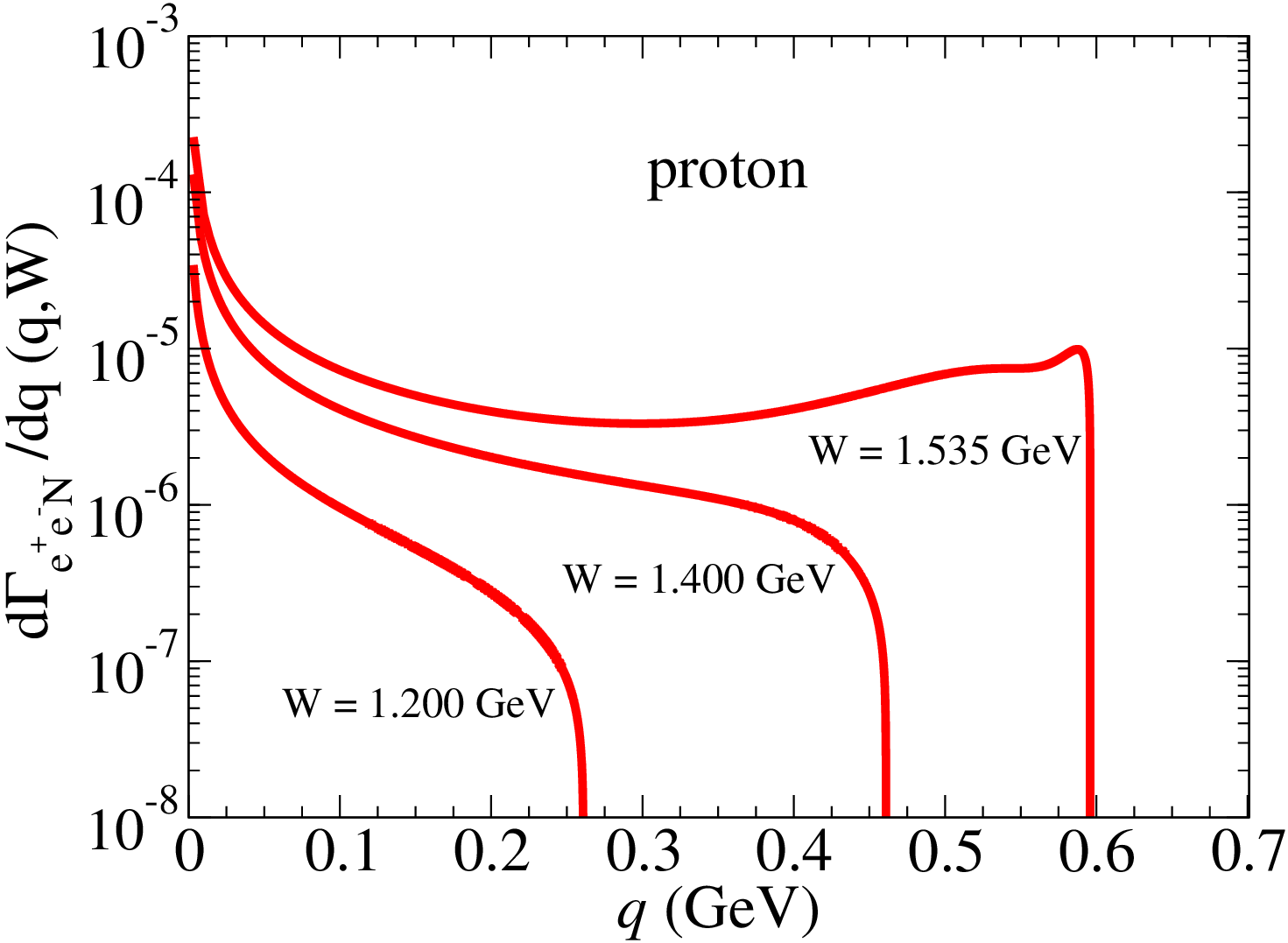}} \hspace{1.cm}
\mbox{
\includegraphics[width=7.5cm]{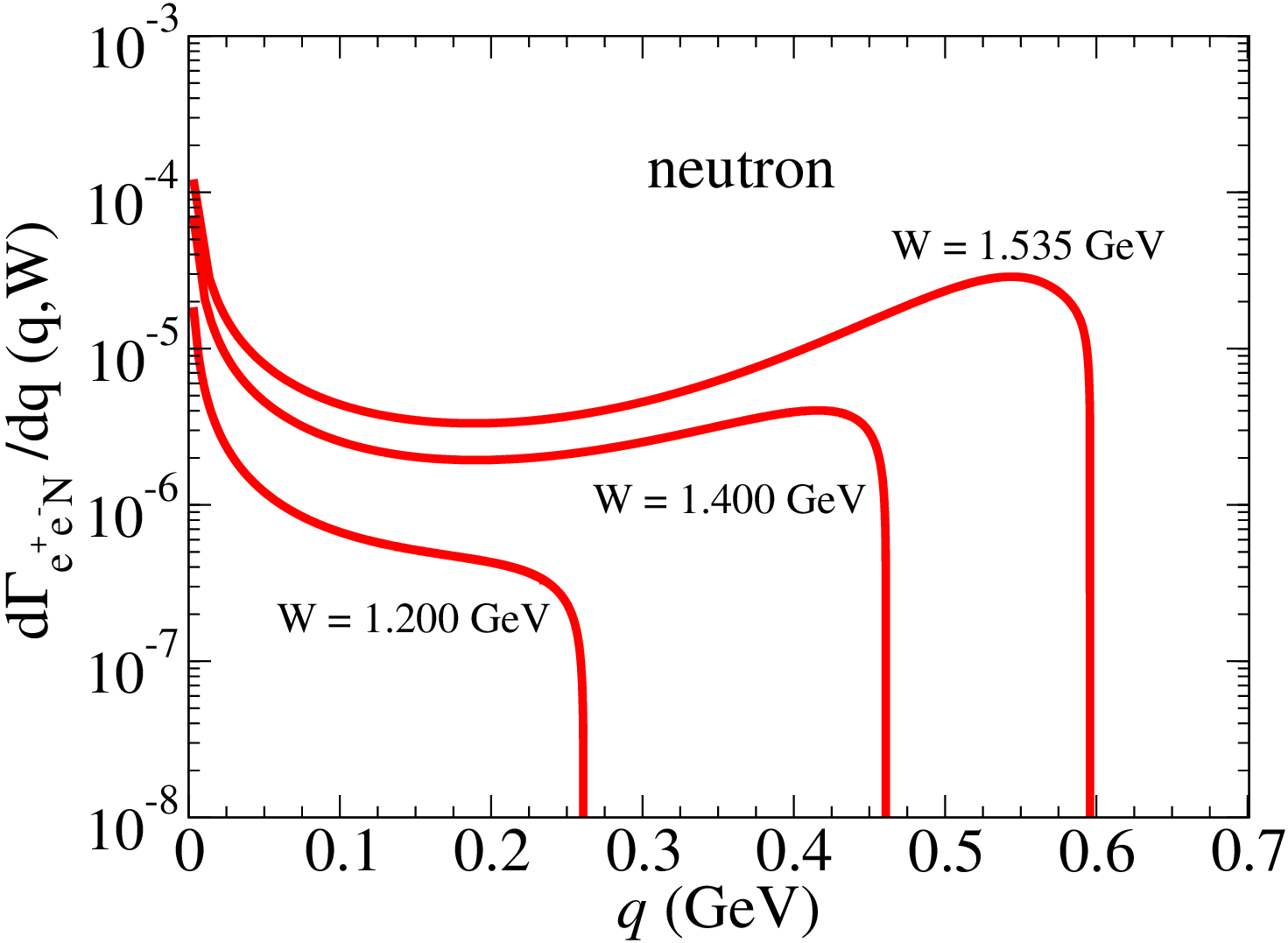}}}
\caption{\footnotesize
Dilepton decay rates $\frac{d }{d q} \Gamma_{e^+e^- N} (q,W)$.
for the cases $W=1.2$, 1.4  and 1.535 GeV.
The upper limit in $q$ is $W-M_N$.}
\label{figDGamma}       
\end{figure*}

We present in this section 
the observables associated with the timelike region.
First, we present our results for the radiative decay widths 
($\Gamma_{\gamma N}$).
Next we discuss our results for the dilepton decay rates 
$\frac{d }{d q} \Gamma_{e^+e^- N} (q,W)$.
We also show the results for the Dalitz decay 
widths ($\Gamma_{e^+e^- N}$), as function of $W$.
We consider the proton and neutron cases.


\begin{table}[t]
\begin{center}
\begin{tabular}{c |c c | c c c }
\hline
\hline
   &   $A_{1/2} (0)$  [GeV$^{-1/2}$]  &   &  $\Gamma_{\gamma N}$ [MeV]  &  \\
   &   Data       &   Model     &  Estimate & PDG limits  & Model  \\
\hline
\hline
$p$  & $\;$ 0.105$\pm$0.015  &  $\;$ 0.101 & 0.49$\pm$0.14  &  0.19--0.53  &  0.503 \\  
$n$  & $-0.075\pm$0.020  & $-0.074$ & 0.25$\pm$0.13  &  0.013--0.44 &  0.240 \\
\hline
\hline
\end{tabular}
\end{center}
\caption{\footnotesize
$N(1535) \to \gamma N$ decay widths.
Estimate represent the PDG result calculated from  
the amplitude $A_{1/2}(0)$.
PDG limits labels result obtained
from the branching ratios.} 
\label{tabN1535-2}
\end{table}

\subsection{Radiative decay widths}

The radiative decay widths for the proton and neutron 
are determined by the function $\Gamma_{\gamma^\ast N} (q,W)$
as defined by Eq.~(\ref{eqGammaS1}) 
in the limit $q^2=0$, when the virtual photon  became real.

The results for $\Gamma_{\gamma N}$ are presented 
in Fig.~\ref{figGamma-gN}, for the proton and neutron cases.
Our results differ significantly 
from the results of a model with constant form factors.

Notice that the result for  $\Gamma_{\gamma N}$
is related to  $|G_T (0,W)|^2$. 
The results of the  function $|G_T (0,W)|$
are presented in  Fig.~\ref{figGT}.
From the figure it is clear that the constant form factor, i.e.~a 
$W$ independent form factor, is a bad approximation.

The results for  $\Gamma_{\gamma N} (W)$ for the 
physical point ($W=M_R$) compare well with
experimental values  presented in Table~\ref{tabN1535-2}.
The data presented in Fig.~\ref{figGamma-gN}
are PDG results based on the amplitudes $A_{1/2}(0)$
(fourth column of Table~\ref{tabN1535-2}).
The uncertainties in the widths are the consequence 
of limits on $A_{1/2}(0)$ [proportional to $G_E(0)$].
Note that there is some overlap between 
the data results for the proton and neutron,
meaning that the data are compatible with 
an identical result for both decays (exact isospin symmetry).

In our model, the isospin symmetry  is clearly broken 
in the $N(1535) \to \gamma N$ decay.
The good agreement between model and data is a consequence of the 
accurate description of the transition form factor $G_E$ at $q^2=0$, 
for both isospin channels.


\subsection{Dalitz decay rates}

The dilepton decay rate
$\frac{d }{d q} \Gamma_{e^+e^- N} (q,W)$
can be calculated combining Eq.~(\ref{eqDGamma})
with Eq.~(\ref{eqGammaS1}).
The results for $W=1.2$, 1.4  and 1.535 GeV 
are presented in Fig.~\ref{figDGamma} for the 
proton (left panel) and neutron (right panel) cases.
The upper limit in $q$ is determined by $q=W-M_N$, as before.

From Fig.~\ref{figDGamma}, we can conclude that the 
more relevant kinematic regions, for both channels,  
is the low-$q$ region or near the pseudothreshold for large $W$,
where there is a substantial enhancement of the decay rate.
In the figure,
one can also notice that the magnitude 
of the decay rates near $q^2=0$ is larger for the proton.

\subsection{Dalitz decay widths}

\begin{figure}[t] 
\vspace{.5cm}
\centerline{\mbox{
\includegraphics[width=7.5cm]{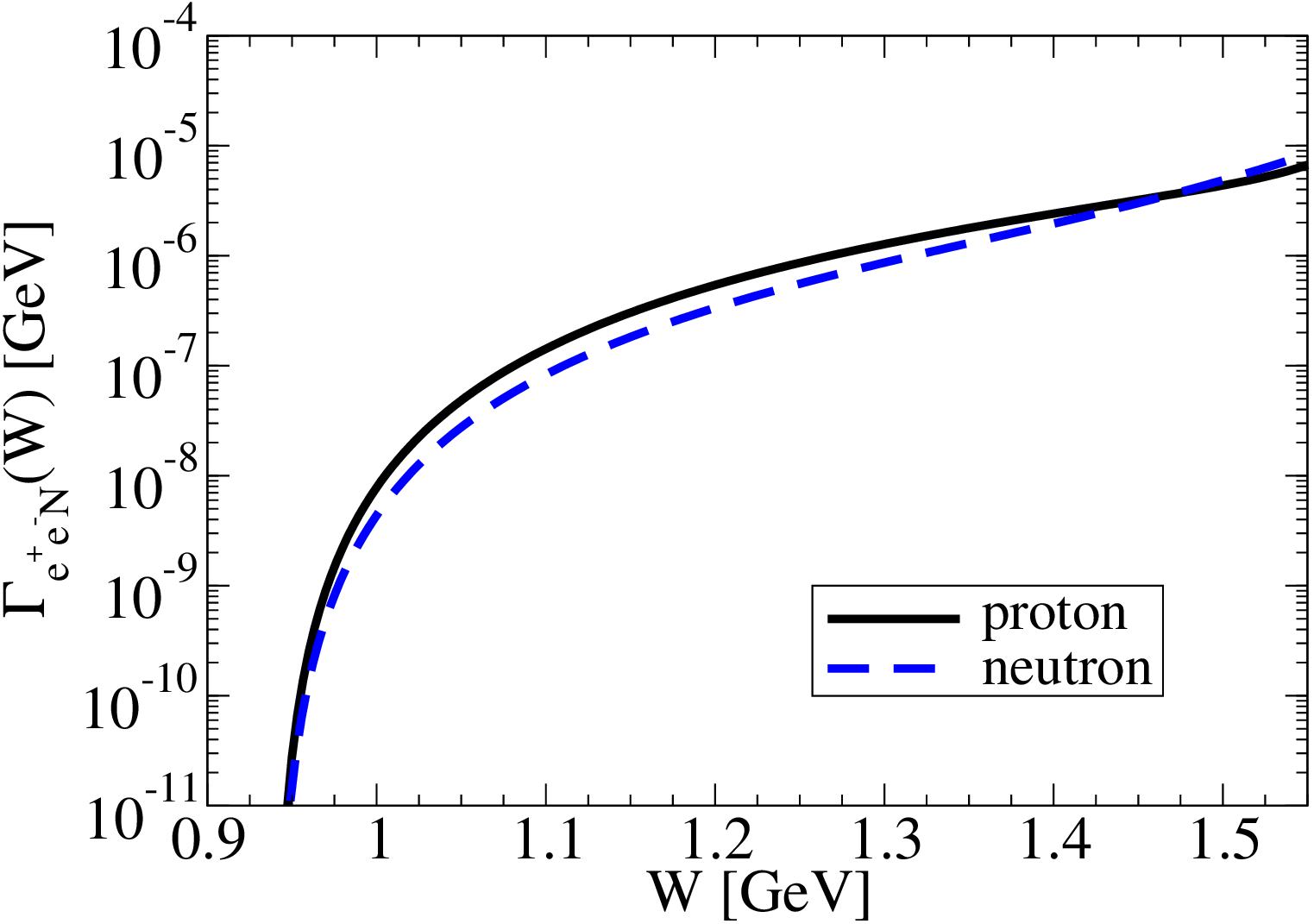}}} 
\caption{\footnotesize
Dalitz decay widths as function of $W$
for the proton and neutron.}
\label{figGamma_eeN}       
\end{figure}

The function $\Gamma_{e^+ e^- N} (W)$
is determined by the integral of the 
dilepton decay rate according to Eq.~(\ref{eqGeeNint}).
The results for the proton and neutron cases 
are presented in Fig.~\ref{figGamma_eeN}.

In the figure we can notice a dominance of the 
proton decay width up to $W=1.4$ GeV 
and very close values for proton and neutron cases near \mbox{$W=1.5$ GeV.}
Above 1.5 GeV, close to the 
$\rho$ meson mass pole ($W = M_N + m_\rho \simeq 1.7$ GeV) 
the effect of the corresponding pole starts to manifest.
The main effect is the enhancement of $\Gamma_{e^+ e^- N} (W)$. 
We have a  glimpse of this effect  in 
the graph for the neutron decay (dashed-line).

The dominance of the Dalitz decay 
width for proton decay over the results 
for neutron decay is explained by the dominance 
of the dilepton decay rates \mbox{near $q=0$}, 
as can be confirmed by Fig.~\ref{figDGamma}
(right panel versus left panel).
For larger values of $q$ (and larger $W$) 
the magnitude of the neutron dilepton decay rates 
increases more in comparison to the 
proton dilepton decay rates (see Fig.~\ref{figDGamma}).
When we integrate on $q$ to obtain $\Gamma_{e^+ e^- N} (W)$,
the impact of the large $q$ region 
on the dilepton decay rate is larger,
and the neutron Dalitz decay width is enhanced.

Since we aim at the range of the HADES experiments, 
we do not go beyond $W  \simeq 1.55$ GeV.
The values of the function $\Gamma_{e^+ e^- N} (W)$, 
at $W=M_R$ are given in Table~\ref{tabN1535-3}.
From the table we can conclude that 
the results for proton and neutron decays are 
very close, $\Gamma_{e^+ e^- N} (M_R) \simeq 6$--7 keV.
This result contrasts with what occurs in the radiative decay, $\Gamma_{\gamma N}(M_R)$,
where the widths for the two isospin channels differ much more.

In a model where we reduce 
the isoscalar component $A(0)$ by in about 0.05, which
as discussed in Sec.~\ref{secTotal}
($A (0) \to A(0)  - 0.05 \simeq 0.075$) is still 
well within the experimental limits,
the results for $\Gamma_{e^+ e^- N} (W)$ are 
almost indistinguishable in the two channels.

The timelike data about the neutron decays 
is very important because they provide information  
about the neutron structure which
is not available at the moment from spacelike experiments.
For this reason pion-induced reactions 
at HADES~\cite{ColeTL,Ramstein18a}
are fundamental to pin down to electromagnetic 
structure of the neutron and 
complement the information from the spacelike region.

\begin{figure}[t] 
\vspace{.5cm}
\centerline{\mbox{
\includegraphics[width=7.5cm]{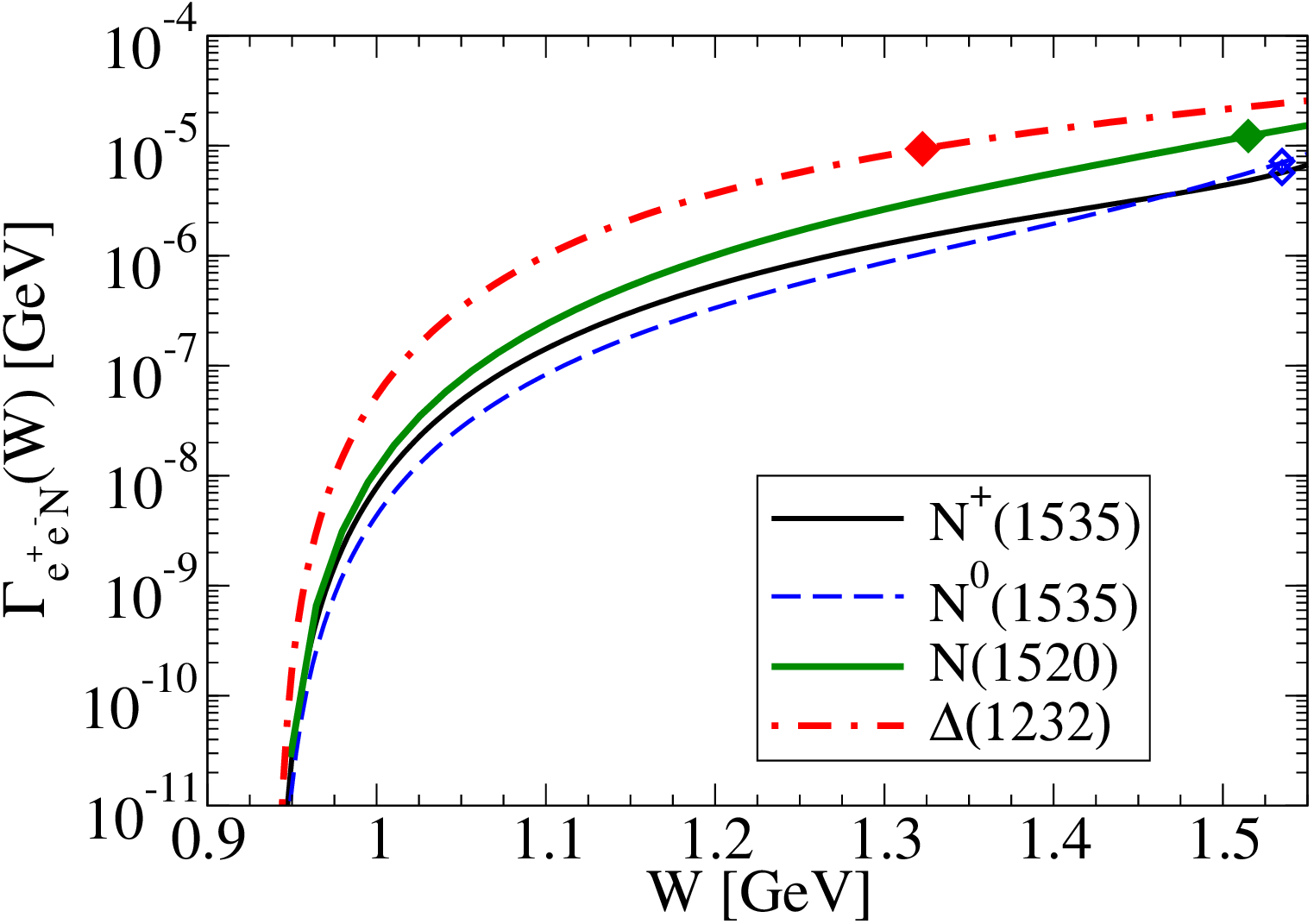}}} 
\caption{\footnotesize
Comparison between Dalitz decay widths $\Gamma_{e^+ e^- N} (W)$
for different resonances. 
Models from Refs.~\cite{Timelike2,N1520TL}.
The diamonds indicate the Dalitz decay 
widths at the physical point ($W= M_R$).}
\label{figGamma_eeN2}       
\end{figure}

In Fig.~\ref{figGamma_eeN2}, we compare 
the $N(1535)$ Dalitz widths with estimates for
other light mass resonances,
based on the covariant spectator quark model.
We show the results for the $\Delta(1232)\frac{3}{2}^+$, 
where the pion cloud contributes 
with about 45\% to the transition form factors 
at the photon point~\cite{Timelike2}, and 
also the results  for $N(1520)\frac{3}{2}^-$\cite{N1520TL}.

Figure~\ref{figGamma_eeN2} shows that the
 $\Delta(1232)\frac{3}{2}^+$ dominates 
within the range of $W$ considered,
although the  $\Delta(1232)$ Dalitz decay at the pole 
is measured for smaller values of $W$.
The Dalitz decay branching ratio for this resonance is consistent with
the value recently extracted from the dilepton production 
spectrum data~\cite{HADES17a}.


\begin{table}[t]
\begin{center}
\begin{tabular}{c | c}
\hline
\hline
   & $\Gamma_{e^+ e^- N}$ (keV)\\
\hline
\hline
$p$  & 5.7 \\  
$n$  & 7.2 \\
\hline
\hline
\end{tabular}
\end{center}
\caption{\footnotesize
$N(1535) \to \gamma N$ Dalitz decay widths,
estimated by the present model.} 
\label{tabN1535-3}
\end{table}


\section{Outlook and conclusions}
\label{secConclusions}

Theoretical models for the electromagnetic structure of the $N^\ast$ 
resonances in the timelike region are necessary for the interpretation 
of $N^\ast$ Dalitz decays measured currently 
in experiments at HADES~\cite{HADES17a,HADES-dilepton,Ramstein18a}.

The structure of the $\gamma^\ast N \to N(1535)$ transition 
given by the experimental data for spacelike form factors is non-trivial and
suggests that it results from a combination of 
valence quark and meson cloud effects.
Valence quark models describe well the Dirac 
form factor for $Q^2 > 1.5$ GeV$^2$
but they fail to describe the Pauli form factor data.
In contrast, chiral models predict important meson cloud contributions 
to the Pauli form factor in the low-$Q^2$ region.

In this work, we developed a model 
for the $\gamma^\ast N \to N(1535)$ transition in the timelike region.
The model is based on the covariant spectator quark model 
in the spacelike region, which is here  combined 
with the semirelativistic approximation that neglects 
baryon mass difference in the overlap integral of the initial and final state.
This approximation guarantees the orthogonality of the initial 
and final wave functions as well as current conservation. 
It also enables us to use the same radial wave function for the nucleon and the $N(1535)$.
Therefore all the estimates for the valence quark contributions 
in the spacelike region 
are true predictions of a parametrization fixed from nucleon 
elastic form factors.
We also modify the behavior of the form factors  at low $Q^2$
in order to obtain the correct experimental behavior of the 
Dirac form factor  near $Q^2=0$.

In the present work, we use the available data (proton and neutron targets)
to infer the effect of the meson cloud contributions within the spacelike regime.
The meson cloud contributions are 
parametrized according to the observed meson decay 
rates  of the $N(1535)$ resonance, 
dominated by the $\pi N$ and $\eta N$ channels.
The meson cloud parametrization for 
the Dirac and Pauli form factors 
are dominated by the isovector component, as
suggested by the photo-production data (amplitude $A_{1/2}(0)$),
and some other theoretical models. 
In the case of the Pauli form factor we consider 
also a small isoscalar component.

We extended our parametrizations of the 
$\gamma^\ast N \to N(1535)$ transition to the timelike region,
considering analytic continuations of  
the  valence quark and meson cloud contributions 
from the spacelike region to the timelike region.
The transition form factors are calculated 
in terms of $q^2$ and the invariant mass of the $\gamma^\ast N$ system $W$,
and used to estimate the radiative and Dalitz decay widths. 

We separated the proton and neutron cases,  
since the results reveal an important isospin dependence.
Our estimates for neutron targets are poorly constrained by the spacelike data,
but alternative estimates can be performed adjusting one single 
parameter (isoscalar coefficient $A(0)$), 
when more accurate data will become available.
Timelike experiments provide an alternative method 
to probe the physics associated with the neutron targets,
where contrary to spacelike experiments, 
the channels associated to neutrons are directly 
accessed by pion-induced reactions.

We compared our $N(1535)$ Dalitz decay results as a function of $W$ 
with previous results for the $\Delta(1232)$ and $N(1520)$ resonances 
(which have almost no isospin dependence).
Our calculation can be in a near future compared 
with the dilepton decay rates measured at HADES.


\begin{acknowledgments}
G.~R.~was supported by the Funda\c{c}\~ao de Amparo \`a
Pesquisa do Estado de S\~ao Paulo (FAPESP):
Project No.~2017/02684-5, Grant No.~2017/17020-BCO-JP. 
M.~T.~Pe\~na was supported in part by Funda\c c\~ao para a Ci\^encia 
e a Tecnologia (FCT), Grant No.~CFTP-FCT (UID/FIS/00777/2015).
\end{acknowledgments}

\appendix

\section{$\gamma^\ast N \to N(\frac{1}{2}^-)$ 
form factors and helicity amplitudes}
\label{appS11}

We discuss here the generic expressions 
for the form factors in different representations 
and their relations with the helicity amplitudes.
The  discussion follows 
Refs.~\cite{Devenish76,Krivoruchenko02},
but is based on  the current (\ref{eqJS11}),
where the form factors $F_i^\ast$ ($i=1,2$)
are dimensionless.
To compare with the results from Ref.~\cite{Devenish76}
one uses $G_1= q^2 F_1^\ast$ and $G_2 = \frac{2}{(M_R + M_N)^2} F_2^\ast$.

The expressions associated with the decay widths 
can be expressed directly in terms of 
the helicity amplitudes, as in Eq.~(\ref{eqGamma1}),
or in terms of multipole form factors.
Those form factors can be defined 
using different conventions as the ones 
proposed in Refs.~\cite{Devenish76,Krivoruchenko02}.
We use here a representation equivalent
to those authors, but with different normalizations.
More specifically, we use 
the following representation of  
the electric and Coulomb form factors:
\ba
G_E &=& 
F_1^\ast + \eta F_2^\ast, 
\label{eqGEb}\\
G_C &=& - \frac{M_R}{2}
\frac{(M_R+M_N)}{Q^2} 
\left[\eta F_1^\ast  - \tau F_2^\ast\right].
\label{eqGCb}
\ea
The conversion to the 
form factors from Ref.~\cite{Krivoruchenko02},
$\bar G_E$ and $\bar G_C$ 
can be performed 
using $\bar G_E = - \frac{\sqrt{2}}{M_R} G_E$
and  $\bar G_C = - \frac{\sqrt{2}}{M_R} G_C$.
To compare with the form factors from 
Ref.~\cite{Devenish76} we can use 
$h_1 = - \frac{2}{M_R} G_C$ and $h_3 = - \frac{2}{M_R} G_E$. 
An advantage in the use of the form factors 
(\ref{eqGEb})--(\ref{eqGCb}) is that they are dimensionless.

The motivation to the identification 
with the electric and Coulomb form factors 
are the result of the resemblance 
with the nucleon Sachs form factors 
$G_C \propto F_1 - \frac{Q^2}{4 M_N^2} F_2$
and $G_M \propto F_1 + F_2$, 
combined with the connection between 
negative parity and positive parity multipole amplitudes.
Recall that in the change from 
positive parity and negative parity states 
we should replace 
$G_E \leftrightarrow G_M$~\cite{Aznauryan12a,Devenish76,Krivoruchenko02}.

The multipole form factors can be related 
to the helicity amplitudes (\ref{eqA12s})--(\ref{eqS12s}) by
\ba
& &
G_E = \frac{1}{\cal B} A_{1/2},  
\label{eqGEc} \\
& &
G_C = \frac{1}{\sqrt{2}{\cal B}} \frac{M_R}{|{\bf q}|} S_{1/2},  
\label{eqGCc}
\ea
where 
\ba
{\cal B} = \frac{e}{2} \sqrt{\frac{Q_+^2}{M_N M_R K}}.
\ea

A simple consequence of the relations
(\ref{eqGEc})--(\ref{eqGCc}) is that,
according to Siegert's theorem,
near the  pseudothreshold, 
one has~\cite{Siegert1}:
 $S_{1/2} \simeq \frac{1}{\sqrt{2}} \frac{|{\bf q}|}{M_R -M_N} A_{1/2}$,
which imply that
\ba
G_C = 2 \frac{M_R-M_N}{M_R} G_E,
\label{eqST1}
\ea 
when $Q^2= -(M_R -M_N)^2$. 

Another important remark is that Eq.~(\ref{eqST1}) 
is valid also for complex form factors, $G_E$ and $G_C$.
One has then a constraint for the real 
part and another for the imaginary part.

\section{Estimate of $G_C(0)$}
\label{appGC0}

In the present appendix we 
calculate the value of $G_C$ at $Q^2=0$ 
for proton target, based on the valence quark 
and meson cloud parametrization
for the $F_1^\ast$ and $F_2^\ast$ form factors.

Starting from the definition (\ref{eqGC0}) 
we can write
\ba
G_C = - \frac{1}{2}M_R (M_R -M_N) \frac{F_1^\ast}{Q^2}
+ \frac{M_R}{2 (M_R + M_N)} F_2^\ast.
\ea
We can then consider the limit $Q^2=0$:

\ba
G_C (0) &=& - \frac{1}{2}M_R (M_R -M_N) 
\left. \frac{F_1^\ast}{Q^2} \right|_{Q^2=0} \nonumber \\
& & + \frac{M_R}{2 (M_R + M_N)} F_2^\ast (0).
\ea

The value of $F_2^\ast (0)$ is determined 
exclusively by the meson cloud contribution 
$F_2^{\rm mc} (0) = A(0) + B(0)$ according 
with Eq.~(\ref{eqF2mc1}).
Recall that $A(0)$ and $B(0)$ are determined 
by by the fit to the data.

We focus now on the calculation 
of $\frac{F_1^\ast}{Q^2}$ in the limit $Q^2=0$.
We recall that $F_1^\ast$ can be decomposed 
in two components ($F_1^{\rm B} + F_1^{\rm mc}$),
and that both components scale with $Q^2$
near the photon point.
Based on our parametrizations 
of the bare and meson cloud components,
we can write:
\ba
& &
\left. \frac{F_1^{\rm B}}{Q^2} \right|_{Q^2=0} = \frac{1}{3} 
\left. 
\frac{{\cal I}_R^\prime}{Q^2}
\right|_{Q^2=0}, \\
& &
 \left. \frac{F_1^{\rm mc}}{Q^2} \right|_{Q^2=0} = - \frac{C_0}{\Lambda_C^2}, 
\ea
using the notation of Eq.~(\ref{eqIntRp}).
In the first equation, we used also 
the result $\left.(3 j_1^S + j_1^A)\right|_{Q^2=0} = \frac{2}{3}$.

The factor $\left. {\cal I}_R^\prime\right|_{Q^2=0}$  
can be calculated based on the definition 
\ba
{\cal I}_R^\prime& =& {\cal I}_R \sqrt{\frac{Q^2}{m_\rho^2 + Q^2}}
\nonumber \\
&=& \frac{G_D (Q^2) Q^2}{M_N(M_R + M_N)}   
\sqrt{\frac{(M_R +M_N)^2 + Q^2}{m_\rho^2 + Q^2}},
\ea
using the dipole approximation for ${\cal I}_R/|{\bf q}|$,
according with Eq.~(\ref{eqGD}).
From this previous relation, we conclude that
${\cal I}_R^\prime/Q^2 \to G_D(0)/(M_N m_\rho)$ in the limit $Q^2=0$.
As a consequence:
\ba
\left. \frac{F_1^{\rm B}}{Q^2} \right|_{Q^2=0} =  
\frac{1}{3} \frac{G_D(0)}{M_N m_\rho}.
\ea

Combined the previous results, we obtain
\ba
G_C(0) & =  &
\frac{1}{2}
\frac{M_R}{M_R + M_N} F_2^\ast (0) 
+ \frac{1}{2} M_R (M_R -M_N) \frac{C_0}{\Lambda_C^2} \nonumber \\ 
& &
- 
\frac{1}{6} M_R (M_R -M_N) \frac{G_D(0)}{M_N m_\rho}.
\ea

The corollary of this analysis is that 
the use of a small regulator (mass $m_\rho$) 
tends to reduce the magnitude of $G_C (0)$
(enhancement of the last term, more significant cancellation of terms).
In alternative, the use of a large value for $m_\rho$
tend to increase the value of   $G_C (0)$
(less significant cancellation of terms).

\section{Analytic extension of the overlap integrals 
to the timelike region}
\label{appIntR}

In the present appendix, we describe how we 
estimate the overlap integral ${\cal I}_R$ 
and the function ${\cal I}_R^\prime$ 
in the timelike region ($Q^2 < 0$).
For the purpose of the discussion we recall 
that in general ${\cal I_R}$ is proportional to $|{\bf q}|$,
and that in the semirelativistic approximation 
$|{\bf q}| \equiv |{\bf q}|_{\rm sr} = Q \sqrt{1 + \tau}$.
The effect of the factor $Z$ is discussed later.


Since, in the context of the semirelativistic approximation, 
we cannot extrapolate the overlap integral for $|{\bf q}|_{\rm sr} < 0$, 
because the radial wave functions cannot be defined 
below $Q^2=0$, we use an analytic continuation 
of the overlap integral  ${\cal I}_R$  defined 
in the spacelike region.

Our analytic continuation of ${\cal I}_R$ is based 
on the observation that ${\cal I}_R/|{\bf q}|$  
is finite in the limit $Q^2=0$, 
and in the realization that  ${\cal I}_R/|{\bf q}|$  
can for small $Q^2$ be approximated by a dipole form:
\ba
\frac{{\cal I}_R^{\rm sr}}{|{\bf q}|_{\rm sr}} \simeq \frac{G_D (q^2)}{M_N}
= 
\frac{{\cal C}}{M_N}\left( \frac{\Lambda_D^2}{\Lambda_D^2 - q^2} \right)^2,
\label{eqDipole1}
\ea
where ${\cal C}$ is a constant with no dimensions 
and $\Lambda_D$ a cutoff parameter.
The upper index on ${\cal I}_R$ indicate the 
result of the integral in the semirelativistic approximation.

Our analytic extension is then based on the 
replacement 
\ba
\frac{{\cal I}_R}{|{\bf q}|} \to 
\frac{G_D(q^2)}{M_N} 
\label{eqDiopole1a}
\ea
in the $Q^2 < 0$ region.
In simple words, we replace the 
numeric result $\frac{{\cal I}_R^{\rm sr}}{|{\bf q}|_{\rm sr}}$
from spacelike by a simple expression for 
$\frac{{\cal I}_R}{|{\bf q}|}$ for $Q^2 < 0$.
We consider then an analytic continuation of 
the results for $Q^2 > 0$.

The consequence of this extension is that 
we estimate  ${\cal I}_R$,
using the relation (\ref{eqDipole1}):
\ba
{\cal I}_R = G_D (q^2) \frac{|{\bf q}|}{M_N}, 
\label{eqDipole2}
\ea
where $|{\bf q}|$ represent the magnitude 
of the transition momentum in the general case [see Eq.~(\ref{eqQR})].
With this simple procedure, 
we obtain an analytic continuation 
of the function ${\cal I}_R/|{\bf q}|$  
in the region $0 \le q^2  \le (M_R-M_N)^2$.
The transition form factors are then 
defined by continuity in the timelike region.

\begin{figure}[t] 
\vspace{.5cm}
\centerline{\mbox{
\includegraphics[width=7.5cm]{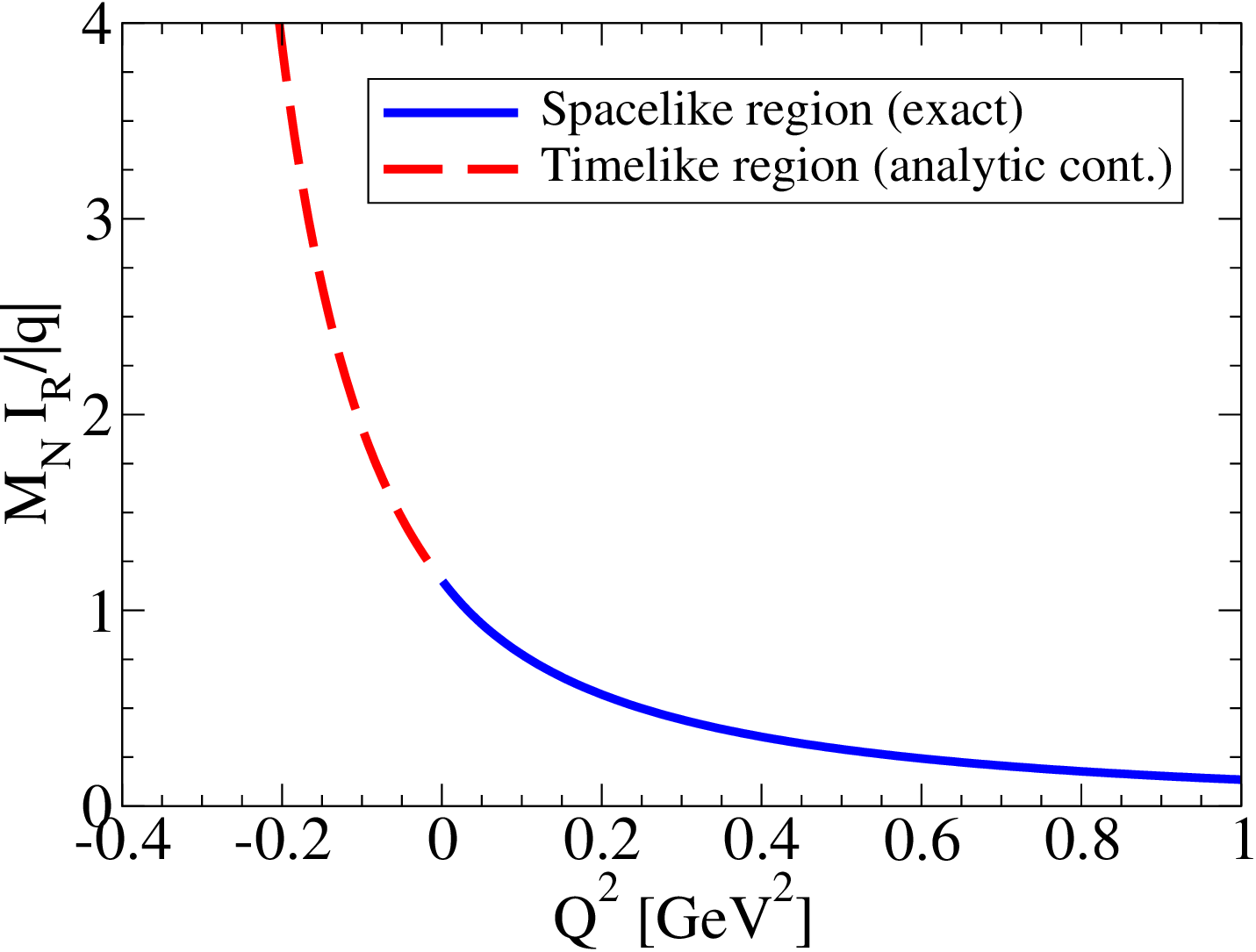}}}
\caption{\footnotesize 
Representation of the function 
 ${\cal I}_R/|{\bf q}|$ in terms of $Q^2$ 
in the spacelike and in the timelike regions,
normalized by the factor $M_N$ (nucleon mass),
for the case $W=1.535$ GeV.
The results in spacelike region are determined 
by the quark model (SR approach). 
The results in the timelike region are 
determined by the analytic continuation based 
on the dipole form $G_D= {\cal C}/(1 + Q^2/\Lambda_D^2)^2$,
with the parameters ${\cal C}=1.154$ and $\Lambda_D^2= 0.4396$ GeV$^2$. }
\label{figInt}       
\end{figure}

Since the dipole approximation (\ref{eqDipole1})
generate necessarily singularities in the 
timelike region at $q^2= \Lambda_D^2$,
it is necessary to regularize the expression 
including some effective width $\Gamma_D$,
according to the replacement
$\Lambda_D^2 \to \Lambda_D^2 - i \Lambda_D \Gamma_D$.
For simplicity, we approximate the dipole function $G_D$,
by the magnitude of $G_D$:
\ba
G_D (q^2)\to {\cal C}
\frac{\Lambda_D^4}{(\Lambda_D^2 - q^2) + \Lambda_D^2 \Gamma_D^2}. 
\ea
The explicit form of the function $\Gamma_D (q^2)$
is discussed in the next subsection.

Combining the expression of  ${\cal I}_R$
in the semirelativistic approximation, 
based on Eq.~(\ref{eqDipole1}):
${\cal I}_R = G_D Q\sqrt{1 + \tau}/M_N$, with 
$Z= Q/\sqrt{m_\rho^2 + Q^2}$,
we obtain
\ba
{\cal I}_R^{\prime}
\to  \frac{G_D Q^2}{M_N(M_R + M_N)} \sqrt{\frac{Q_+^2}{m_\rho^2 + Q^2}}. 
\label{eqInt2}
\ea
This expression is consistent with 
the results from spacelike, 
and ensures then the continuity between spacelike and timelike regions,
provided that the normalization of $G_D$ is correct.

Note, however, that the relation (\ref{eqInt2}) include a singularity 
for $q^2 = m_\rho^2$.
Since in the present study, our applications 
are restricted to the region $M_R < M_N + m_\rho$,
we do not need to deal with the singularity  $q^2 = m_\rho^2$ directly.
Nevertheless, we recall that the singularity $q^2= m_\rho^2$ 
is already present in the quark current (VMD parametrization).
For consistence we regularize also the factor $1/\sqrt{m_\rho^2 -q^2}$,
as the remaining multipoles, according to Eq.~(\ref{eqRegulator}),
using $n=\frac{1}{2}$.

\subsection*{Explicit form for $\Gamma_D(q^2)$}
\label{secGammaD}

For the effective width $\Gamma_D$, 
we follow the regularization of previous works~\cite{Timelike2}
and use the form 
\ba
\Gamma_D (q^2)= \Gamma_D^0 \left( \frac{q^2}{\Lambda_D^2 + q^2}\right)^2
\theta(q^2),
\ea
where $\theta$ is the Heaviside step function.
The parameter $\Gamma_D^0$ define the range 
of influence of the regularization pole $q^2= \Lambda_D^2$.
For very large values of $\Lambda_D$ 
we do not need to worry about the 
regularization and we can use 
the dipole form with $\Gamma_D=0$.
However, for values of  $\Lambda_D$ such that:
$\Lambda_D \approx m_\rho$, as in the present case, 
the results can depend critically of the magnitude of $\Gamma_D^0$.

In order to obtain a timelike extension 
closer to the natural extension of the 
transition form factors (when $\Gamma_D^0 =0$), 
and because the form factors are dominated 
by the imaginary part of the vector meson poles 
($\rho$ and $\omega$) near $q^2= (W-M_N)^2$, 
we choose $\Gamma_D^0 = n \Gamma_\rho^0$,
where $\Gamma_\rho^0$ is the $\rho$-meson physical width,
and $n=1,2,...$
Larger values of $n$ lead to a significant reduction of the 
transition form factors, comparatively 
to an extension with $\Gamma_D^0 =0$, near the pseudothreshold.
Small values of $n$ ($n=1,2$) modify 
the transition form factors only slightly,
except near the pseudothreshold.
Comparing the final results for 
the Dalitz decay width, after the integration on $q$,
we conclude that the results with $n=1,2,3,4$ 
are almost indistinguishable,
showing that the results are almost independent 
of the regulator $\Gamma_D^0$.

In these conditions, we use $\Gamma_D^0 = 2 \Gamma_\rho^0$.
One obtains then transition form factors 
that are not very large near the pseudothreshold 
(as in the cases $n=0$ and $n=1$), 
generating smother functions for the Dalitz decay rates.
As for the Dalitz decay widths, 
the results are almost insensitive to the value of $\Gamma_D^0$.

\section{Regularization of multipoles in the timelike region}
\label{appRegulariza}

To regularize the multipole factors associated 
with a effective cutoff (regulator) $\Lambda$,
based on Eq.~(\ref{eqRegulator})
we follow the procedure from Ref.~\cite{N1520TL},
and include the effective width
\ba
\Gamma_\Lambda (q^2) =
4 \Gamma_\Lambda^0 \left(  
\frac{q^2}{\Lambda^2 + q^2 }
\right)^2 \theta(q^2),
\ea
where $\theta$ is the Heaviside step function,
and $\Gamma_\Lambda^0$ is a constant 
given by $\Gamma_\Lambda^0 = 4 \Gamma_\rho^0 \simeq 0.6$ GeV 
($\Gamma_\rho^0$ is the $\rho$ physical decay constant).

The previous definition ensures that $\Gamma_\Lambda (q^2) =0$
when $q^2 < 0$ and that $\Gamma_\Lambda (q^2)$ 
is continuously extended for $q^2 > 0$.
As a consequence, the results in the spacelike region (where there
are no singularities) are kept unchanged.
The factor  $4 \Gamma_\Lambda^0$ 
was chosen in order to obtain $\Gamma_\Lambda (q^2)=  \Gamma_\Lambda^0$
for $q^2= \Lambda^2$ 
and $\Gamma_\Lambda (q^2)= 4 \Gamma_\Lambda^0$ for very large $q^2$.
Finally, the value of $\Gamma_\Lambda^0$ was chosen 
to avoid very narrow peaks around $q^2= \Lambda^2$.

\end{document}